\documentclass{JHEP3}
\usepackage[english]{babel}
\usepackage{cite}

\usepackage{gensymb}

\usepackage{epsf}
\usepackage{amssymb}
\usepackage{amsmath}
\usepackage{amsfonts}
\usepackage{psfrag,epsfig,graphicx,graphics}

\usepackage{float}

\def\slashchar#1{\setbox0=\hbox{$#1$}
   \dimen0=\wd0
   \setbox1=\hbox{/} \dimen1=\wd1
   \ifdim\dimen0>\dimen1
      \rlap{\hbox to \dimen0{\hfil/\hfil}}
      #1
   \else
      \rlap{\hbox to \dimen1{\hfil$#1$\hfil}}
      /
   \fi}

\def\vsp{\vspace{.5cm}}

\def\bei{\begin{itemize}}
\def\ei{\end{itemize}}

\def\beeq{\begin{eqnarray}} 
\def\beqa{\begin{eqnarray}}
\def\bea{\begin{eqnarray}}

\def\eea{\end{eqnarray}}
\def\eqa{\end{eqnarray}}
\def\eeeq{\end{eqnarray}}

\def\eqar{\end{array}}
\def\beqar{\begin{array}}

\def\beas{\begin{eqnarray*}}
\def\beqas{\begin{eqnarray*}}

\def\eqas{\end{eqnarray*}}
\def\eeas{\end{eqnarray*}}

\def\beq{\begin{equation}} 
\def\be{\begin{equation}}

\def\ee{\end{equation}}
\def\eq{\end{equation}}
\def\eeq{\end{equation}}

\def\beqd{\begin{displaymath}}
\def\eeqd{\end{displaymath}}
\def\eqd{\end{displaymath}}

\def\beeq{\begin{eqnarray}} \def\eeeq{\end{eqnarray}}

\newcommand{\fin}{\end{document}}

\def\pv{\vec{p}_t}
\def\dv{\vec{\Delta}_t}

\def\ar{\alpha_\rho}

\def\mr{m_\rho}

\def\fin{\end{document}}

\newcommand{\alb}{\bar{\alpha}}

\title{Exclusive photoproduction of a $\gamma\,\rho$ pair \\with a large invariant mass}
\author{R.~Boussarie\\
LPT, Universit{\'e} Paris-Sud, CNRS, Universit\'e Paris-Saclay, 91405, Orsay, France\\
Email: \email{renaud.boussarie@th.u-psud.fr}}
\author{B. Pire\\
 Centre de Physique Th\'eorique, Ecole polytechnique, CNRS, Universit\'e Paris-Saclay, 91128 Palaiseau, France  \\
Email: \email{bernard.pire@polytechnique.edu}}
\author{ L. Szymanowski\\
National Center for Nuclear Research (NCBJ), Warsaw, Poland\\
Email: \email{Lech.Szymanowski@ncbj.gov.pl}}
\author{S. Wallon\\
LPT, Universit{\'e} Paris-Sud, CNRS, Universit\'e Paris-Saclay, 91405, Orsay, France {\em \&} \\
UPMC Univ. Paris 06, facult\'e de physique, 4 place Jussieu, 75252 Paris Cedex 05, France\\
Email: \email{wallon@th.u-psud.fr}}

\abstract{Exclusive photoproduction of a $\gamma\,\rho$ pair  in the kinematics where the pair has a large invariant mass and the final nucleon has a small transverse momentum is described in the collinear factorization framework. The scattering amplitude is calculated at leading order in $\alpha_s$ and  the  differential cross sections for the process  where the $\rho-$meson is either longitudinally or transversely polarized are estimated in the kinematics of the JLab~12-GeV experiments.
}

\date{\today}
\begin{document}

\pagestyle{empty}
\newpage

\mbox{}

\pagestyle{plain}

\setcounter{page}{1}
\section{ Introduction}
\label{Sec:Introduction}

The near forward photoproduction of a pair of particles with a large invariant mass is a case for a natural extension of collinear QCD factorization theorems which have been much studied for near forward deeply virtual Compton scattering (DVCS) and deeply virtual meson production~\cite{Goeke:2001tz,Diehl:2003ny,Belitsky:2005qn,Boffi:2007yc,Burkert:2007zz,Guidal:2008zza}. 
In the present paper, we study the case where a wide angle Compton scattering subprocess $\gamma (q\bar q) \to \gamma \rho $ characterized by the large scale $M_{\gamma \rho}$ (the invariant mass of the final state) factorizes from generalized parton distributions (GPDs). This large scale $M_{\gamma \rho}$ is related to the large transverse momenta transmitted to  the final photon and to  the final meson, the pair having an overall small transverse momentum.  This opens a new way to the extraction of these  GPDs 
and thus to check their universality.

The study of such processes was initiated in Ref.~\cite{Ivanov:2002jj,Enberg:2006he}, where the process under study was 
the high energy diffractive photo- (or electro-) production
 of two vector mesons, the hard probe being the virtual "Pomeron" exchange (and the hard scale being the virtuality of this pomeron), in analogy with the virtual photon exchange occuring in the deep inelastic electroproduction of a meson. A similar strategy has also been advocated in Ref.~\cite{Beiyad:2010cxa,Kumano:2009he,Larionov:2016mim} to enlarge the number of processes which could be used to extract information on GPDs.

The process we study here\footnote{Some of the results presented
here have been reported previously
\cite{Boussarie:2015tdh,Boussarie:2016aoq}.
}
\begin{equation}
\gamma^{(*)}(q) + N(p_1) \rightarrow \gamma(k) + \rho^0(p_\rho,\varepsilon_\rho) + N'(p_2)\,,
\label{process1}
\end{equation}
is  sensitive  to both chiral-even and chiral-odd GPDs due to the chiral-even (resp. chiral-odd) character of the leading twist distribution amplitude (DA) of $\rho_L$ (resp. $\rho_T$). 

 Its experimental study should not present major difficulties to large acceptance detectors such as those developed for the 12 GeV upgrade of JLab. The estimated rate depends of course much on the magnitude of the  GPDs, 
 but we show that the experiment is feasible under reasonable assumptions based on
  their relations to usual parton distributions and to  lattice~\cite{Hagler:2003jd,Gockeler:2003jfa,Gockeler:2005cj,Gockeler:2006zu}
 calculations. 

Let us briefly comment on the extension of the existing factorization proofs in the framework of QCD to our process. The  argument is two-folded.

\begin{figure}[h]

\psfrag{TH}{$\Large T_H$}
\psfrag{Pi}{$\pi$}
\psfrag{P1}{$\,\phi$}
\psfrag{P2}{$\,\phi$}
\psfrag{Phi}{$\,\phi$}
\psfrag{Rho}{$\rho$}
\psfrag{tp}{$t'$}
\psfrag{s}{$s$}
\psfrag{x1}{$\!\!\!\!\!\!x+\xi$}
\psfrag{x2}{$\!x-\xi$}
\psfrag{RhoT}{$\rho_T$}
\psfrag{t}{$t$}
\psfrag{N}{$N$}
\psfrag{Np}{$N'$}
\psfrag{M}{$M^2_{\gamma \rho}$}
\psfrag{GPD}{$\!GPD$}

\centerline{
\raisebox{1.6cm}{\includegraphics[width=14pc]{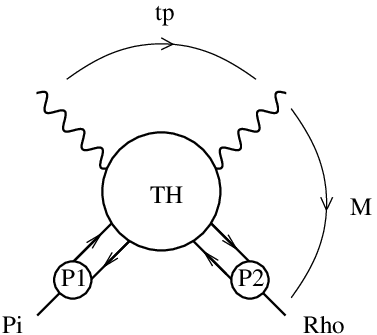}}~~~~~~~~~~~~~~
\psfrag{TH}{$\,\Large T_H$}
\includegraphics[width=14pc]{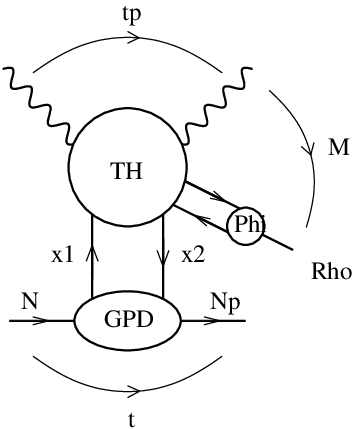}}

\caption{a) Factorization of the amplitude for the process $\gamma + \pi \rightarrow \gamma + \rho $ at large $s$ and fixed angle (i.e. fixed ratio $t'/s$); b) replacing one DA by a GPD leads to the factorization of the amplitude  for $\gamma + N \rightarrow \gamma + \rho +N'$ at large $M_{\gamma\rho}^2$\,.}
\label{Fig:feyndiag}
\end{figure}
The now classical proof of factorization of exclusive scattering at fixed angle and large energy~\cite{Lepage:1980fj} allows to write the leading twist
 amplitude for the process $\gamma + \pi \rightarrow \gamma + \rho $ as the convolution of a mesonic distribution amplitude and a hard scattering subprocess amplitude $\gamma  +( q + \bar q) \rightarrow \gamma + (q + \bar q) $ with the meson state replaced by a collinear quark-antiquark pair. This is described in Fig.~\ref{Fig:feyndiag}a. The demonstration of the absence of any pinch singularity (which is the weak point of the proof for the generic case $A+B\to C+D$) has been proven in the case  of interest here~\cite{Farrar:1989wb}.

We extract from the factorization procedure of the exclusive meson electroproduction amplitude near the forward region~\cite{Collins:1996fb} the right to replace in Fig.~\ref{Fig:feyndiag}a the lower left meson distribution amplitude by a $N \to N'$ GPD, and thus get Fig.~\ref{Fig:feyndiag}b. Indeed the same collinear factorization property bases the validity of the leading twist approximation which replaces either the meson wave function by its distribution amplitude or the $N \to N'$ transition to its GPDs. A slight difference is that light cone fractions ($z, 1- z$) leaving the DA are positive, but the corresponding fractions ($x+\xi,\xi-x$) may be positive or negative in the case of the GPD. Our calculation will show that this difference does not spoil the factorization property, at least at the (leading) order at which we are working here.

The analogy to the timelike Compton scattering process~\cite{Mueller:1998fv,Berger:2001xd,Pire:2011st}:
\begin{equation}
\gamma^{(*)} N  \to \gamma^* N' \to \mu^+ \mu^- N' \,,
\label{process2}
\end{equation}
where the lepton pair has a large squared  invariant mass $Q^2$, is quite instructive.  Although the photon-meson pair in our process  (\ref{process1}) has a more complex momentum flow, one may draw on this analogy to ascribe the role of the hard scale to the photon-meson pair invariant  mass.  

In order for the factorization of a partonic amplitude to be legitimate, one should avoid the dangerous kinematical regions where a small momentum transfer is exchanged in the upper blob, namely small $t' =(k -q)^2$ or small $u'=(p_\rho-q)^2$, and the region where strong  final state interactions between the $\rho$ meson and the nucleon  are dominated by resonance effects, namely where the invariant mass $M^2_{\rho N'} = (p_\rho +p_{N'})^2$ is not large enough.

Our paper is organized as follows. In Sec.~\ref{Sec:Kinematics}, we clarify the  kinematics we are interested in and set our conventions. 
Sec.~\ref{Sec:DAs-GPDs} is devoted to the presentation of our model for DAs and GPDs.
Then, in Sec.~\ref{Sec:Scattering_Amplitude}, we  describe the scattering amplitude of the process under study  in the framework of  QCD factorization.  Sec.~\ref{Sec:Cross-Section-and-Rates} presents our results for  the unpolarized  differential cross section in the kinematics of  quasi-real photon beams at JLab where $S_{\gamma N} \sim$ 6-22 GeV$^2$\,. Finally, in Sec.~\ref{Sec:rates} we give estimates of expected rates at JLab.
In appendices, we describe several technical details required by analytical and numerical aspects of our calculations.

As a final remark in this introduction, let us stress that our discussion applies as well to the case of electroproduction where a moderate virtuality of the initial photon may help to access the perturbative domain with a lower value of the hard scale $M_{\gamma\rho}$.

\section{Kinematics}
\label{Sec:Kinematics}

We study the exclusive photoproduction of a vector meson $\rho^0$ and a real photon on a  polarized or unpolarized proton or neutron target

\begin{equation}
\gamma(q, \varepsilon_q) + N(p_1,\lambda) \rightarrow \gamma(k, \varepsilon_k) + \rho^0(p_\rho,\varepsilon_\rho) + N'(p_2,\lambda')\,,
\label{process}
\end{equation}
 in the kinematical regime of large invariant mass $M_{\gamma\rho}$ of the final photon and meson pair and small momentum transfer $t =(p_2-p_1)^2$ between the initial and the final nucleons. Roughly speaking, these kinematics mean  moderate to large, and approximately opposite, transverse momenta of the final photon and  meson.
Our conventions  are the following. We define
\begin{equation}
P^\mu = \frac{p_1^\mu + p_2^\mu}{2} ~,~ \Delta^\mu = p_2^\mu - p_1^\mu\,,
\end{equation}
and decompose momenta on a Sudakov basis  as
\begin{equation}
\label{sudakov1}
v^\mu = a \, n^\mu + b \, p^\mu + v_\bot^\mu \,,
\end{equation}
with $p$ and $n$ the light-cone vectors
\begin{equation}
\label{sudakov2}
p^\mu = \frac{\sqrt{s}}{2}(1,0,0,1)\qquad n^\mu = \frac{\sqrt{s}}{2}(1,0,0,-1) \qquad p\cdot n = \frac{s}{2}\,,
\end{equation}
and
\begin{equation}
\label{sudakov3}
v_\bot^\mu = (0,v^x,v^y,0) \,, \qquad v_\bot^2 = -\vec{v}_t^2\,.
\end{equation}
The particle momenta read
\begin{equation}
\label{impini}
 p_1^\mu = (1+\xi)\,p^\mu + \frac{M^2}{s(1+\xi)}\,n^\mu~, \quad p_2^\mu = (1-\xi)\,p^\mu + \frac{M^2+\vec{\Delta}^2_t}{s(1-\xi)}n^\mu + \Delta^\mu_\bot\,, \quad q^\mu = n^\mu ~,
\end{equation}
\beqa
\label{impfinc}
k^\mu &=& \alpha \, n^\mu + \frac{(\vec{p}_t-\vec\Delta_t/2)^2}{\alpha s}\,p^\mu + p_\bot^\mu -\frac{\Delta^\mu_\bot}{2}~,\nonumber \\
 p_\rho^\mu &=& \alpha_\rho \, n^\mu + \frac{(\vec{p}_t+\vec\Delta_t/2)^2+m^2_\rho}{\alpha_\rho s}\,p^\mu - p_\bot^\mu-\frac{\Delta^\mu_\bot}{2}\,,
\eqa
with  
$M$, $m_\rho$ the masses of the nucleon and the $\rho$ meson.
From these kinematical relations it follows that
\beq
\label{2xi}
2 \, \xi = \frac{(\pv -\frac{1}2 \dv)^2 }{s \, \alpha} +
\frac{(\pv +\frac{1}2 \dv)^2 + \mr^2}{s \, \ar}
\eq
and
\beq
\label{exp_alpha}
1-\alpha-\ar = \frac{2 \, \xi \, M^2}{s \, (1-\xi^2)} + \frac{\dv^2}{s \, (1-\xi)}\,.
\eq
The total squared center-of-mass energy of the $\gamma$-N system is
\begin{equation}
\label{energysquared}
S_{\gamma N} = (q + p_1)^2 = (1+\xi)s + M^2\,.
\end{equation}
On the nucleon side, the squared transferred momentum is
\begin{equation}
\label{transfmom}
t = (p_2 - p_1)^2 = -\frac{1+\xi}{1-\xi}\vec{\Delta}_t^2 -\frac{4\xi^2M^2}{1-\xi^2}\,.
\end{equation}
The other useful Mandelstam invariants read
\begin{eqnarray}
\label{M_gamma_rho}
s'&=& ~(k +p_\rho)^2 = ~M_{\gamma\rho}^2= 2 \xi \, s \left(1 - \frac{ 2 \, \xi \, M^2}{s (1-\xi^2)}  \right) - \dv^2 \frac{1+\xi}{1-\xi}\,, \\
\label{t'}
- t'&=& -(k -q)^2 =~\frac{(\vec p_t-\vec\Delta_t/2)^2}{\alpha} \;,\\
\label{u'}
- u'&=&- (p_\rho-q)^2= ~\frac{(\vec p_t+\vec\Delta_t/2)^2+(1-\alpha_\rho)\, m_\rho^2}{\alpha_\rho}
 \; ,
\end{eqnarray}
and
\beqa
\label{M_rho_N}
M_{\rho N'}^2 = s\left(1-\xi+ \frac{(\pv+\dv/2)^2+ \mr^2}{s\, \ar}\right)
\left(\ar + \frac{M^2 + \dv^2}{s \, (1-\xi)}  \right) - \left(\pv - \frac{1}2 \dv \right)^2\,.
\eqa

The hard scale $M^2_{\gamma\rho}$ is the invariant squared mass of the ($\gamma$ $\rho^0$) system.  The leading twist calculation of the hard part only involves the approximated kinematics in the generalized Bjorken limit: neglecting $\dv$ in front of $\pv$ as well as hadronic masses, it amounts to
\beqa
\label{skewness2}
M^2_{\gamma\rho} &\approx & \frac{\vec{p}_t^2}{\alpha\bar{\alpha}} \,, 
\\
\ar &\approx& 1-\alpha \equiv \alb \,,\\
\xi &= & \frac{\tau}{2-\tau} ~~~~,~~~~\tau \approx 
\frac{M^2_{\gamma\rho}}{S_{\gamma N}-M^2}\,,\\
-t' & \approx & \bar\alpha\, M_{\gamma\rho}^2  ~~~~,~~~~ -u'  \approx  \alpha\, M_{\gamma\rho}^2 \,.
\eqa
For further details on kinematics, we refer to appendix~\ref{App:kinematics}.

The typical cuts that one should apply are $-t', -u' > \Lambda^2$ and
 $M_{\rho N'}^2= (p_\rho +p_{N'})^2 > M_R^2$ where $\Lambda \gg \Lambda_{QCD}$
and $M_R$ is a typical baryonic resonance mass. This amounts to cuts in
$\alpha $ and $\bar\alpha$ at fixed $M_{\gamma\rho}^2$, which can
be
 translated in terms of $u'$ at fixed $M_{\gamma\rho}^2$ and $t$.
These conditions boil down to a safe kinematical domain $(-u')_{min} \leqslant -u' \leqslant (-u')_{max} $ which we will discuss in more details in Sec.~\ref{Sec:Cross-Section-and-Rates}.

In the following, we will choose as independent kinematical  variables $t, u', M^2_{\gamma \rho}\,.$

Due to electromagnetic gauge invariance, the scattering amplitude for the production of a $\rho_T$ meson with chiral-odd GPDs and the scattering amplitude for the production of a $\rho_L$ meson with chiral-even GPDs are separately gauge invariant, up to the well known corrections of order $\frac{\Delta_T}{\sqrt{s}}$ which have been much studied for the DVCS case \cite{Anikin:2000em,Braun:2012hq}. We choose the  axial gauge $p_\mu\,\varepsilon^\mu_k=0$  and parametrize  the polarization vector of the final photon in terms of its transverse components
\begin{equation}
\label{eps_k}
\varepsilon^\mu_k=\varepsilon^\mu_{k\perp} - \frac{\varepsilon_{k\perp} \cdot k_{\bot} }{p\cdot k}p^\mu\,,
\end{equation}
while the initial photon polarization is simply written as
\begin{equation}
\label{eps_q}
\varepsilon^\mu_q=\varepsilon^\mu_{q\perp} \,.
\end{equation}
We will use the transversity relation $p_\rho \cdot \varepsilon_\rho=0$ to express the polarization of the $\rho$
meson in terms of only its transverse components and its component along $n$, using
\begin{equation}
\label{n.eps}
n \cdot \varepsilon_\rho = \frac{1}{\alpha_\rho} \left[ \frac{p_\perp^2}{\alpha_\rho s} (p \cdot \varepsilon_\rho) +
(p_\perp \cdot \varepsilon_{\rho\perp}) \right].
\end{equation}

\section{Non-perturbative ingredients: DAs and GPDs}
\label{Sec:DAs-GPDs}

In this section, we describe the way the non-perturbative quantities which enter the scattering amplitude are parametrized.

\subsection{Distribution amplitudes for the $\rho$ meson}
\label{SubSec:DAs}

The chiral-even light-cone DA for the longitudinally polarized meson vector $\rho^0_L$  is defined, at the leading twist 2, by the matrix element~\cite{Ball:1996tb}
\begin{equation}
\langle 0|\bar{u}(0)\gamma^\mu u(x)|\rho^0(p_\rho,\varepsilon_{\rho L}) \rangle = \frac{1}{\sqrt{2}}p_\rho^\mu f_{\rho^0}\int_0^1dz\ e^{-izp_\rho\cdot x}\ \phi_{\parallel}(z),
\label{defDArhoL}
\end{equation}
with $f_{\rho^0}=216\,\mbox{MeV}$,
while the chiral-odd light-cone DA for the transversely polarized meson vector $\rho^0_T$  is defined as:
\begin{equation}
\langle 0|\bar{u}(0)\sigma^{\mu\nu}u(x)|\rho^0(p_\rho,\varepsilon_{\rho\pm}) \rangle = \frac{i}{\sqrt{2}}(\varepsilon^\mu_{\rho \pm}\, p^\nu_\rho - \varepsilon^\nu_{\rho \pm}\, p^\mu_\rho)f_\rho^\bot\int_0^1dz\ e^{-izp_\rho\cdot x}\ \phi_\bot(z),
\label{defDArho}
\end{equation}
where $\varepsilon^\mu_{\rho \pm}$ is the $\rho$-meson transverse polarization and with $f_\rho^\bot$ = 160 MeV. The factor $\frac{1}{\sqrt{2}}$ takes into account  the quark structure of the $\rho^0-$meson: $|\rho^0\rangle =\frac{1}{\sqrt{2}}(|u\bar u\rangle -|d\bar d\rangle)$.
We shall use the asymptotic form for the normalized functions $\phi_{\parallel}$ and $\phi_\bot$
\beqa
\label{DA-asymp}
\phi_{\parallel}(z)&=&6 z (1-z)\,, \nonumber \\
\phi_{\bot}(z)&=&6 z (1-z)\,.
\eqa

\subsection{Generalized parton distributions}
\label{SubSec:GPDs}

The chiral-even GPDs of a parton $q$ (here $q = u,\ d$) in the nucleon target   ($\lambda$ and $\lambda'$ are the light-cone helicities of the nucleons with the momenta $p_1$ and $p_2$) are  defined by~\cite{Diehl:2001pm}:
\beqa
&&\langle p(p_2,\lambda')|\, \bar{q}\left(-\frac{y}{2}\right)\,\gamma^+q \left(\frac{y}{2}\right)|p(p_1,\lambda) \rangle \\ \nonumber 
&&= \int_{-1}^1dx\ e^{-\frac{i}{2}x(p_1^++p_2^+)y^-}\bar{u}(p_2,\lambda')\, \left[ \gamma^+ H^{q}(x,\xi,t)   +\frac{i}{2m}\sigma^{+ \,\alpha}\Delta_\alpha  \,E^{q}(x,\xi,t) \right]
u(p_1,\lambda)\,,
\label{defGPDEvenV}
\eqa
and
\beqa
&&\langle p(p_2,\lambda')|\, \bar{q}\left(-\frac{y}{2}\right)\,\gamma^+ \gamma^5 q\left(\frac{y}{2}\right)|p(p_1,\lambda)\rangle \\ \nonumber
&&= \int_{-1}^1dx\ e^{-\frac{i}{2}x(p_1^++p_2^+)y^-}\bar{u}(p_2,\lambda')\, \left[ \gamma^+ \gamma^5 \tilde H^{q}(x,\xi,t)   +\frac{1}{2m}\gamma^5 \Delta^+  \,\tilde E^{q}(x,\xi,t) \right]
u(p_1,\lambda)\,.
\label{defGPDEvenA}
\eqa

The transversity GPD of a quark $q$   is defined by:
\beqa
&&\langle p(p_2,\lambda')|\, \bar{q}\left(-\frac{y}{2}\right)i\,\sigma^{+j} q \left(\frac{y}{2}\right)|p(p_1,\lambda)\rangle \\ \nonumber
&&= \int_{-1}^1dx\ e^{-\frac{i}{2}x(p_1^++p_2^+)y^-}\bar{u}(p_2,\lambda')\, \left[i\,\sigma^{+j}H_T^{q}(x,\xi,t)
+\dots
\right]u(p_1,\lambda)\,,
\label{defGPD}
\eqa
where $\dots$ denote the remaining three chiral-odd GPDs which contributions are omitted in the present analysis.

We parametrize the GPDs in terms of double distributions (DDs)
\cite{Radyushkin:1998es}
\begin{equation}
\label{DDdef}
H^q(x,\xi,t=0) = \int_\Omega d\beta\, d\alpha\ \delta(\beta+\xi\alpha-x) \,{\cal F}^q(\beta,\alpha,t=0) \,,
\end{equation}
where ${\cal F}^q$ is a generic  quark  DD and $\Omega = \{|\beta|+|\alpha| \leqslant 1\}$ is its support domain. A $D$-term contribution, necessary to be completely  general while fulfilling the polynomiality constraints, could be added. In our parameterization, we do not include such an arbitrary term. Note that similar GPD parameterizations have been used in Ref.~\cite{Goloskokov:2011rd}.

As shown in Sec.~\ref{SubSec:Square}, with a good approximation we will only
use the three GPDs $H,$ $\tilde{H}$ and $H_T$.
We adhere on Radyushkin-type parameterization and write the unpolarized DD
 $f^q$ and the transversity DD $f_T^q$ in the form
\begin{equation}
\label{DD-t}
f^q(\beta,\alpha,t=0) = \Pi(\beta,\alpha)\, q(\beta)\Theta(\beta) - \Pi(-\beta,\alpha)\,\bar{q}(-\beta)\,\Theta(-\beta)\,,
\end{equation}
and~\cite{Beiyad:2010cxa}
\begin{equation}
\label{DD-fT}
f_T^q(\beta,\alpha,t=0) = \Pi(\beta,\alpha)\, \delta q(\beta)\,\Theta(\beta) - \Pi(-\beta,\alpha)\,\delta\bar{q}(-\beta)\,\Theta(-\beta)\,,
\end{equation}
while the polarized DD $\tilde{f}^q$ reads
\begin{equation}
\label{DD-ftilde}
\tilde{f}^q(\beta,\alpha,t=0) = \Pi(\beta,\alpha)\, \Delta q(\beta)\,\Theta(\beta) + \Pi(-\beta,\alpha)\,\Delta\bar{q}(-\beta)\,\Theta(-\beta)\,,
\end{equation}
where $ \Pi(\beta,\alpha) = \frac{3}{4}\frac{(1-\beta)^2-\alpha^2}{(1-\beta)^3}$ is a profile function and $q$, $\bar{q}$
are the quark and antiquark unpolarized parton distribution functions (PDFs), $\Delta q$, $\Delta \bar{q},$
are the quark and antiquark polarized PDFs and 
$\delta q$, $\delta \bar{q}$
are the quark and antiquark transversity PDFs.

We now give specific formulas for the three GPDs which we use in the present study.
The GPD $H^q$ reads
\begin{eqnarray}
H^q(x,\xi,t=0) &=& \Theta(x>\xi)\int_{\frac{-1+x}{1+\xi}}^{\frac{1-x}{1-\xi}}dy\ \frac{3}{4}\frac{(1-x+\xi y)^2-y^2}{(1-x+\xi y)^3} q(x-\xi y) \nonumber \\
&+&  \Theta(\xi>x>-\xi)\left[\int_{\frac{-1+x}{1+\xi}}^{\frac{x}{\xi}}dy\ \frac{3}{4}\frac{(1-x+\xi y)^2-y^2}{(1-x+\xi y)^3} q(x-\xi y) \right. \nonumber \\
&-& \left. \int_{\frac{x}{\xi}}^{\frac{1+x}{1+\xi}}dy\ \frac{3}{4}\frac{(1+x-\xi y)^2-y^2}{(1+x-\xi y)^3} \bar{q}(-x+\xi y) \right] \nonumber \\
&-& \Theta(-\xi>x)\int_{-\frac{1+x}{1-\xi}}^{\frac{1+x}{1+\xi}}dy\ \frac{3}{4}\frac{(1+x-\xi y)^2-y^2}{(1+x-\xi y)^3} \bar{q}(-x+\xi y)\,.
\label{DD-H}
\end{eqnarray}
Similarly, the transversity GPD $H_T^q$ reads
\begin{eqnarray}
H_T^q(x,\xi,t=0) &=& \Theta(x>\xi)\int_{\frac{-1+x}{1+\xi}}^{\frac{1-x}{1-\xi}}dy\ \frac{3}{4}\frac{(1-x+\xi y)^2-y^2}{(1-x+\xi y)^3}\delta q(x-\xi y) \nonumber \\
&+&  \Theta(\xi>x>-\xi)\left[\int_{\frac{-1+x}{1+\xi}}^{\frac{x}{\xi}}dy\ \frac{3}{4}\frac{(1-x+\xi y)^2-y^2}{(1-x+\xi y)^3}\delta q(x-\xi y) \right. \nonumber \\
&-& \left. \int_{\frac{x}{\xi}}^{\frac{1+x}{1+\xi}}dy\ \frac{3}{4}\frac{(1+x-\xi y)^2-y^2}{(1+x-\xi y)^3}\delta \bar{q}(-x+\xi y) \right] \nonumber \\
&-& \Theta(-\xi>x)\int_{-\frac{1+x}{1-\xi}}^{\frac{1+x}{1+\xi}}dy\ \frac{3}{4}\frac{(1+x-\xi y)^2-y^2}{(1+x-\xi y)^3}\delta \bar{q}(-x+\xi y)\,,
\label{DD-HT}
\end{eqnarray}
while the GPD $\tilde{H}^q$ reads
\begin{eqnarray}
\tilde{H}^q(x,\xi,t=0) &=& \Theta(x>\xi)\int_{\frac{-1+x}{1+\xi}}^{\frac{1-x}{1-\xi}}dy\ \frac{3}{4}\frac{(1-x+\xi y)^2-y^2}{(1-x+\xi y)^3}\Delta q(x-\xi y) \nonumber \\
&+&  \Theta(\xi>x>-\xi)\left[\int_{\frac{-1+x}{1+\xi}}^{\frac{x}{\xi}}dy\ \frac{3}{4}\frac{(1-x+\xi y)^2-y^2}{(1-x+\xi y)^3}\Delta q(x-\xi y) \right. \nonumber \\
&+& \left. \int_{\frac{x}{\xi}}^{\frac{1+x}{1+\xi}}dy\ \frac{3}{4}\frac{(1+x-\xi y)^2-y^2}{(1+x-\xi y)^3}\Delta \bar{q}(-x+\xi y) \right] \nonumber \\
&+& \Theta(-\xi>x)\int_{-\frac{1+x}{1-\xi}}^{\frac{1+x}{1+\xi}}dy\ \frac{3}{4}\frac{(1+x-\xi y)^2-y^2}{(1+x-\xi y)^3}\Delta \bar{q}(-x+\xi y)\,.
\label{DD-Htilde}
\end{eqnarray}
Since our process selects the exchange of charge conjugation $C=-1$ in the $t-$channel, we now consider the corresponding valence GPDs
\beqa
\label{def:H-}
H^{q(-)}(x,\xi,t)=H^{q}(x,\xi,t)+H^{q}(-x,\xi,t)
\eqa
and 
\beqa
\label{def:HT-}
H_T^{q(-)}(x,\xi,t)=H_T^{q}(x,\xi,t)+H_T^{q}(-x,\xi,t)
\eqa
which have the symmetry properties
$H^{q(-)}(x,\xi,t)=H^{q(-)}(-x,\xi,t)$ and $H_T^{q(-)}(x,\xi,t)=H_T^{q(-)}(-x,\xi,t),$
as well as the valence GPD
\beqa
\label{def:HTilde-}
\tilde{H}^{q(-)}(x,\xi,t)=\tilde{H}^{q}(x,\xi,t)-\tilde{H}^{q}(-x,\xi,t)\,,
\eqa
which has the antisymmetry property $\tilde{H}^{q(-)}(x,\xi,t)=-\tilde{H}^{q(-)}(-x,\xi,t)\,.$

Introducing the 
symmetric valence distributions 
\beqa
\label{Def:q-valence}
q_{\rm val}(x) = \theta(x) [q(x) - \bar{q}(x)] 
+ \theta(-x) [q(-x) - \bar{q}(-x)]\, 
\eqa
and
\beqa
\label{Def:deltaq-valence}
\delta q_{\rm val}(x) = \theta(x) [\delta q(x) - \delta \bar{q}(x)] 
+ \theta(-x) [\delta q(-x) - \delta \bar{q}(-x)]\,, 
\eqa
and the antisymmetric valence distribution
\beqa
\label{Def:Deltaq-valence}
\Delta q_{\rm val}(x) = \theta(x) [\Delta q(x) - \Delta \bar{q}(x)] 
- \theta(-x) [\Delta q(-x) - \Delta \bar{q}(-x)]\,,
\eqa
the set of GPDs which we use in our computation of the scattering amplitude reads
\begin{eqnarray}
\label{DD-H-val}
\frac{1}2 H^{q(-)}(x,\xi,t=0) &=& \frac{1}2 \left[\left(\Theta(x>\xi)\int_{\frac{-1+x}{1+\xi}}^{\frac{1-x}{1-\xi}}dy\ \frac{3}{4}\frac{(1-x+\xi y)^2-y^2}{(1-x+\xi y)^3} q_{\rm val}(x-\xi y) \right. \right. \nonumber \\
&+& \left. \left. \Theta(\xi>x>-\xi) \int_{\frac{-1+x}{1+\xi}}^{\frac{x}{\xi}}dy\ \frac{3}{4}\frac{(1-x+\xi y)^2-y^2}{(1-x+\xi y)^3} \, q_{\rm val}(x-\xi y)\right) \right.\nonumber \\
&+& \left. (x \leftrightarrow  -x) \right]\,,
\end{eqnarray}
\begin{eqnarray}
\label{DD-HT-val}
\frac{1}2 H_T^{q(-)}(x,\xi,t=0) &=& \frac{1}2 \left[\left(\Theta(x>\xi)\int_{\frac{-1+x}{1+\xi}}^{\frac{1-x}{1-\xi}}dy\ \frac{3}{4}\frac{(1-x+\xi y)^2-y^2}{(1-x+\xi y)^3}\delta  q_{\rm val}(x-\xi y) \right. \right. \nonumber \\
&+& \left. \left. \Theta(\xi>x>-\xi) \int_{\frac{-1+x}{1+\xi}}^{\frac{x}{\xi}}dy\ \frac{3}{4}\frac{(1-x+\xi y)^2-y^2}{(1-x+\xi y)^3} \, \delta q_{\rm val}(x-\xi y)\right) \right.\nonumber \\
&+& \left. (x \leftrightarrow  -x) \right]\,,
\end{eqnarray}
and
\begin{eqnarray}
\label{DD-Htilde-val}
\frac{1}2\tilde{H}^{q(-)}(x,\xi,t=0) &=& \frac{1}2 \left[\left(\Theta(x>\xi)\int_{\frac{-1+x}{1+\xi}}^{\frac{1-x}{1-\xi}}dy\ \frac{3}{4}\frac{(1-x+\xi y)^2-y^2}{(1-x+\xi y)^3}\Delta  q_{\rm val}(x-\xi y) \right. \right. \nonumber \\
&+& \left. \left. \Theta(\xi>x>-\xi) \int_{\frac{-1+x}{1+\xi}}^{\frac{x}{\xi}}dy\ \frac{3}{4}\frac{(1-x+\xi y)^2-y^2}{(1-x+\xi y)^3} \, \Delta q_{\rm val}(x-\xi y)\right) \right.\nonumber \\
&-& \left. (x \leftrightarrow  -x) \right]\,.
\end{eqnarray}

\subsection{Numerical modeling}

For the various PDFs, we neglect any QCD evolution (in practice, we take a fixed  factorization scale $\mu_F^2=10~{\rm GeV}^2$) and we
use the following models:
\begin{itemize}
\item 
For $x q(x)$, we rely on the GRV-98 parameterization~\cite{Gluck:1998xa}, as made available from the Durham database. 
To evaluate the uncertainty of this parameterization, we also consider a few other sets of PDFs, namely MSTW2008lo and MSTW2008nnlo~\cite{Martin:2009iq}, ABM11nnlo~\cite{Alekhin:2012ig}, CT10nnlo~\cite{Gao:2013xoa}.

\psfrag{H}{$x$}

\begin{figure}[!h]
\begin{center}
\psfrag{T}{}
\psfrag{V}{\raisebox{.3cm}{\scalebox{1}{$\hspace{-.4cm}\displaystyle H^{u(-)}(x,\xi)$}}}
\hspace{.2cm}\includegraphics[width=7.3cm]{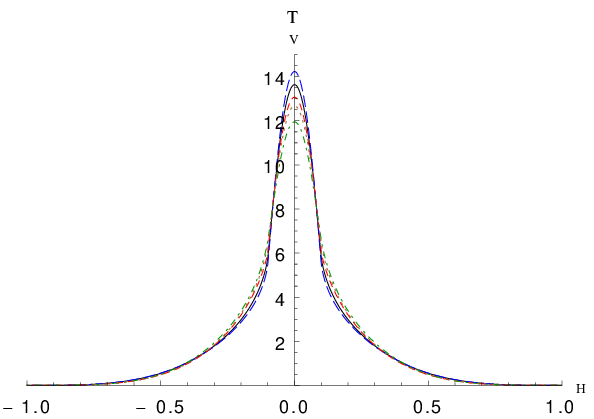}
\psfrag{T}{}
\psfrag{V}{\raisebox{.3cm}{\scalebox{1}{$\hspace{-.4cm}\displaystyle H^{d(-)}(x,\xi)$}}}
\hspace{0.1cm}\includegraphics[width=7.3cm]{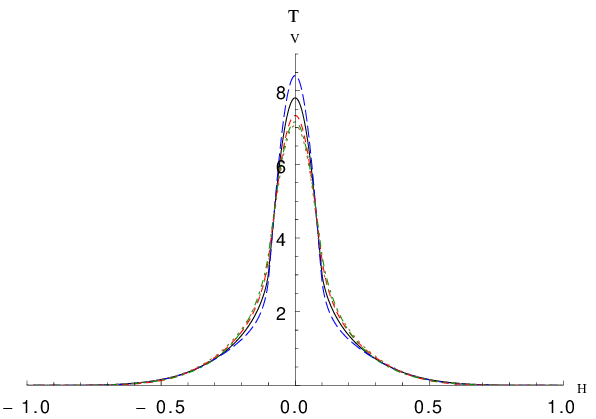}
\vspace{.2cm}
\caption{Models for the GPDs $H^{u(-)}$ and $H^{d(-)}$ for $\xi=.1$,
a value corresponding to $S_{\gamma N}=20~{\rm GeV}^2$ and $M^2_{\gamma\rho}=3.5~{\rm GeV}^2$. 
The various curves differ with respect to the ans\"atze for the PDFs $q$: GRV-98 (solid black), MSTW2008lo (long-dashed blue), MSTW2008nnlo (short-dashed red), ABM11nnlo (dotted-dashed green), CT10nnlo (dotted brown). Note that the two GPDs $H^{d(-)}$ based on these two last ans\"atze are hardly distinguishable.
}
\label{Fig:GPD-H}
\end{center}
\end{figure}
In Fig.~\ref{Fig:GPD-H},
we show the resulting GPDs $H^{u(-)}$ and $H^{d(-)}$ for $\xi=.1$ corresponding in our process to the typical value $S_{\gamma N}=20~{\rm GeV}^2$ and $M^2_{\gamma\rho}=3.5~{\rm GeV}^2$.

\item
For $x \Delta q(x)\,,$ we rely on the  GRSV-2000 parameterization~\cite{Gluck:2000dy}, as made available from the Durham database. Two scenarios are proposed in this parameterization: the ``standard'', {\it i.e.} with 
flavor-symmetric light sea quark and
antiquark distributions, and the ``valence'' scenario with a completely flavor-asymmetric 
light sea densities. 
We use both of them in order to evaluate the order of magnitude of the theoretical uncertainty.
\psfrag{H}{$x$}

\begin{figure}[!h]
\begin{center}
\psfrag{T}{}
\psfrag{V}{\raisebox{.3cm}{\scalebox{1}{$\hspace{-.4cm}\displaystyle \tilde{H}^{u(-)}(x,\xi)$}}}
\hspace{.2cm}\includegraphics[width=7.3cm]{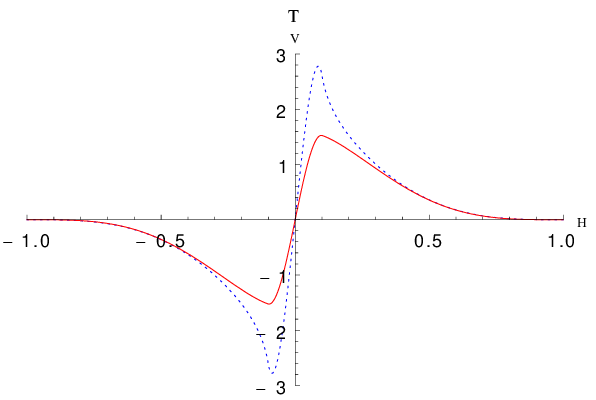}
\psfrag{T}{}
\psfrag{V}{\raisebox{.3cm}{\scalebox{1}{$\hspace{-.4cm}\displaystyle \tilde{H}^{d(-)}(x,\xi)$}}}
\hspace{0.1cm}\includegraphics[width=7.3cm]{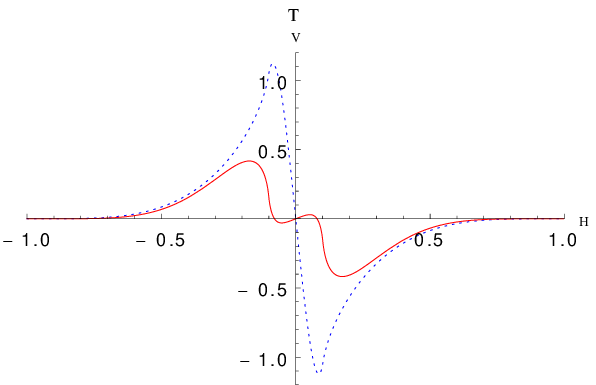}
\vspace{.2cm}
\caption{Models for the GPDs $\tilde{H}^{u(-)}$ and $\tilde{H}^{d(-)}$ for $\xi=.1$,
a value corresponding to $S_{\gamma N}=20~{\rm GeV}^2$ and $M^2_{\gamma\rho}=3.5~{\rm GeV}^2$. In dotted blue, the ``standard'' scenario and in red the ``valence'' scenario.
}
\label{Fig:GPD-HTilde}
\end{center}
\end{figure}

In Fig.~\ref{Fig:GPD-HTilde},
we show the resulting GPDs $\tilde{H}^{u(-)}$ and $\tilde{H}^{u(-)}$ for $\xi=.1$ corresponding in our process to the typical value $S_{\gamma N}=20~{\rm GeV}^2$ and $M^2_{\gamma\rho}=3.5~{\rm GeV}^2$.

\item
For  $x \delta q(x)$ we rely on a parameterization performed for TMDs (based on a fit of azimuthal asymmetries in semi-inclusive deep inelastic scattering), from which the transversity PDFs $x \delta q(x)$ are obtained as a limiting case~\cite{Anselmino:2013vqa}. The parameterization of Ref.~\cite{Anselmino:2013vqa} for TMDs is based on the GRV-98 PDF $x \Delta q(x)$ and GRSV-2000
PDF $x \Delta q(x).$
These transversity PDFs are parametrized as
\beqa
\label{Deltaq}
\delta q(x)=\frac{1}2 {\cal N}^T_q(x) [q(x) + \Delta(x)]
\eqa
with
\beqa
\label{param-Deltaq}
{\cal N}^T_q(x) = N_q^T x^\alpha (1-x)^\beta \frac{(\alpha+\beta)^{(\alpha+\beta)}}{\alpha^\alpha \beta^\beta}\,.
\eqa
%

\psfrag{H}{$x$}
\begin{figure}[!h]
\begin{center}
\psfrag{T}{}
\psfrag{V}{\raisebox{.3cm}{\scalebox{1}{$\hspace{-.4cm}\displaystyle H_T^{u(-)}(x,\xi)$}}}
\hspace{.2cm}\includegraphics[width=7.3cm]{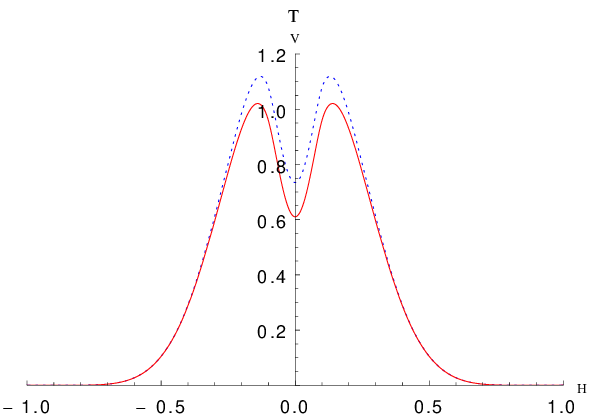}
\psfrag{T}{}
\psfrag{V}{\raisebox{.3cm}{\scalebox{1}{$\hspace{-.4cm}\displaystyle H_T^{d(-)}(x,\xi)$}}}
\hspace{0.1cm}\includegraphics[width=7.3cm]{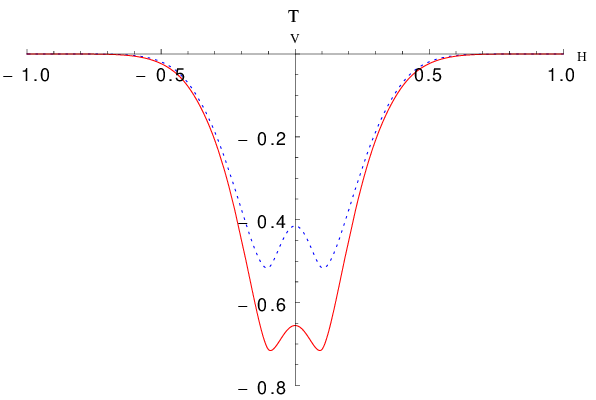}
\vspace{.2cm}
\caption{Models for the GPDs $H_T^{u(-)}$ and $H_T^{d(-)}$ for $\xi=.1$,
a value corresponding to $S_{\gamma N}=20~{\rm GeV}^2$ and $M^2_{\gamma\rho}=3.5~{\rm GeV}^2$.  In blue dotted the ``standard'' scenario and in red the ``valence'' scenario.
}
\label{Fig:GPD-HT}
\end{center}
\end{figure}

Since this parameterization itself relies on the knowledge of  $x q(x)$ and $x \Delta q(x),$ we 
will evaluate the uncertainty on these PDFs by two means: first by passing from the ``standard'' to the ``valence''
polarized PDFs (see above), second by performing a variation of the set of parameters $N_q^T, \alpha, \beta,$ using the  $\chi^2$ distribution of these parameters as used in Ref.~\cite{Anselmino:2013vqa}\footnote{We thank S.~Melis for providing us the complete set of parameters with the corresponding $\chi^2$ distribution.}. We further discuss our procedure in Sec.~\ref{SubSec:chiral-odd}.

In Fig.~\ref{Fig:GPD-HT},
we show the resulting GPDs $H_T^{u(-)}$ and $H_T^{d(-)}$ for $\xi=.1$ corresponding in our process to the typical value $S_{\gamma N}=20~{\rm GeV}^2$ and $M^2_{\gamma\rho}=3.5~{\rm GeV}^2$.

\end{itemize}

In order to evaluate the scattering amplitudes of our process, we calculate, for each of the above three types of GPDs, sets of $u$ and $d$ quarks GPDs indexed by $M^2_{\gamma\rho}$, {\it i.e.} ultimately by $\xi$ given by 
\beqa
\label{rel-xi-M2-S}
\xi = \frac{M^2_{\gamma \rho}}{2(S_{\gamma N}-M^2)-M^2_{\gamma \rho}}
\eqa
We vary $M^2_{\gamma \rho}$ from $2.2$~GeV~$^2$ up to
$10$~GeV~$^2$, with a step of $0.1$~GeV~$^2$, in order to have a full coverage of $M^2_{\gamma\rho}$ for the case $S_{\gamma N}=20$~GeV~$^2$, see appendix~\ref{App:phase}.

For each $M^2_{\gamma \rho}$, the GPDs are computed as tables of 1000 values for $x$ from $-1$ to $1.$ Figs.~\ref{Fig:GPD-H}, \ref{Fig:GPD-HTilde} and \ref{Fig:GPD-HT} are examples of these sets.

\section{The Scattering Amplitude}
\label{Sec:Scattering_Amplitude}

\subsection{Analytical part}
\label{SubSec:Scattering_Amplitude-analytical}

 We now pass to the computation of the scattering amplitude of the process (\ref{process}).
 When the hard scale is large enough, it is possible to study it in the framework of collinear QCD factorization, where the squared invariant mass of the ($\gamma$, $\rho^0$) system $M^2_{\gamma \rho}$ is taken as the factorization scale. We write the scattering amplitude of the process (\ref{process}), taking into account the fact that the  $\rho^0$ meson is described as $\frac{u\bar{u}-d\bar{d}}{\sqrt{2}}$:
\beq
\label{AmplitudeFactorized}
\mathcal{M}_{\parallel, \bot}(t,M^2_{\gamma\rho},u')  =\frac{1}{\sqrt{2}} (\mathcal{M}^u_{\parallel, \bot} - \mathcal{M}^d_{\parallel, \bot})
\eq
where $\mathcal{M}^u_{\parallel, \bot}$ and $\mathcal{M}^d_{\parallel, \bot}$ are expressed in terms of  form factors ${\cal H}^{q}, {\cal E}, \tilde {\cal H}^{q}, \tilde {\cal E}^{q}$ and ${\cal H}_{T\perp j}^{q}, \tilde {\cal H}_{T\perp j}^{q},$  ${\cal E}_{T\perp j}^{q}, \tilde {\cal E}_{T\perp j}^{q}$, analogous to Compton form factors in DVCS, in the factorized form and read
  \begin{eqnarray}
 \mathcal{M}_{\parallel}^q \equiv 
  \frac{1}{n\cdot p}\bar{u}(p_2,\lambda') \!\! \left[\,   \hat n  {\cal H}^{q}(\xi,t) +\frac{i\,\sigma^{n\,\alpha}\Delta_\alpha}{2m}  {\cal E}^{q}(\xi,t) +   \hat n\gamma^5  \tilde {\cal H}^{q}(\xi,t)
  + \frac{n\cdot \Delta}{2m} \,\gamma^5\, \tilde {\cal E}^{q}(\xi,t)
 \right] \!\! u(p_1,\lambda) \!\!\!\!\!\nonumber
 \\
  \label{CEGPD}
  \end{eqnarray}
  in the chiral-even case, and
  
   \begin{eqnarray}
 \mathcal{M}_{\bot}^{q} &\equiv& \frac{1}{n\cdot p}\,\bar{u}(p_2,\lambda')\, \left[i\,\sigma^{nj}{\cal H}_{T\perp j}^{q}(\xi,t)
+ \frac{P\cdot n\;\Delta^j-\Delta\cdot n\;P^j}{m^2} \tilde {\cal H}_{T\perp j}^{q}(\xi,t) \right.
\nonumber \\
&& \left.
 + \frac{\gamma\cdot n\;\Delta^j-\Delta\cdot n\;\gamma^j}{2m}{\cal E}_{T\perp j}^{q}(\xi,t)
 + \frac{\gamma\cdot n\;P^j-P\cdot n\;\gamma^j}{m} \tilde {\cal E}_{T\perp j}^{q}(\xi,t) \right]u(p_1,\lambda)
  \label{COGPD}
  \end{eqnarray}
  in the chiral-odd case.

For convenience, we now define
\begin{eqnarray}
\label{Def:T}
\mathcal{M}^q(t,M^2_{\gamma\rho},p_T) = & \equiv & \int_{-1}^1 \! dx \int_0^1 \! dz \, T^q(t,M^2_{\gamma\rho},p_T,x,z)\,.
\end{eqnarray}

\def\diagici{2.65cm}
\begin{figure}[h]
\begin{center}
\psfrag{z}{\begin{small} $z$ \end{small}}
\psfrag{zb}{\raisebox{0cm}{ \begin{small}$\bar{z}$\end{small}} }
\psfrag{gamma}{\raisebox{+.1cm}{ $\,\gamma$} }
\psfrag{pi}{$\,\pi$}
\psfrag{rho}{$\,\rho$}
\psfrag{TH}{\hspace{-0.2cm} $T_H$}
\psfrag{tp}{\raisebox{.5cm}{\begin{small}     $t'$       \end{small}}}
\psfrag{s}{\hspace{.6cm}\begin{small}$s$ \end{small}}
\psfrag{Phi}{ \hspace{-0.3cm} $\phi$}
\hspace{-.4cm}
\begin{tabular}{ccccc}
\includegraphics[width=\diagici]{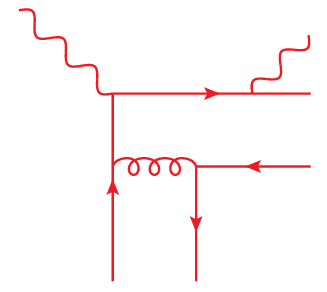}
&
\includegraphics[width=\diagici]{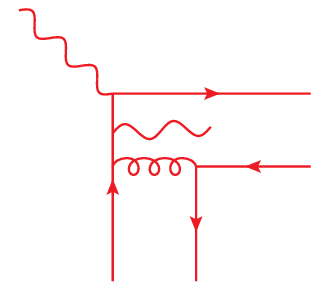}
&
\includegraphics[width=\diagici]{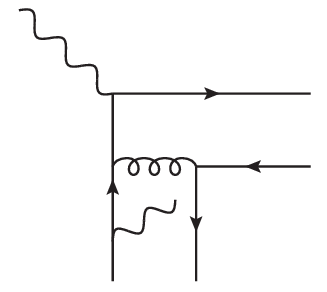}
&
\includegraphics[width=\diagici]{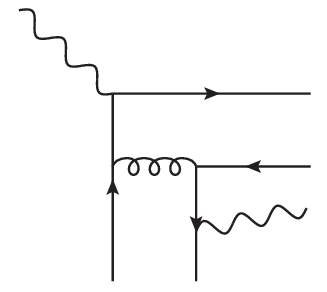}
&
\includegraphics[width=\diagici]{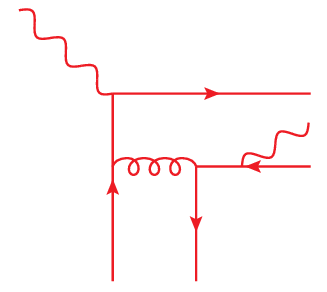} \\
$A_1$ & $A_2$ & $A_3$ & $A_4$ & $A_5$ \\
\\
\includegraphics[width=\diagici]{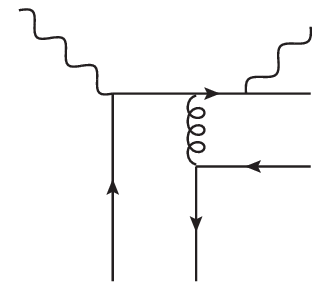}
&
\includegraphics[width=\diagici]{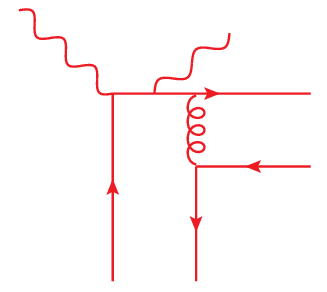}
&
\includegraphics[width=\diagici]{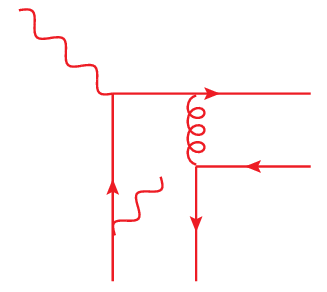}
&
\includegraphics[width=\diagici]{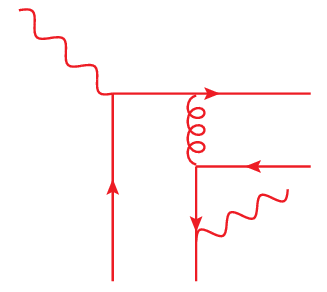}
&
\includegraphics[width=\diagici]{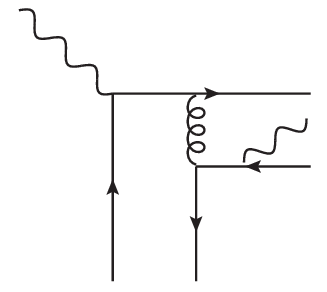}
\\
$B_1$ & $B_2$ & $B_3$ & $B_4$ & $B_5$
\end{tabular}
\caption{Half of the Feynman diagrams contributing to the hard amplitude. In the chiral-odd case, $A_3$, $A_4$ and $B_1$, $B_5$ are the only contributing diagrams (the red diagrams cancel in this case).}
\label{Fig:diagrams}
\end{center}
\end{figure}

The scattering sub-process is described at lower twist by 20 Feynman diagrams, but using the $q \leftrightarrow \bar{q}$ (anti)symmetry properties allows one to only compute 10 of them, shown in Fig.~\ref{Fig:diagrams}, then deduce the remaining contributions by substituting $(x,z) \leftrightarrow (-x, 1-z)$. \\
In the case of $(\gamma, \rho_L )$ production all the diagrams contribute. In the case of $(\gamma, \rho_\bot )$ production, due to the chiral-odd structure of DAs and GPDs, there are only 8 non-vanishing diagrams, out of which one only needs to compute $B_1$, $A_3$, $A_4$ and $B_5$.

 We now discuss diagram $B_1$ in some details, and give the results for the other diagrams in appendix~\ref{App:diagrams}.

 The chiral-even scattering amplitudes for longitudinally polarized $\rho^0$ described by the DA (\ref{defDArhoL}) involve both the vector GPDs (\ref{defGPDEvenV}) and the axial GPDs (\ref{defGPDEvenA}). We now give the detailed expressions for $T^{q\,CE}_V(B_1)$, $T^{q\,CE}_A(B_1),$ for a quark with flavor $q$ and for diagram $B_1$ in Feynman gauge. The vector amplitude reads
\begin{eqnarray}
 && T^{q\,CE}_V(B_1)=  \,\frac{1}{i}\,\frac{tr(t^at^a)}{(4N)^2}\,f_\rho\,\phi_{||}(z)\,(-ieQ_q)^2\,(-ig)^2\,i^2\,(-i)
 \nonumber \\  
 && \times \, tr_D\left[   \hat p_\rho  \hat \varepsilon_k^*\frac{\hat k +z\hat p_\rho}{(k+z p_\rho)^2+i\epsilon } \gamma^\mu \frac{\hat q+(x+\xi)p}{(q+(x+\xi)p)^2+i\epsilon}
 \,\hat \varepsilon_q\,\hat p \,\gamma_\mu\,\frac{1}{(\bar{z} p_\rho+(x-\xi)p)^2+i\epsilon}\right]
 \nonumber \\
 && \times  \, \frac{2}{s}\,\bar{u}(p_2,\lambda')\, \left[ \hat n H^{q}(x,\xi,t) + \frac{i}{2m} \sigma^{n\,\alpha}\Delta_\alpha  \,E^{q}(x,\xi,t) \right]u(p_1,\lambda)
\\ \nonumber 
 && =\,C^{q\,CE}\,tr_D^V\left[ B_1\right]\,\phi_{||}(z)\,\frac{2}{s}\,\bar{u}(p_2,\lambda')\, \left[ \hat n\, H^{q}(x,\xi,t)   + \frac{i}{2m} \sigma^{n\,\alpha}\Delta_\alpha  \,E^{q}(x,\xi,t)   \right]u(p_1,\lambda)\,,
 \label{CEVB_1}
  \end{eqnarray}
which includes all non trivial factors (vertices as well as quark and gluon propagators) of the hard part of diagram $B_1.$
Here, $C^{q\,CE}$ is a common coefficient for  all diagrams involving vector and axial GPDs, reading 
   \begin{equation}
 C^{q\,CE} =  \frac{4}{9}\,f_\rho \, \alpha_{em}\,\alpha_s\,\pi^2\,Q_q^2\; .
 \label{coefCE}
 \end{equation}   
The trace reads:
  \begin{eqnarray}
  tr_D^V\left[ B_1\right] 
&=&   tr_D\left[   \hat p_\rho  \hat \varepsilon_k^*\frac{\hat k +z\hat p_\rho}{(k+z p_\rho)^2+i\epsilon } \gamma^\mu \frac{\hat q+(x+\xi)p}{(q+(x+\xi)p)^2+i\epsilon}
 \,\hat \varepsilon_q\,\hat p \,\gamma_\mu\,\frac{1}{(\bar{z} p_\rho+(x-\xi)p)^2+i\epsilon}\right]
 \nonumber\\
 &=& \frac{    8s\left[    -s\xi \alpha \, (\varepsilon_{q\bot}\cdot \varepsilon_{k\perp}^*) + \frac{z}{\alpha}  \,  (\varepsilon_{q\perp}\cdot p_{\rho \bot})\, ( \varepsilon_{k\perp}^*\cdot p_{\rho \bot} ) \right]                          }{    ( (k+z p_\rho)^2+i\epsilon )       ( (q+(x+\xi)p)^2+i\epsilon )     ( (\bar{z} p_\rho+(x-\xi)p)^2+i\epsilon )        }  \,,  
\\ \nonumber
 &=&\frac{4\left[-\alpha^{2}\xi sT_{A}+zT_{B}\right]}{\alpha\bar{\alpha}\xi s^{2}z\bar{z}\left(x-\xi+i\epsilon\right)\left(x+\xi+i\epsilon\right)} .
 \label{trCEVB_1trace}
  \end{eqnarray}  
We introduced the two tensor structures that will appear in chiral-even diagrams in the vector sector:
\beqa
\label{def:TA-TB}
T_A &=& (\varepsilon_{q\perp} \cdot \varepsilon_{k\perp}^*)\,, \nonumber \\                                                  
T_B &=& (\varepsilon_{q\perp} \cdot p_\perp) (p_\perp \cdot                                      
\varepsilon_{k\perp}^*) .
\eqa
Similarly one can write in the axial sector:
 \begin{eqnarray}
 T^{q\,CE}_A(B_1)&=&  \,\frac{1}{i}\,\frac{tr(t^at^a)}{(4N)^2}\,f_\rho\,\phi_{||}(z)\,(-ieQ_q)^2\,(-ig)^2\,i^2\,(-i)
 \nonumber \\  
 &&\hspace{-1.2cm} \times \, tr_D\left[   \hat p_\rho  \hat \varepsilon_k^*\frac{\hat k +z\hat p_\rho}{(k+z p_\rho)^2+i\epsilon } \gamma^\mu \frac{\hat q+(x+\xi)p}{(q+(x+\xi)p)^2+i\epsilon}
 \,\hat \varepsilon_q\,\hat p\, \gamma^5 \,\gamma_\mu\,\frac{1}{(\bar{z} p_\rho+(x-\xi)p)^2+i\epsilon}\right]
 \nonumber      \nonumber \\
 &&\hspace{-1.2cm}  \times \, \frac{2}{s}\,\bar{u}(p_2,\lambda')\, \left[ \gamma^5\,\hat n\, \tilde H^{q}(x,\xi,t)   - \frac{n\cdot \Delta}{2m} \gamma^5  \,\tilde E^{q}(x,\xi,t)   \right]u(p_1,\lambda)
 \\ \nonumber
 &&\hspace{-1.2cm} = C^{q\,CE}\,tr_D^A\left[ B_1\right]\,\phi_{||}(z)\,\frac{2}{s}\,\bar{u}(p_2,\lambda')\, \left[ \gamma^5\,\hat n\, \tilde H^{q}(x,\xi,t)   - \frac{n\cdot \Delta}{2m} \gamma^5  \,\tilde E^{q}(x,\xi,t)   \right]u(p_1,\lambda)\, ,
 \label{CEAB_1}
  \end{eqnarray} 
 with
 \begin{eqnarray}
   tr_D^A\left[ B_1\right] 
&=&   tr_D\!\left[   \hat p_\rho  \hat \varepsilon_k^*\frac{\hat k +z\hat p_\rho}{(k+z p_\rho)^2\!+\!i\epsilon } \gamma^\mu \frac{\hat q+(x+\xi)p}{(q+(x+\xi)p)^2+i\epsilon}
 \,\hat \varepsilon_q\,\hat p\, \gamma^5 \,\gamma_\mu\,\frac{1}{(\bar{z} p_\rho+(x-\xi)p)^2\!+\!i\epsilon}\right]
 \nonumber
  \\
&=& -\frac{8i}{\alpha \alpha_\rho}\,\frac{  \left[   \alpha\,( \varepsilon_{q\perp}\cdot p_{\rho \bot}) \,\epsilon^{p\,n\,p_{ \rho \bot}\,\varepsilon_{k\perp}^*}  -(\alpha+2z \alpha_\rho) \, (\varepsilon_{k\perp}^*\cdot  p_{\rho \bot})   \,\epsilon^{p\,n\,p_{\rho \bot}\,\varepsilon_{q\perp}}             \right]    }{  (  (k+z p_\rho)^2+i\epsilon        )    (   (q+(x+\xi)p)^2+i\epsilon           )           ( (\bar{z} p_\rho+(x-\xi)p)^2+i\epsilon                )       }\,\nonumber
\\ 
 &=& \frac{-4i\left[-\left(\alpha+2\bar{\alpha}z\right)T_{A_5}-\alpha T_{B_5}\right]}{\alpha\bar{\alpha}^{2}\xi s^{3}z\bar{z}\left(x-\xi+i\epsilon\right)\left(x+\xi+i\epsilon\right)} ,
 \label{trCEAB_1trace}
  \end{eqnarray}
where we introduced the two tensor structures which will appear in chiral-even diagrams in the axial sector:
\beqa
\label{def:TA5-TB5}
T_{A_5} &=& \ \ (p_\perp \cdot                                      
\varepsilon_{k\perp}^*) \,  \epsilon^{n \,p \,\varepsilon_{q\perp}\, p_\perp}\,, \nonumber \\
T_{B_5} &=& -(p_\perp \cdot \varepsilon_{q\perp})\, \epsilon^{n \,p \varepsilon_{k\perp}^*\, p_\perp} .
\eqa

 The chiral-odd (CO) scattering amplitude involving quark of flavor $q$ ($q=u,d$) corresponding to diagram $B_1$ in Feynman gauge has the form:
 \begin{eqnarray}
 && T^{q\,CO}(B_1)= - \,\frac{2}{i}\,\frac{tr(t^at^a)}{(8N)^2}\,i\,2f_\rho^\bot\,\phi_\bot(z)\,(-ieQ_q)^2\,(-ig)^2\,i^2\,(-i)
 \nonumber \\
 && tr_D\left[   \hat p_\rho \hat \varepsilon^*_\rho \hat \varepsilon_k^*\frac{(\hat k +z\hat p_\rho)}{(k+z p_\rho)^2+i\epsilon } \gamma^\mu \frac{\hat q+(x+\xi)p}{(q+(x+\xi)p)^2+i\epsilon}
 \,\hat \varepsilon_q\,\hat p \,\gamma_{\bot j}\,\gamma_\mu\,\frac{1}{(\bar{z} p_\rho+(x-\xi)p)^2+i\epsilon}\right]
 \nonumber \\
 && \frac{2}{s}\,\bar{u}(p_2,\lambda')\, \left[ \sigma^{n j}H_T^{q}(x,\xi,t)  \right]u(p_1,\lambda)
 \nonumber \\
&& =\,C^{q\,CO}\,tr_D^{CO}\left[ B_1\right]_j\,\phi_\bot(z)\,\frac{2}{s}\,\bar{u}(p_2,\lambda')\, \left[ i \sigma^{n j}H_T^{q}(x,\xi,t)  \right]u(p_1,\lambda)
\end{eqnarray}
where
\begin{equation}
C^{q\,CO} = - \frac{4}{9}\,f_\rho^\bot \, \alpha_{em}\,\alpha_s\,\pi^2\,Q_q^2
\label{coefCO}
\end{equation} 
is a common coefficient for all diagrams involving chiral-odd DA and GPD, and
\begin{eqnarray}
&&tr_D^{CO}[B_1]_j\,=
\\
&&  tr_D\left[   \hat p_\rho \hat \varepsilon^*_\rho \hat \varepsilon_k^*\frac{(\hat k +z\hat p_\rho)}{(k+z p_\rho)^2+i\epsilon } \gamma^\mu \frac{\hat q+(x+\xi)p}{(q+(x+\xi)p)^2+i\epsilon}
\,\hat \varepsilon_q\,\hat p \,\gamma_{\bot j}\,\gamma_\mu\,\frac{1}{(\bar{z} p_\rho+(x-\xi)p)^2+i\epsilon}\right]\,,
\nonumber
\label{trCOB_1}
\end{eqnarray}
includes all non trivial factors (vertices as well as quark and gluon propagators) of the hard part of diagram $B_1.$ The calculation of traces over $\gamma$-matrices leads to the expression
\begin{eqnarray}
tr_D^{CO}[B_1]_j\,&=& \frac{8s\left[      
(q \cdot p) \varepsilon_{q\bot\,j} \left(  p_\rho \cdot \varepsilon_k^*\,\varepsilon_\rho^*\cdot k  -   s\xi\, \varepsilon^*_k \cdot \varepsilon^*_\rho  \right) 
-\epsilon^{k\, \varepsilon^*_k\, p_\rho \, \varepsilon^*_\rho}\, \epsilon^{q\,\varepsilon_q\,p\,\nu} g_{\bot \nu j}
\right]}{((k+zp_\rho)^2+i\epsilon) \,   ((\bar{z} p_\rho+(x-\xi)p)^2+i\epsilon)      ((q+(x+\xi)p)^2+i\epsilon)  }\nonumber
\\ 
&=& \frac{T_{B\perp j}}{2\bar{\alpha}\xi s^{3}z\bar{z}\left(x+\xi+i\epsilon\right)\left(x-\xi+i\epsilon\right)}\,.
\label{trCOB_1trace}
\end{eqnarray} 
Here 
$T_{B\perp j}$ is one of the two tensor structures which will appear in chiral-odd diagrams,
\begin{eqnarray} 
\label{def-TiAperp}
\nonumber
T_{A\perp}^i & \equiv &  (p \cdot k) \,\varepsilon^{i*}_{k \bot} \left[ ( \varepsilon_q\cdot p_\rho ) \, ( q\cdot \varepsilon^*_\rho ) - (q \cdot p_\rho) \, (\varepsilon_q\cdot \varepsilon^*_\rho)  \right]
   - \epsilon^{p_\rho\, \varepsilon^*_\rho\,q\, \varepsilon_q }   \,\epsilon^{p\, \nu\, k\, \varepsilon_k^*}g_{\bot \nu}^i      
\\ \nonumber
& = & \frac{-8s}{\bar{\alpha}}\left\{ \alpha\varepsilon_{k\perp}^{i*}\left[-2\alpha\xi\left(p\cdot\epsilon_{\rho}^*\right)\left(p_{\perp}\cdot\varepsilon_{q\perp}\right)+\left(p_{\perp}\cdot\varepsilon_{q\perp}\right)\left(p_{\perp}\cdot\varepsilon_{\rho\perp}^*\right)+\alpha\bar{\alpha}\xi s\left(\varepsilon_{q\perp}\cdot\varepsilon_{\rho\perp}^*\right)\right]\right.\\
 &  & \!\!-\bar{\alpha}\varepsilon_{\rho\perp}^{i*}\left[\alpha\left(\alpha-2\right)\xi s\left(\varepsilon_{q\perp}\cdot\varepsilon_{k\perp}^*\right)-\left(p_{\perp}\cdot\varepsilon_{q\perp}\right)\left(p_{\perp}\cdot\varepsilon_{k\perp}^*\right)\right]\\ \nonumber
 &  &\!\! +p_{\perp}^{i}\left[-2\alpha^{2}\xi p_{\perp}^{i}\left(p\cdot\epsilon_{\rho}^*\right)\left(\varepsilon_{q\perp}\cdot\varepsilon_{k\perp}^*\right)+\left(p_{\perp}\cdot\varepsilon_{\rho\perp}^*\right)\left(\varepsilon_{q\perp}\cdot\varepsilon_{k\perp}^*\right)-\bar{\alpha}\left(\varepsilon_{k\perp}^*\cdot\varepsilon_{\rho\perp}^*\right)\left(p_{\perp}\cdot\varepsilon_{q\perp}\right)\right]\nonumber\\ 
 &  &\!\! \left.+\varepsilon_{q\perp}^i\left[2\alpha^{2}\xi\left(p\cdot\epsilon_{\rho}^*\right)\left(p_{\perp}\cdot\varepsilon_{k\perp}^*\right)-\left(p_{\perp}\cdot\varepsilon_{\rho\perp}^*\right)\left(p_{\perp}\cdot\varepsilon_{k\perp}^*\right)+\alpha\bar{\alpha}\left(\alpha-2\right)\xi s\left(\varepsilon_{k\perp}^*\cdot\varepsilon_{\rho\perp}^*\right)\right]\right\} \,,\nonumber
\end{eqnarray} 
the other one being 
\begin{eqnarray} 
\label{def-TiBperp}
T_{B\perp}^i & \equiv & (q \cdot p)\,\varepsilon_{q\bot}^i \left[ (p_\rho \cdot \varepsilon_k^*) \, (\varepsilon_\rho^* \cdot k)  -   s\xi\, (\varepsilon^*_k \cdot \varepsilon^*_\rho ) \right]  -\epsilon^{k\, \varepsilon^*_k\, p_\rho \, \varepsilon^*_\rho}\, \epsilon^{q\,\varepsilon_q\,p\,\nu} g_{\bot \nu}^i \\ \nonumber
& = & \frac{8s}{\alpha\bar{\alpha}}\left\{ \bar{\alpha}\varepsilon_{\rho\perp}^{i*}\left[\left(p_{\perp}\cdot\varepsilon_{q\perp}\right)\left(p_{\perp}\cdot\varepsilon_{k\perp}^*\right)-\alpha\left(2\alpha-1\right)\xi s\left(\varepsilon_{q\perp}\cdot\varepsilon_{k\perp}^*\right)\right]\right.\nonumber\\
 &  & +\alpha\varepsilon_{k\perp}^{i*}\left[\bar{\alpha}\left(2\alpha-1\right)\xi s\left(\varepsilon_{q\perp}\cdot\varepsilon_{\rho\perp}^*\right)+2\xi\left(p\cdot\epsilon_{\rho}^*\right)\left(p_{\perp}\cdot\varepsilon_{q\perp}\right)+\left(p_{\perp}\cdot\varepsilon_{q\perp}\right)\left(p_{\perp}\cdot\varepsilon_{\rho\perp}^*\right)\right]\nonumber\\ 
 &  & +\varepsilon_{q\perp}^i\left[2\alpha\xi\left(p\cdot\epsilon_{\rho}^*\right)\left(p_{\perp}\cdot\varepsilon_{k\perp}^*\right)-\left(p_{\perp}\cdot\varepsilon_{\rho\perp}^*\right)\left(p_{\perp}\cdot\varepsilon_{k\perp}^*\right)-\alpha\bar{\alpha}\xi s\left(\varepsilon_{k\perp}^*\cdot\varepsilon_{\rho\perp}^*\right)\right] \nonumber \\ 
 &  & \left.+p_{\perp}^{i}\left[2\alpha\xi\left(p\cdot\epsilon_{\rho}^*\right)\left(\varepsilon_{q\perp}\cdot\varepsilon_{k\perp}^*\right)-\alpha\left(p_{\perp}\cdot\varepsilon_{\rho\perp}^*\right)\left(\varepsilon_{q\perp}\cdot\varepsilon_{k\perp}^*\right)-\bar{\alpha}\left(\varepsilon_{q\perp}\cdot\varepsilon_{\rho\perp}^*\right)\left(p_{\perp}\cdot\varepsilon_{k\perp}^*\right)\right]\right\}\,.
 \nonumber
\end{eqnarray}
 Here, we expressed these two tensor structures in terms of the transverse polarization vectors and of $(p \cdot \varepsilon_\rho)$, using Eqs.~(\ref{eps_k}-\ref{n.eps}), for later convenience. 
 
 At the dominant twist, the sum over the transverse polarizations of the $\rho$ meson can be written as
\begin{equation}
\sum_{pol} \varepsilon_{\rho}^\mu \varepsilon_{\rho}^{\nu\ast} = -g^{\mu\nu} + \frac{p_{\rho}^\mu
p_{\rho}^\nu}{m_\rho^2},
\end{equation}
when computing the square of the chiral odd amplitude. The second term of this sum, which arises mainly from
the longitudinal polarization, does not contribute at leading twist.
We thus note that $(p\cdot\varepsilon_\rho)$ terms in the tensor structures will not contribute to the cross
section since when summed over the transverse polarizations at the dominant twist they will produce terms
involving the scalar product of $p$ either with a transverse vector or with itself, which is null in both
cases.

 In a similar way we obtain the  expressions for the remaining independent diagrams: $A_1,$ $A_2,$ $A_3,$ $A_4,$ $A_5,$  $B_2,$ $B_3$, $B_4$, $B_5$ in the chiral-even sector and  $A_3,$ $A_4$ and $B_5$ in the chiral-odd sector. We show these results in  appendix~\ref{App:diagrams}.

The integral with respect to $z$ is trivially performed in the case of a DA
expanded in the basis of Gegenbauer polynomials. 
The expressions for the case of two
asymptotical DAs $\phi_\parallel$ and $\phi_\perp$, which we only consider in the present article, are given explicitly in appendix~\ref{App:z-integration}, and expressed as linear combination of building blocks.

The integration with respect to $x$, for a given set of GPDs,  (which can be our model described in Sec.~\ref{Sec:DAs-GPDs} or any other model), is then reduced to the numerical
evaluation of these building block integrals.

\subsection{Square of $\mathcal{M}_{\parallel}$ and  $\mathcal{M}_{\bot}$}
\label{SubSec:Square}

In the forward limit $\Delta_{\bot} = 0 = P_{\bot}$, one can show that the squares of $\mathcal{M}_{\parallel}$ and of $\mathcal{M}_{\bot}$ read after summing over nucleon helicities:
\begin{eqnarray}
\label{squareCEresult}
&&\mathcal{M}_{\parallel}^q \mathcal{M}_{\parallel}^{q'*} \equiv  
 \sum_{\lambda ',\, \lambda}
 \mathcal{M}_{\parallel}^q (\lambda,\lambda')\,
 \mathcal{M}_{\parallel}^{q'*}(\lambda,\lambda') \\  
 &=&   8(1-\xi^2) 
 \left(  {\cal H}^{q}(\xi,t)  {\cal H}^{q'*}(\xi,t)    +  \tilde {\cal H}^{q}(\xi,t) \tilde {\cal H}^{q'*}(\xi,t)  \right) \nonumber \\
 &-& 4\,\xi^2   
\left(  {\cal E}^{q}(\xi,t)
{\cal E}^{q'*}(\xi,t)
+  \tilde {\cal E}^{q}(\xi,t)
\tilde {\cal E}^{q'*}(\xi,t)
\right)\nonumber
 \\ 
&-& 8\, \xi^2   \left(  {\cal H}^{q}(\xi,t) {\cal E}^{q' *}(\xi,t) + 
{\cal H}^{q'*}(\xi,t) {\cal E}^{q}(\xi,t)
+
\tilde {\cal H}^q(\xi,t)\tilde {\cal E}^{q'*}(\xi,t) 
+
\tilde {\cal H}^{q'*}(\xi,t)\tilde {\cal E}^{q}(\xi,t)
\right) ,\nonumber
\end{eqnarray}
  and
\begin{eqnarray}
\label{squareCOresult}
&& \mathcal{M}_{\bot}^q \mathcal{M}_{\bot}^{q'*}  \equiv  
 \sum_{\lambda ',\, \lambda}
 \mathcal{M}_{\bot}^q (\lambda,\lambda')\,
 \mathcal{M}_{\bot}^{q'*}(\lambda,\lambda') 
 \\ \nonumber 
  &=& 8\left[\!   -(1-\xi^2) {\cal H}_T^{q \, i}(\xi,t)
  {\cal H}_T^{q'  j \,*}(\xi,t)
  - \frac{\xi^2}{1-\xi^2} [ \xi \,{\cal E}_T^{q \, i}(\xi,t) - \tilde {\cal E}_T^{q \, i}(\xi,t)       ]  [ \xi \,{\cal E}_T^{q' j *}(\xi,t) - \tilde {\cal E}_T^{q' j *}(\xi,t) ] 
\right. \\ \nonumber
&& \left.
+\, \xi  \left\{ {\cal H}_T^{q i}(\xi,t)[\xi \,{\cal E}_T^{q' j}(\xi,t) - \tilde {\cal E}_T^{q ' j}(\xi,t)  ]^* +
{\cal H}_T^{q' i *}(\xi,t)[\xi \, {\cal E}_T^{q j}(\xi,t) - \tilde {\cal E}_T^{q j}(\xi,t)  ]
\right\}\right] g_{\perp ij}.
\end{eqnarray}
For moderately small values of $\xi$, these become:
\begin{eqnarray}
\label{squareCEresultsmallxi}
\mathcal{M}_{\parallel}^q \mathcal{M}_{\parallel}^{q'*} &=&   8
 \left(  {\cal H}^{q}(\xi,t) \, {\cal H}^{q'*}(\xi,t)    +  \tilde {\cal H}^{q}(\xi,t)\, \tilde {\cal H}^{q'*}(\xi,t)  \right),
\\
\label{squareCOresultsmallxi}
\mathcal{M}_{\bot}^q \mathcal{M}_{\bot}^{q'*}  
  &=& -8\, {\cal H}_T^{q \, i}(\xi,t)\,
  {\cal H}_T^{q'  j \,*}(\xi,t) \,  g_{\perp ij}\,.
\end{eqnarray}
Hence we will restrict ourselves to $H^q$, $\tilde{H}^q$ and $H_T^q$ to perform our estimates of the cross section\footnote{In practice, we keep the first line in the r.h.s. of Eq.~(\ref{squareCEresult}) and the first term
in the r.h.s of Eq.~(\ref{squareCOresult}).}.

\section{Unpolarized Differential Cross Section and Rate Estimates}
\label{Sec:Cross-Section-and-Rates}

\subsection{From amplitudes to cross sections}
\label{SubSec:amplitude-to-cross sections}

We isolate the tensor structures of the form factors as
\begin{eqnarray}
\label{dec-tensors-quarks}
\mathcal{H}^q(\xi , t) &=&  \mathcal{H}^q_A (\xi , t) T_A + \mathcal{H}^q_B (\xi , t) T_B \,,\\
\mathcal{\tilde{H}}^q(\xi , t) &=& \mathcal{\tilde{H}}^q_{A} (\xi , t) T_{A_5} + \mathcal{\tilde{H}}^q_B (\xi , t) T_{B_5} \,,\\
\mathcal{H}^{q\,i}_T(\xi , t) &=& \mathcal{H}^q_{T_A} (\xi , t) T_{A\perp}^i + \mathcal{H}^q_{T_B} (\xi , t) T_{B\perp}^i.
\end{eqnarray}
These coefficients can be expressed in terms  of the sum over diagrams of the integral of the product of their traces, of GPDs and DAs, as defined and given explicitly in  
appendix~\ref{App:z-integration}.
They reads
\beqa
\label{form-factors-N}
\mathcal{H}_A^q = \frac{1}s C^{q\,CE} N_A^q \,, \\
\mathcal{H}_B^q = \frac{1}{s^2} C^{q\,CE} N_B^q\,,
\eqa
\beqa
\label{form-factors-TildeN}
\tilde{\mathcal{H}}_{A_5}^q = \frac{1}{s^3}C^{q\,CE} N_{A_5}^q \,, \\
\tilde{\mathcal{H}}_{B_5}^q = \frac{1}{s^3}C^{q\,CE} N_{B_5}^q\,,
\eqa
and
\beqa
\label{form-factors-NT}
\mathcal{H}_{T \,A}^q = \frac{1}{s^3}C^{q\,CO} N_{T\, A}^q \,, \\
\mathcal{H}_{T \,B}^q = \frac{1}{s^3}C^{q\,CO} N_{T\, B}^q\,.
\eqa

For the specific case of our process, it is convenient
to define the total form factors as follows:
\begin{eqnarray}
\mathcal{H}(\xi, t) &\equiv & \mathcal{H}^u(\xi, t) - \mathcal{H}^d(\xi, t)\,, \\
\tilde{\mathcal{H}}(\xi, t) &\equiv & \tilde{\mathcal{H}}^u(\xi, t) - \tilde{\mathcal{H}}^d(\xi, t) \,,\\
\mathcal{H}_T^i(\xi, t) &\equiv & \mathcal{H}_T^{u \, i}(\xi, t) - \mathcal{H}_T^{d \, i}(\xi, t) \,,
\end{eqnarray}
from which we isolate the tensor structures
\begin{eqnarray}
\mathcal{H}(\xi , t) &=&  \mathcal{H}_A (\xi , t) \,  T_A + \mathcal{H}_B (\xi , t) \, T_B \,,\\
\mathcal{\tilde{H}}(\xi , t) &=& \mathcal{\tilde{H}}_{A_5} (\xi , t) \,  T_{A_5} + \mathcal{\tilde{H}}_{B_5} (\xi , t)  \, T_{B_5} \,,\\
\mathcal{H}^{i}_T(\xi , t) &=& \mathcal{H}_{T_A} (\xi , t)  \, T_{A\perp}^i + \mathcal{H}_{T_B} (\xi , t) \,  T_{B\perp}^i.
\end{eqnarray}

In this paper, we are interested in the unpolarized cross section. As a result, we will need the squared form factors after summation over all the polarizations (outgoing $\gamma$ and $\rho$, incoming $\gamma$):
\begin{eqnarray}
\label{FF-squared-H}
|\mathcal{H}(\xi , t)|^2 & \equiv & \sum_{\lambda_k \lambda_q} \mathcal{H}(\xi , t, \lambda_k, \lambda_q) \, \mathcal{H}(\xi , t, \lambda_k, \lambda_q) \\ \nonumber
 &=& 2|\mathcal{H}_A (\xi , t)|^2 + p_\bot^4 | \mathcal{H}_B (\xi , t)|^2 + p_\bot^2 \left[ \mathcal{H}_A (\xi , t)\mathcal{H}^{\ast}_B (\xi , t) + \mathcal{H}^{\ast}_A (\xi , t)\mathcal{H}_B (\xi , t) \right], \\ \nonumber \\
|\mathcal{\tilde{H}}(\xi , t)|^2 &\equiv & \sum_{\lambda_k \lambda_q} 
\mathcal{\tilde{H}}(\xi , t, \lambda_k, \lambda_q) \, \mathcal{\tilde{H}}^*(\xi , t, \lambda_k, \lambda_q)
\\ \nonumber
\label{FF-squared-HTilde}
 &=& \frac{s^2 p_\bot^4}{4} \left(| \mathcal{\tilde{H}}_{A_5} (\xi , t)|^2 + | \mathcal{\tilde{H}}_{B_5} (\xi , t) |^2\right), \\ \nonumber \\
 \label{FF-squared-HT}
|\mathcal{H}_T(\xi , t)|^2 &\equiv & -g_{\perp i\, j} \sum_{\lambda_k \lambda_q \lambda_\rho} \mathcal{H}_{T}^{i}(\xi , t, \lambda_k, \lambda_q, \lambda_\rho) \mathcal{H}_T^{j*}(\xi , t, \lambda_k, \lambda_q, \lambda_\rho) \\ \nonumber
&=& 512\xi^2 s^4 \left(\alpha^4 | \mathcal{H}_{T_A} (\xi , t)|^2 +  |\mathcal{H}_{T_B} (\xi , t)|^2 \right).
\end{eqnarray}

We now define the averaged amplitude squared $|\mathcal{\overline{M}}|^2,$ which includes 
the factor 1/4 coming from the averaging over the 
polarizations of the initial particles.
Collecting all prefactors
(including a factor of $2^2$ for the missing half of the set of diagrams and a factor of 1/2 from the square of the $\rho^0$ wave function, see Eq.~(\ref{AmplitudeFactorized})), which reads 
$$
\frac{1}{s^2} 2^2  8 (1-\xi^2)  \left(C^{q\,CE(OD)}\right)^2 \frac{1}{2^3}\,,
$$
we have
the net result (factorizing out the coefficient for the $u-$quark), for the chiral-even case
\beqa
\label{all-CE}
&&|\mathcal{\overline{M}}^{CE}|^2 = \frac{4}{s^2}   (1-\xi^2)  \left(C^{u\,CE}\right)^2 
\left\{ 
2 \left|N_A^u - \frac{1}4 N_A^d\right|^2 
+ \frac{p_\perp^4}{s^2} \left|N_B^u - \frac{1}4 N_B^d\right|^2 
\right.\\
&&
\left.
+ \frac{p_\perp^2}s \left(\left[N_A^u - \frac{1}4 N_A^d\right]
\left[N_B^u - \frac{1}4 N_B^d\right]^* + c.c. \right)
+ \frac{p_\perp^4}{4 s^2} 
\left|\tilde{N}_A^u - \frac{1}4 \tilde{N}_A^d\right|^2 
+ \frac{p_\perp^4}{4 s^2} \left|\tilde{N}_B^u - \frac{1}4 \tilde{N}_B^d\right|^2
\right\},\nonumber
\eqa
while
for the chiral-odd case, we get
\beqa
\label{all-CO}
|\mathcal{\overline{M}}^{CO}|^2 = \frac{2048}{s^2}  \xi^2  (1-\xi^2)  \left(C^{u\,CO}\right)^2 
\left\{ 
\alpha^4 \left|N_{T\,A}^u - \frac{1}4 N_{T\,A}^d\right|^2
+
\left|N_{T\,B}^u - \frac{1}4 N_{T\,B}^d\right|^2
\right\}.
\eqa

The differential cross section as a function of $t$, $M^2_{\gamma\rho},$ $-u'$ then reads
\begin{equation}
\label{difcrosec}
\left.\frac{d\sigma}{dt \,du' \, dM^2_{\gamma\rho}}\right|_{\ -t=(-t)_{min}} = \frac{|\mathcal{\overline{M}}|^2}{32S_{\gamma N}^2M^2_{\gamma\rho}(2\pi)^3}\,.
\end{equation}

\subsection{Numerical evaluation of the scattering amplitudes and of cross sections}
\label{SubSec:Scattering_Amplitude-numerical}

Above, we have reduced the calculation 
of the cross sections, see Eq.~(\ref{difcrosec}),
to the numerical evaluation of the coefficients 
(\ref{Def:NTildeA-axial}), (\ref{Def:NTildeB-axial}), (\ref{Def:NA-vector}), (\ref{Def:NB-vector}), (\ref{Def:N_A}), (\ref{Def:N_B}). These coefficients are expressed as linear combinations of numerical integrals, listed in appendix~\ref{App:z-integration}.

Our central set of curves, displayed below, is obtained
for $S_{\gamma N}=20~$GeV$^2$, with $M^2_{\gamma \rho}$ varying in the range $2.10~{\rm GeV}^2~<~M^2_{\gamma \rho}~<~9.47~{\rm GeV}^2$ (this latter value comes from the vanishing of the phase-space in $-t$, as shown in appendix~\ref{App:phase}, see Eq.~(\ref{Def:MaxtminmaxM2})) with a $0.1$~{\rm GeV}$^2$ step.

For each of these $M^2_{\gamma \rho}$ values, 
we chose 100 values of  $-u'$, linearly varying from $(-u')_{min}=1~{\rm GeV}^2$ up to 
$(-u')_{maxMax}$ as defined by Eq.~(\ref{Def:mupmaxMax}).

For each of these couples of values of ($M^2_{\gamma \rho}, -u'\,,)$
 we compute each of the numerical coefficients $N_A^u,\,
N_A^d,\,
N_B^u,\,
N_B^d$ and $\tilde{N}_A^u,\,
\tilde{N}_A^d,\,
\tilde{N}_B^u,\,
\tilde{N}_B^d$ for the chiral-even case, as well as the coefficients
$N_{T\,A}^u,\,
N_{T\,A}^d,\,
N_{T\,B}^u,\,
N_{T\,B}^d$ for the chiral-odd case,
using the sets of GPDs indexed by $M^2_{\gamma \rho}$ and computed as explained in 
Sec.~\ref{SubSec:GPDs}.

This whole set of dimensionless
numerical coefficients allows us to perform the various 
phenomenological studies discussed in the next subsections.

\subsection{Fully differential cross sections}
\label{SubSec:3-diff}

 Let us first discuss chiral-even results,  showing in parallel the proton  and neutron target cases.

%
%
\psfrag{H}{\hspace{-1.5cm}\raisebox{-.6cm}{\scalebox{.7}{$-u' ({\rm GeV}^{2})$}}}
\psfrag{V}{\raisebox{.3cm}{\scalebox{.7}{$\hspace{-.4cm}\displaystyle\frac{d \sigma_{\rm even}}{d M^2_{\gamma \rho} d(-u') d(-t)}~({\rm pb} \cdot {\rm GeV}^{-6})$}}}
\begin{figure}[!h]
\begin{center}
\psfrag{T}{}
\hspace{.2cm}\includegraphics[width=7.3cm]{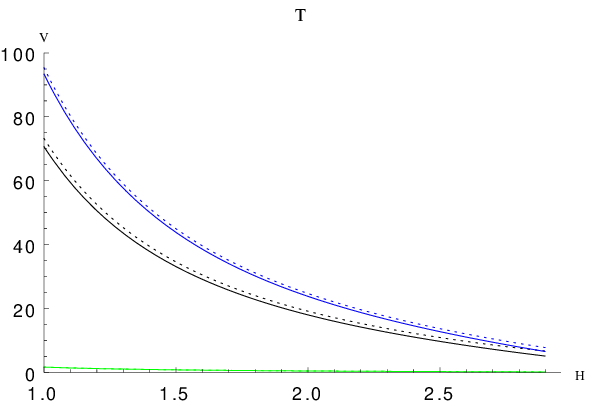}
\psfrag{T}{}
\hspace{0.1cm}\includegraphics[width=7.3cm]{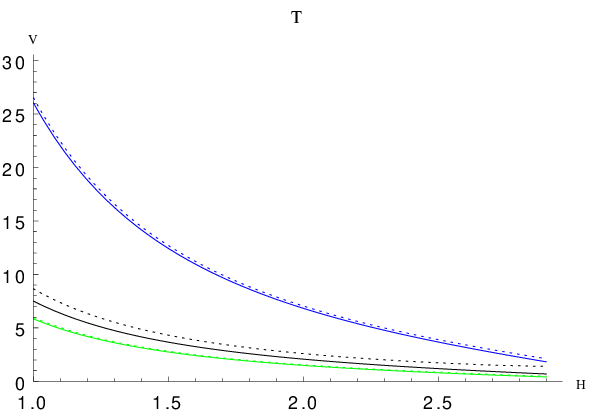}
\vspace{.2cm}
\caption{Differential cross section for a photon and a  longitudinally polarized $\rho$ meson production, for the proton (left) and the neutron (right), at $M^2_{\gamma \rho}=4~{\rm GeV}^2$. Both vector and axial GPDs are included.
In black (middle curves) the contributions of both $u$ and $d$ quarks, in blue (top curves) the contribution of the $u$ quark, and in green (bottom curves) the contribution of the $d$ quark. Solid: ``valence'' model, dotted: ``standard'' model.
This figure shows the dominance of the u-quark contribution due to the charge effect. Note that the interference between $u-$quark and $d-$quark contributions is important and negative.}
\label{Fig:dsigmaEVENdM2dupdtSgN20M4U+D_U_D}
\end{center}
\end{figure}
We first analyze the various contributions to the differential cross section in the specific kinematics: $M^2_{\gamma \rho} = 4$~GeV$^2$, $S_{\gamma N}= 20~{\rm GeV}^2$, $-t=(-t)_{min}$ as a function of $-u'$. 
 The dependency with respect to $S_{\gamma N}$
 will be discussed in Sec.~\ref{SubSec:integrated-cross section}.

%
%
\psfrag{H}{\hspace{-1.5cm}\raisebox{-.6cm}{\scalebox{.7}{$-u' ({\rm GeV}^{2})$}}}
\psfrag{V}{\raisebox{.3cm}{\scalebox{.7}{$\hspace{-.4cm}\displaystyle\frac{d \sigma_{\rm even}}{d M^2_{\gamma \rho} d(-u') d(-t)}~({\rm pb} \cdot {\rm GeV}^{-6})$}}}
\begin{figure}[!h]
\begin{center}
\psfrag{T}{}
\hspace{.2cm}\includegraphics[width=7.3cm]{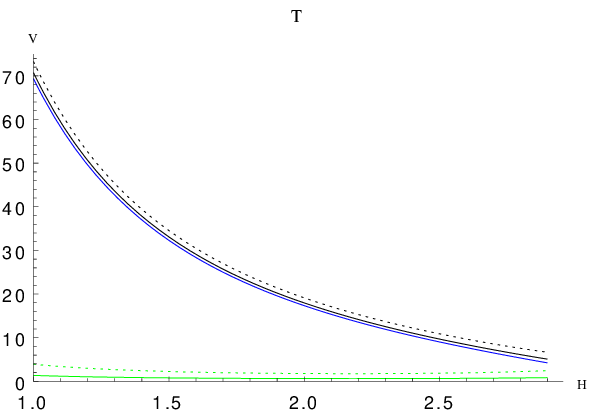}
\psfrag{T}{}
\hspace{0.1cm}\includegraphics[width=7.3cm]{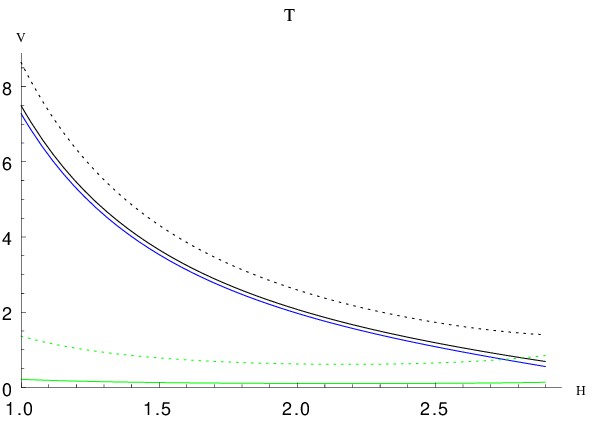}
\vspace{.2cm}
\caption{
Differential cross section for a photon and a  longitudinally polarized $\rho$ meson production, for the proton (left) and the neutron (right), at $M^2_{\gamma \rho}=4~{\rm GeV}^2$. Both $u$ and $d$ quark contributions are included.
In black (two top curves) the contributions of  both vector and axial amplitudes, in blue (middle curve) the contribution of the vector amplitude, and in green (two bottom curves) the contribution of the axial amplitude. Solid: ``valence'' model, dotted: ``standard'' model.
This figure shows the dominance of the vector GPD contributions. There is no interference between the vector and axial amplitudes.  }
\label{Fig:dsigmaEVENdM2dupdtSgN20M4V+A_V_A}
\end{center}
\end{figure}

\psfrag{H}{\hspace{-1.5cm}\raisebox{-.6cm}{\scalebox{.7}{$-u'~({\rm GeV}^{2})$}}}
\psfrag{V}{\raisebox{.3cm}{\scalebox{.7}{$\hspace{-.4cm}\displaystyle\frac{d \sigma_{\rm even}}{d M^2_{\gamma \rho} d(-u') d(-t)}~({\rm pb} \cdot {\rm GeV}^{-6})$}}}
\begin{figure}[!h]
\begin{center}
\psfrag{T}{}
\hspace{.2cm}\includegraphics[width=7.3cm]{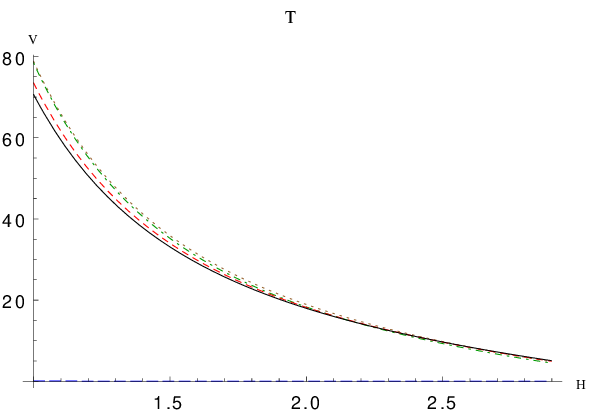}
\psfrag{T}{}
\hspace{0.1cm}\includegraphics[width=7.3cm]{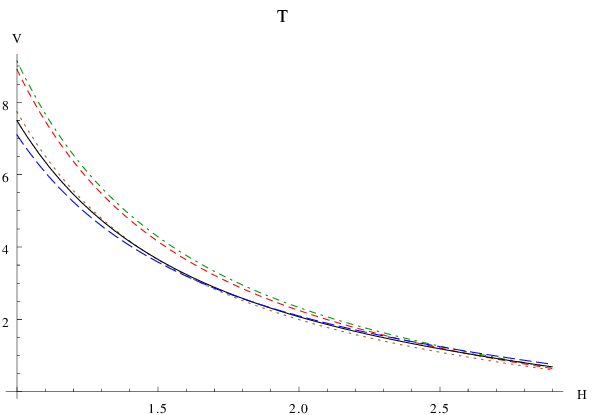}
\vspace{.2cm}
\caption{
Differential cross section for a photon and a  longitudinally polarized $\rho$ meson production, for the proton (left) and the neutron (right), as a function of $-u'$, for $M_{\gamma \rho}^2=4~{\rm GeV}^2$.
The various curves differ with respect to the ans\"atze for the PDFs $q$, and thus for the GPDs $H^u$ and $H^d$: GRV-98 (solid black), MSTW2008lo (long-dashed blue), MSTW2008nnlo (short-dashed red), ABM11nnlo (dotted-dashed green), CT10nnlo (dotted brown).}
\label{Fig:dsigmaEVENdM2dupdtSgN20M4PDFs}
\end{center}
\end{figure}

\psfrag{H}{\hspace{-1.5cm}\raisebox{-.6cm}{\scalebox{.7}{$-u'~({\rm GeV}^{2})$}}}
\psfrag{V}{\raisebox{.3cm}{\scalebox{.7}{$\hspace{-.4cm}\displaystyle\frac{d \sigma_{\rm even}}{d M^2_{\gamma \rho} d(-u') d(-t)}~({\rm pb} \cdot {\rm GeV}^{-6})$}}}
\begin{figure}[!h]
\begin{center}
\psfrag{T}{}
\hspace{.2cm}\includegraphics[width=7.3cm]{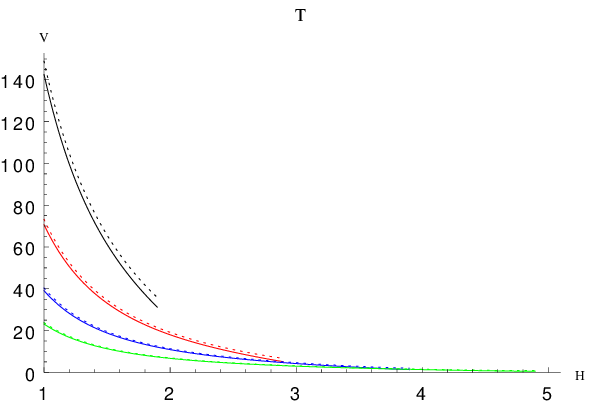}
\psfrag{T}{}
\hspace{0.1cm}\includegraphics[width=7.3cm]{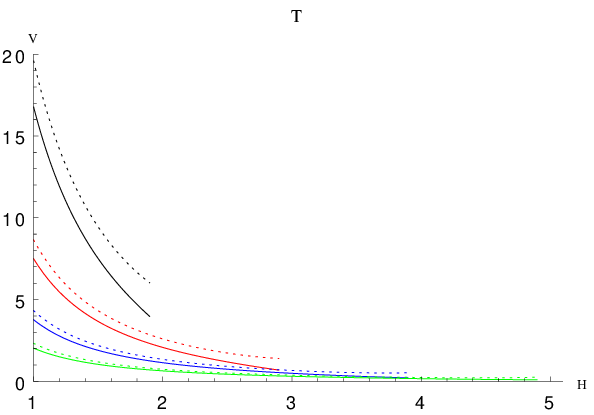}
\vspace{.2cm}
\caption{
Differential cross section for a photon and a  longitudinally polarized $\rho$ meson production, for the proton (left) and the neutron (right), as a function of $-u'$, for $M_{\gamma \rho}^2=3,4,5,6~{\rm GeV}^2$
(resp. in black, red, blue, green, from top to down). Solid: ``valence'' model, dotted: ``standard'' model.}
\label{Fig:dsigmaEVENdM2dupdtSgN20M3,4,5,6}
\end{center}
\end{figure}

In Fig.~\ref{Fig:dsigmaEVENdM2dupdtSgN20M4U+D_U_D}, we show the relative contributions of the $u-$ and $d-$quark GPDs (adding the vector and axial contributions), which interfere in a destructive way because of the flavor structure of the $\rho^0 = \frac{u \bar u - d \bar d}{\sqrt 2}$. The $d-$quark contribution is of course more important for the neutron target case. 

In Fig.~\ref{Fig:dsigmaEVENdM2dupdtSgN20M4V+A_V_A}, we show the relative contributions of the GPDs $H$ and $\tilde H$ involving vector and axial correlators. The vector contribution dominates. The two parameterizations of the axial GPD $\tilde H^q(x,\xi,t)$ give similar results for proton target and slightly different results for neutron target, the one corresponding to the unbroken sea (``standard'') scenario being less negligible than the other one (``valence''). As a simple calculation shows, there is no interference effect between $H$ and $\tilde H$ contributions due to lack of a sufficient number of transverse momenta in the tensor structures.

In Fig.~\ref{Fig:dsigmaEVENdM2dupdtSgN20M4PDFs} we display the effect on
 the differential cross section
of changing the ans\"atze for the PDFs $q$, and thus for the GPDs $H^u$ and $H^d$. For $\tilde{H}^u$ and $\tilde{H}^d$ we rely on the ``valence'' scenario for $\Delta q$. This figure shows that the effect is moderate, of the order of $10\%.$ In the rest of this paper we will neglect this variation, and we will only use the uncertainty on $\tilde H$ to get an order of magnitude of the precision of our predictions for the cross-sections.

Fig.~\ref{Fig:dsigmaEVENdM2dupdtSgN20M3,4,5,6} shows the dependence on $M^2_{\gamma \rho}$. The production of the $\gamma \rho$ pair with a large value of $M^2_{\gamma \rho}$ is severely suppressed as anticipated. However, the $-u'$ range allowed by our kinematical requirements is narrower for smaller values of  $M^2_{\gamma \rho}$. The two curves for each value of  $M^2_{\gamma \rho}$ correspond to the two parameterizations of $\tilde H(x,\xi,t)$, the lines corresponding to the unbroken sea scenario lying above the other one.

\subsection{Single differential cross sections}
\label{SubSec:single-differential-cross section}

 To get an estimate of the total rate of events of interest for our analysis, we first get the $M^2_{\gamma\rho}$ dependence of the differential cross section integrated over $u'$ and $t$,
\begin{equation}
\label{difcrosec2}
\frac{d\sigma}{dM^2_{\gamma\rho}} = \int_{(-t)_{min}}^{(-t)_{max}} \ d(-t)\ \int_{(-u')_{min}}^{(-u')_{max}} \ d(-u') \ F^2_H(t)\times\left.\frac{d\sigma}{dt \, du' d M^2_{\gamma\rho}}\right|_{\ -t=(-t)_{min}} \,.
\end{equation}
Since this is mostly an order of magnitude estimate, we  use a simple universal dipole factorized $t-$dependence of GPDs, 
\beq
\label{dipole}
F_H(t)= \frac{C^2}{(t-C)^2}\,,
\eq
with $C=0.71~{\rm GeV}^2.$
For a more precise study dedicated to an impact picture of the nucleon \cite{Burkardt:2000za,Ralston:2001xs,Diehl:2002he,Burkardt:2005hp,Diehl:2005jf,Mukherjee:2009nw}, a more sophisticated approach \cite{Diehl:2013xca} should be used. 
The domain of integration over $u'$ and $t$ is discussed 
in detail in
appendix~\ref{App:phase}.

The obtained differential cross section $d\sigma/dM^2_{\gamma \rho}$ is shown in Fig.~\ref{Fig:dsigmaEVENdM2SgN8,10,12,14,16,18,20} for various values of $S_{\gamma N}$ covering the JLab-12 energy range. These cross sections show a maximum around $M^2_{\gamma \rho}\approx 3~$GeV$^2$, for most energy values.

\psfrag{H}{\hspace{-1.5cm}\raisebox{-.6cm}{\scalebox{.7}{$M^2_{\gamma \rho}~({\rm GeV}^{2})$}}}
\psfrag{V}{\raisebox{.3cm}{\scalebox{.7}{$\hspace{-.4cm}\displaystyle\frac{d\sigma_{even}}{d M^2_{\gamma\rho}}~({\rm pb} \cdot {\rm GeV}^{-2})$}}}
\begin{figure}[!h]
\begin{center}
\psfrag{T}{}
\hspace{.2cm}\includegraphics[width=7.3cm]{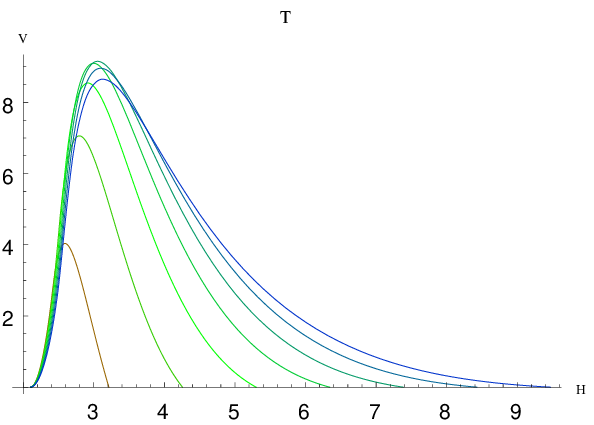}
\psfrag{T}{}
\hspace{0.1cm}\includegraphics[width=7.3cm]{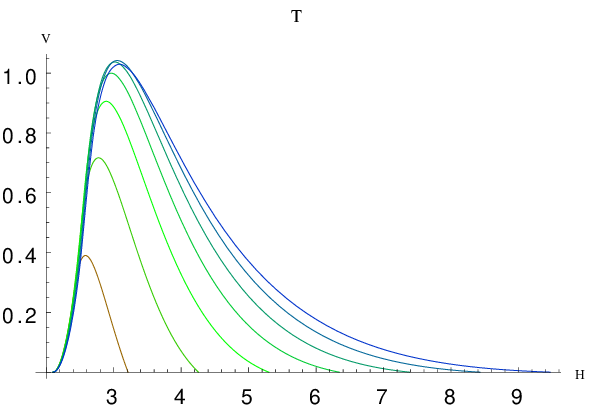}
\vspace{.2cm}
\caption{
Differential cross section $d\sigma/dM^2_{\gamma \rho}$ for a photon and a  longitudinally polarized $\rho$ meson production, on a proton (left) or neutron (right) target. The values of $S_{\gamma N}$ vary in the set 8, 10, 12, 14, 16, 18, 20 ${\rm GeV}^{2}.$ (from 8: left, brown to 20: right, blue), covering the JLab energy range. We use here the ``valence'' scenario.
}
\label{Fig:dsigmaEVENdM2SgN8,10,12,14,16,18,20}
\end{center}
\end{figure}

\subsection{Integrated cross sections and variation with respect to $S_{\gamma N}$}
\label{SubSec:integrated-cross section}

For $S_{\gamma N}=20~$GeV$^2$, the integration over $M^2_{\gamma \rho}$ of our above results within our allowed kinematical region, here $2.10~{\rm GeV}^2~<~M^2_{\gamma \rho}~<~9.47~{\rm GeV}^2$ (see appendix~\ref{App:phase}), allows to obtain the
cross sections 
$\sigma^{\rm proton}_{\rm odd}\simeq 0.54~{\rm pb}$ and
$\sigma^{\rm proton}_{\rm even}\simeq 21~{\rm pb}$ for the proton, and 
$\sigma^{\rm neutron}_{\rm odd}\simeq 0.42~{\rm pb}$ and
$\sigma^{\rm neutron}_{\rm even}\simeq 2.3~{\rm pb}$ for the neutron.

The variation with respect to $S_{\gamma N}$ could be obtained by following the whole chain of steps described above. However, this can be obtained almost directly. 
Our aim is now to show that the only knowledge of the set of numerical results computed for a given value of 
$S_{\gamma N}$, which we take in practice
as $S_{\gamma N}=20~$GeV$^2,$ is sufficient to deduce a whole set of results for any arbitrary smaller values of 
$\tilde{S}_{\gamma N}$.
The key points are the following.

First, the amplitudes only depend on $\alpha$, $\xi$ and on the GPDs (which are computed as grids indexed by $\xi$).
Since $\alpha = -u'/M^2_{\gamma \rho}\,,$
it is thus possible to use exactly the set of already computed amplitudes if one select the same set of $(\alpha,\xi)$

Second, one should note that
to a given value of
\beqa
\label{xi-M2-S}
\xi = \frac{M^2_{\gamma \rho}}{2(S_{\gamma N}-M^2)-M^2_{\gamma \rho}}
\eqa
corresponds an infinite set of couples of values
$(M^2_{\gamma \rho},S_{\gamma N})\,.$

In practice, we use our set of results obtained for $S_{\gamma N}=20~$GeV$^2\,,$ indexed by
$M^2_{\gamma \rho}$ and $-u'$.

Then, choosing a new value of $\tilde{S}_{\gamma N}$,
we obtain a set of values of $\tilde{M}^2_{\gamma \rho}$ indexed by the set of values of $M^2_{\gamma \rho}$ (which vary from 2.2 up to 10 GeV$^2$, with a 0.1~GeV$^2$ step), through the relation
\beqa
\label{set-M2new}
\tilde{M}^2_{\gamma \rho} = M^2_{\gamma \rho} \frac{\tilde{S}_{\gamma N}-M^2}{S_{\gamma N}-M^2}\,,
\eqa
which is deduced from Eq.~(\ref{xi-M2-S}),
and for each of these $\tilde{M}^2_{\gamma \rho}$ a set of values of $-\tilde{u}'\,,$ using the relation
\beq
\label{set-u'new}
-\tilde{u}'= \frac{\tilde{M}^2_{\gamma \rho}}{M^2_{\gamma \rho}} (-u')\,.
\eq
which gives the indexation of allowed values of
$-\tilde{u}'$ as function of known values of $(-u').$

It is now easy to check that this mapping from a given $S_{\gamma N}$ to a lower 
$\tilde{S}_{\gamma N}$ provides a set of 
$(\tilde{M}^2_{\gamma \rho},-\tilde{u}')$ which exhaust the required domain. 

Consider first the range in $\tilde{M}^2_{\gamma \rho}.$
From Eq.~(\ref{Def:M2crit}), which defines the minimal value of $M^2_{\gamma \rho}$, independent of $S_{\gamma N},$ this value is mapped to a smaller value than required, when passing from $S_{\gamma N}$ to $\tilde{S}_{\gamma N}$.
From Eq.~(\ref{Def:MaxtminmaxM2}), it is possible to show that 
$M^2_{\gamma \rho \, {\rm Max}}$ is mapped to a value $\tilde{M}^2_{\gamma \rho \, {\rm Max}}$ slightly larger than the new required value $M'^2_{\gamma \rho \, {\rm Max}}$
(this comes from the little dependency of $\bar{M}$ with respect to $S_{\gamma N}$).
Thus, the mapping covers the whole required domain in $\tilde{M}^2_{\gamma \rho}$ (with a negligible loss of precision since a few points are mapped outside the domain and thus cut).

Now, let us consider the range in $-u'.$
Again, since the minimal value $(-u')_{\rm min}$
is fixed, this value is mapped to a smaller value than required, when passing from $S_{\gamma N}$ to $\tilde{S}_{\gamma N}$.
Concerning the maximal value $(-u')_{\rm maxMax},$ from Eq.~(\ref{Def:mupmaxMax}) it is a linear function of $M^2_{\gamma \rho}$ of the form 
\beqa
(-u')_{\rm maxMax}= -A + M^2_{\gamma \rho}\,,
\eqa
with $A>0.$
The mapping of $M^2_{\gamma \rho}$ leads to the maximal required value
\beqa
(-u')_{\rm maxMax}'= -A + \tilde{M}^2_{\gamma \rho}\,.
\eqa
But the mapping in $-u'$ will transform 
$(-u')_{\rm maxMax}$ to
\beqa
(-\tilde{u}')_{\rm maxMax}= \frac{\tilde{M}^2_{\gamma \rho}}{M^2_{\gamma \rho}} (-A + M^2_{\gamma \rho})=
-A \frac{\tilde{M}^2_{\gamma \rho}}{M^2_{\gamma \rho}}  + \tilde{M}^2_{\gamma \rho}
\,,
\eqa
which shows that the maximal value $(-\tilde{u}')_{\rm maxMax}$ of $(-\tilde{u}')$ obtained from the mapping is larger than the needed $(-u')_{\rm maxMax}'$, since $-A< -A \frac{\tilde{M}^2_{\gamma \rho}}{M^2_{\gamma \rho}} <0.$

We have thus shown that one can obtain the dependency of amplitudes and thus of cross sections for the whole range in $S_{\gamma N}$
from a single set of computation (at $S_{\gamma N}=20~$GeV$^2$), thus avoiding the use of a very  large amount of CPU time.

\begin{figure}[!h]
\begin{center}
\psfrag{T}{}
\hspace{.2cm}\includegraphics[width=7.3cm]{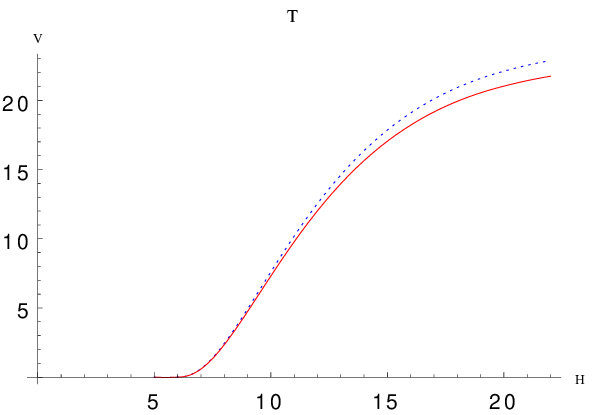}
\psfrag{T}{}
\hspace{0.1cm}\includegraphics[width=7.3cm]{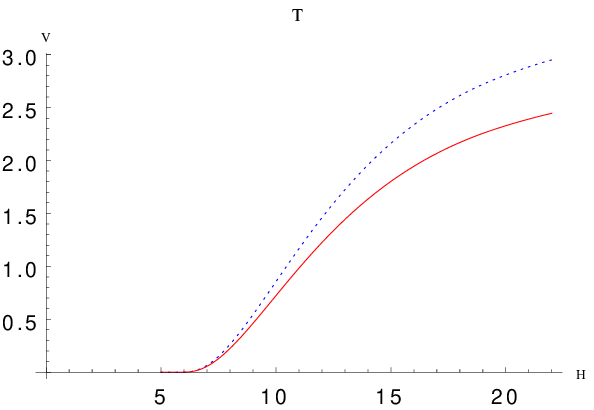}
\vspace{.2cm}
\caption{Integrated cross section for a photon and a  longitudinally polarized $\rho$ meson production, on a proton (left) or neutron (right) target. The solid red curves correspond to the ``valence'' scenario while the
dotted blue curves correspond to the ``standard'' one. 
}
\label{Fig:sigmaEVEN}
\end{center}
\end{figure}

Then, for the obtained cross section which was obtained at a given value of  
$\tilde{S}_{\gamma N}$, the integration over the $(-t,-u')$ phase-space and then over $M^2_{\gamma \rho}$ is performed similarly to $S_{\gamma N}=20~$GeV$^2$ case.
One finally gets the integrated cross section shown in Fig.~\ref{Fig:sigmaEVEN} for both the proton and neutron target, and for both parameterization of the axial GPDs\footnote{A quadratic extrapolation is performed for the small domain above $S_{\gamma N}=20~{\rm GeV}^2.$}.  These cross sections prove that our process is measurable in the typical kinematical conditions and integrated luminosity of a JLab experiment. Counting rates on a proton target are predicted to be one order of magnitude larger than on a neutron target.

\psfrag{H}{\hspace{-1.5cm}\raisebox{-.6cm}{\scalebox{.7}{$S_{\gamma N} ({\rm GeV}^{2})$}}}
\psfrag{V}{\raisebox{.3cm}{\scalebox{.7}{$\hspace{-.4cm}\displaystyle\sigma_{even}~({\rm pb})$}}}

\subsection{Results for the chiral-odd case}
\label{SubSec:chiral-odd}

Let us now pass to the chiral-odd case,  where a transversely polarized $\rho$ meson is produced together with the photon. This process now probes the chiral-odd transversity quark distributions which are connected to the transversity PDFs.

In order to evaluate the theoretical uncertainty in the chiral-odd sector, for each of the two parameterizations of the transversity PDFs, we use 
a set of 1500 trials with their value of the $\chi^2$ test, as provided by the authors of Ref.~\cite{Anselmino:2013vqa}, between $-2 \sigma$ and $+2 \sigma.$ Their 9-parameters $\chi^2$ distribution (see the appendix of Ref.~\cite{Anselmino:2008sga} for details) is given by
\beqa
\label{Pchi2}
P_{\chi^2}(x)=\frac{e^{-x/2} x^{7/2}}{105 \sqrt{2 \pi}}
\eqa
We further renormalize this distribution in order to include on one hand the fact that the 1500 trials only cover the $[ -2 \sigma, +2 \sigma]$ interval, and on the other hand discretization corrections. We then create a histogram of these configurations, with a distribution weighted by the above described renormalized $\chi^2$ distribution. This weighted histogram allows us to finally compute the $-2 \sigma$ and  $+2 \sigma$
values of the cross-section. We perform this analysis at $-u'=1~{\rm GeV}^2$ and for three typical values of $M^2_{\gamma \rho}$ (2.2, 4, 6~{\rm GeV}$^2$), for the ``standard'' scenario. We then extract the two typical configurations which gives cross-section close to the $-2 \sigma$ and  $+2 \sigma$ values, which we now use both for the ``standard'' and ``valence'' scenarios in order to evaluate the typical theoretical uncertainty.

Fig.~\ref{Fig:dsigmaODDdM2dupdtSgN20M3,4,5,6} shows the  $M^2_{\gamma \rho}$ dependence of this cross section, both for the proton and the neutron. Similarly to the chiral-even case, the production of the $\gamma \rho$ pair with a large value of $M^2_{\gamma \rho}$ is severely suppressed. Similarly, the $-u'$ range allowed by our kinematical requirements is narrower for smaller values of  $M^2_{\gamma \rho}$. Comparing the chiral-even case, see Figs.~\ref{Fig:dsigmaEVENdM2dupdtSgN20M4U+D_U_D},
\ref{Fig:dsigmaEVENdM2dupdtSgN20M4V+A_V_A},
\ref{Fig:dsigmaEVENdM2dupdtSgN20M3,4,5,6}
and the chiral-odd case,
see Fig.~\ref{Fig:dsigmaODDdM2dupdtSgN20M3,4,5,6},
one should note the very different behavior of the  differential cross section when varying $-u'$.
In the case of a proton probe, we show in Fig.~\ref{Fig:dsigmaODDdM2dupdtSgN20M3,4,5,6} (left) as error bands
the maximal and minimal values of the cross-section (the maximal values are obtained with the ``standard'' trial at $+2 \sigma$ and the minimal values with the ``valence'' trial at 
$-2 \sigma$).

%
%
\psfrag{H}{\hspace{-1.5cm}\raisebox{-.6cm}{\scalebox{.7}{$-u' ({\rm GeV}^{2})$}}}
\psfrag{V}{\raisebox{.3cm}{\scalebox{.7}{$\hspace{-.4cm}\displaystyle\frac{d \sigma_{\rm odd}}{d M^2_{\gamma \rho} d(-u') d(-t)}~({\rm pb} \cdot {\rm GeV}^{-6})$}}}
\psfrag{T}{}
\begin{figure}[!h]
\begin{center}
\psfrag{vide}{}
\hspace{.2cm}\includegraphics[width=7.3cm]{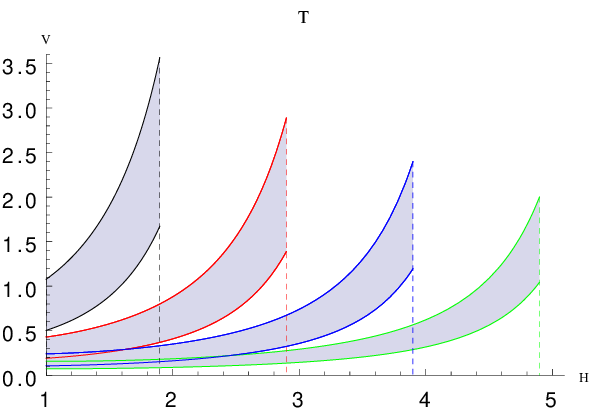}
\hspace{0.1cm}\includegraphics[width=7.3cm]{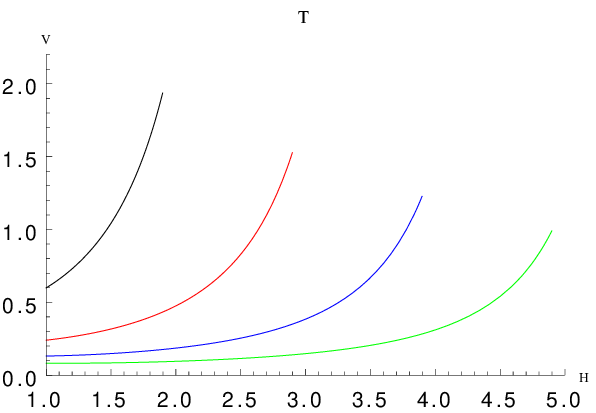}
\vspace{.2cm}
\caption{
Differential cross section for a photon and a  transversally polarized $\rho$ meson production, for the proton (left) and the neutron (right), as a function of $-u'$, for $M_{\gamma \rho}^2=3,4,5,6~{\rm GeV}^2$
(resp. in black, red, blue, green, from left to right). The error bands on the l.h.s. panel (proton) correspond to the procedure discussed in the text. For the neutron, we only show the results for the ``valence'' case.
}
\label{Fig:dsigmaODDdM2dupdtSgN20M3,4,5,6}
\end{center}
\end{figure}

\psfrag{H}{\hspace{-1.5cm}\raisebox{-.6cm}{\scalebox{.7}{$M^2_{\gamma \rho}~({\rm GeV}^{2})$}}}
\psfrag{V}{\raisebox{.3cm}{\scalebox{.7}{$\hspace{-.4cm}\displaystyle\frac{d\sigma_{odd}}{d M^2_{\gamma\rho}}~({\rm pb} \cdot {\rm GeV}^{-2})$}}}
\begin{figure}[!h]
\begin{center}
\psfrag{T}{}
\hspace{.2cm}\includegraphics[width=7.3cm]{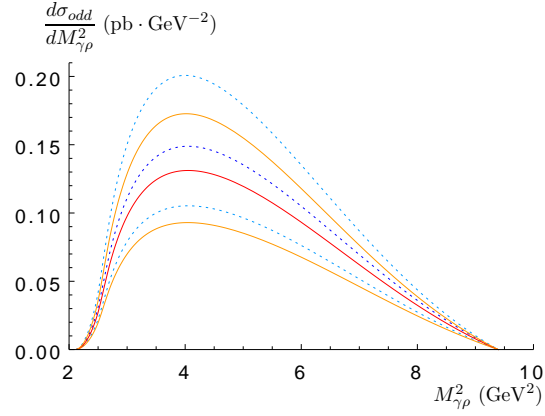}
\vspace{.2cm}
\caption{
Differential cross section $d\sigma/dM^2_{\gamma \rho}$ for a photon
and a  transversally polarized $\rho$ meson production on a proton target for $S_{\gamma N}=20~{\rm GeV}^{2}$. The various curves differ with respect to the ans\"atze for the PDFs $\delta q$ used to build the GPD $H_T$. The dotted curves correspond to the ``standard'' polarized PDFs while the solid curves use the ``valence'' polarized PDFs.
The deep-blue and red curves are central values while the light-blue and orange ones are the results obtained at $\pm 2 \sigma.$}
\label{Fig:dsigmaODDdM2SgN20-incertitudes}
\end{center}
\end{figure}

Similarly to the chiral-even case, we perform the integration in the $(-t,-u')$ phase-space. The obtained differential cross section $d\sigma_{\rm odd}/dM^2_{\gamma \rho}$ is shown in
Fig.~\ref{Fig:dsigmaODDdM2SgN20-incertitudes}
for $S_{\gamma N}=20~{\rm GeV}^{2},$ with the different sets of results depending on the sets of transversity PDFs which we use, as explained above.

In 
Fig.~\ref{Fig:dsigmaODDdM2SgN8,10,12,14,16,18,20},
we show the differential cross section $d\sigma_{\rm odd}/dM^2_{\gamma \rho}$ 
for various values of $S_{\gamma N} $ covering the JLab-12 energy range. These cross sections show a maximum around a similar range of $M^2_{\gamma \rho}\approx 3~$GeV$^2$, for most energy values.

\psfrag{H}{\hspace{-1.5cm}\raisebox{-.7cm}{\scalebox{.7}{$M^2_{\gamma \rho} ({\rm GeV}^{2})$}}}
\psfrag{V}{\raisebox{.3cm}{\scalebox{.7}{$\hspace{-.7cm}\displaystyle\frac{d \sigma_{\rm odd}}{d M^2_{\gamma \rho}}~({\rm pb} \cdot {\rm GeV}^{-2})$}}}
\psfrag{T}{}
\begin{figure}[!h]
\begin{center}
\includegraphics[width=10cm]{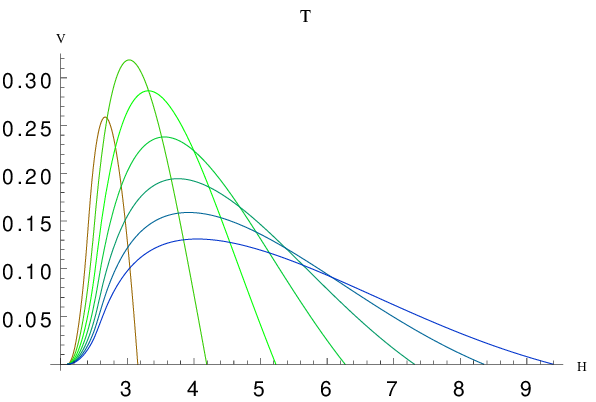}
\vspace{.4cm}

\caption{
Differential cross section $d\sigma/dM^2_{\gamma \rho}$ for a photon and a  transversally polarized $\rho$ meson production on a proton target. The values of $S_{\gamma N}$ vary in the set 8, 10, 12, 14, 16, 18, 20 ${\rm GeV}^{2}.$ (from 8: left, brown to 20: right, blue), covering the JLab energy range. We use here the ``valence'' scenario.
}
\label{Fig:dsigmaODDdM2SgN8,10,12,14,16,18,20}
\end{center}
\end{figure}

\begin{figure}[!h]
\begin{center}
\includegraphics[width=7cm]{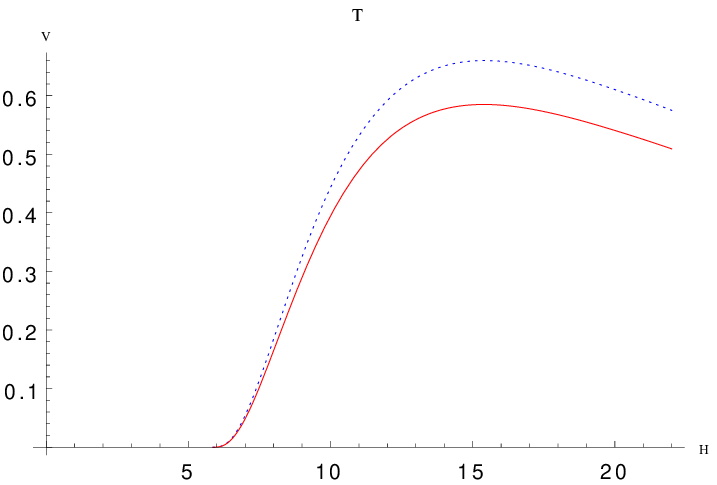}
\includegraphics[width=7cm]{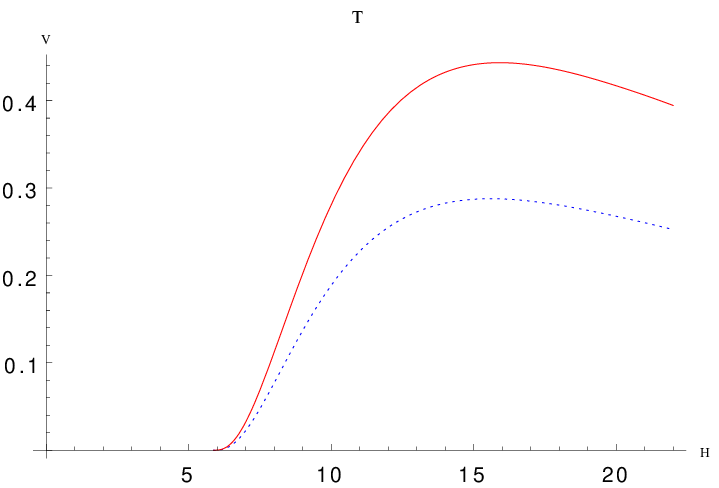}
\vspace{.4cm}

\caption{Integrated cross section for a photon and a  transverse $\rho$ meson production, on a proton (left) or neutron (right) target, as a function of $S_{\gamma N}.$ 
The solid red curves correspond to the ``valence'' scenario while the
dotted blue curves correspond to the ``standard'' one. 
}
\label{Fig:sigmaODD}
\end{center}
\end{figure}

Finally, the dependency of the integrated cross section with respect to $S_{\gamma N}$ is shown in Fig.~\ref{Fig:sigmaODD},
both for proton and neutron, for the two ``valence'' and ``standard'' scenarios.

\psfrag{H}{\hspace{-1.5cm}\raisebox{-.6cm}{\scalebox{.7}{$S_{\gamma N}~({\rm GeV}^{2})$}}}
\psfrag{V}{\raisebox{.3cm}{\scalebox{.7}{$\hspace{-.4cm}\displaystyle\sigma_{odd}~({\rm pb})$}}}

\section{Counting rates}
\label{Sec:rates}

\begin{figure}[!h]
\begin{center}
\vspace{1cm}

\hspace{.1cm}
\psfrag{V}{\raisebox{.3cm}{\scalebox{.7}{$\hspace{-.4cm}\displaystyle F_{even}(S_{\gamma N})~({\rm pb \cdot {\rm GeV}^{-2}})$}}}
\includegraphics[width=7.2cm]{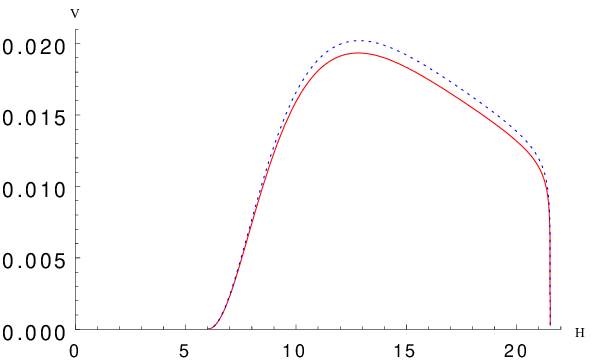}
\hspace{.2cm}
\psfrag{V}{\raisebox{.3cm}{\scalebox{.7}{$\hspace{-.4cm}\displaystyle F_{odd}(S_{\gamma N})~({\rm pb \cdot {\rm GeV}^{-2}})$}}}
\includegraphics[width=7.2cm]{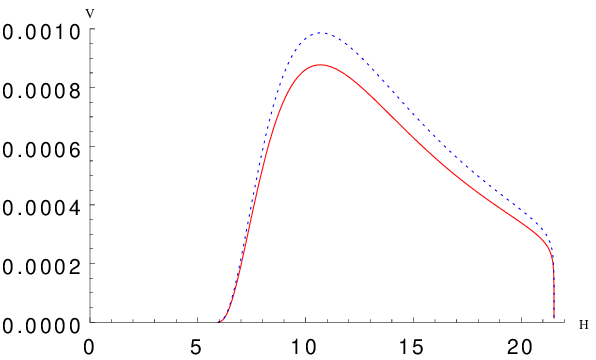}
\vspace{.4cm}

\caption{Shape of the integrand of $\sigma_{e N}$, as a function of the invariant mass of the hadronic produced state, on a proton target. Left: longitudinally polarized $\rho$ meson production. Right: transverse $\rho$ meson production. In 
solid-red: ``valence''. In dotted-blue: ``standard''.}
\label{Fig:F-WW}
\end{center}
\end{figure}

Using the Weizs\"acker-Williams distribution, one can obtain counting rates. This distribution is given by~\cite{Kessler:1975hh,Frixione:1993yw}
\beqa
\label{WW}
f(x)=\frac{\alpha_{\rm em}}{2 \pi}
\left\{2 m_e^2 x
\left(\frac{1}{Q^2_{\rm max}} -\frac{1-x}{m_e^2 x^2}  \right)
+ \frac{\left((1-x)^2+1\right) 
\ln \frac{Q^2_{\rm max}(1-x)}{m_e^2 x^2}}x
\right\},
\eqa
where $x$ is the fraction of energy lost by the incoming electron, $m_e$ is the electron mass and $Q^2_{\rm max}$ is the typical maximal value of the virtuality of the echanged photon, which we take to be $0.1$~GeV$^{2}.$
Using the expression for $x$ as a function of the incoming electron energy $E_e$ 
\beqa
\label{xSgammaN_Ee}
x[S_{\gamma N}] = \frac{S_{\gamma N} - M^2}{2 E_e M},
\eqa
one can easily obtain integrated cross sections at the level
of the $e N$ process, using the relation
\beqa
\label{sigma-WW}
\sigma_{e N} = \int \sigma_{\gamma N}(x)\, f(x)\, dx = \int_{S_{\gamma N \, {\rm crit}}}^{S_{\gamma N \, {\rm max}}} \frac{1}{2 E_e M} 
 \, \sigma_{\gamma N}(x[S_{\gamma N}]) \, f(x[S_{\gamma N}])
 \,dS_{\gamma N} \,.
\eqa
The shape of the integrand 
\beqa
\label{def-F-WW}
F(S_{\gamma N})=\frac{1}{2 E_e M} 
 \, \sigma_{\gamma N}(x[S_{\gamma N}]) \, f(x[S_{\gamma N}])
 \eqa
 of Eq.~(\ref{sigma-WW})
 is shown in Fig.~\ref{Fig:F-WW}.
 
\psfrag{H}{\hspace{-1.5cm}\raisebox{-.6cm}{\scalebox{.7}{$S_{\gamma N}~({\rm GeV}^{2})$}}}

 Up to now we discussed photoproduction of $\gamma \rho$ pair without paying attention to the origin of the quasi-real
initial photon. If it is emitted by a lepton beam, like in electroproduction of photon via DVCS, one should also consider Bethe-Heitler-type processes, in which the final real photon is emitted by the lepton beam.
Let us however note that this mechanism involves an off-shell photon of momentum $q$, since in this case $q^2=(p_\rho +
\Delta)^2 \approx -2\xi s \alpha_\rho $ is large. Thus the Bethe-Heitler mechanism involves scattering
amplitudes with  four hard propagators, whereas the photoproduction mechanism considered so far involves
only three hard propagators. We therefore expect the Bethe-Heitler contribution to be suppressed.
A more precise discussion is left for the future.

At this point, we did not include any experimental constraint on the angular coverage of the final state particles. We discuss this issue in appendix~\ref{App:theta-cut}, taking the constraints of JLab Hall B and showing that this does not affect our predictions. We also show that a binning in the 
outgoing photon angle could help to enhance the chiral-odd versus chiral-even ratio, in particular for observables sensitive to the interference of the two amplitudes, which are beyond the scope of the present paper.

We can now give our predictions for the counting rates. With an expected luminosity ${\cal L}=100~{\rm nb}^{-1}s^{-1}$ we obtain for 100 days of run:
$7.5~10^3$ $\rho_T$ and $1.9~10^5$ $\rho_L\,.$ 

\section{Conclusion}

The analysis of the process $\gamma N \to \gamma \rho^0 N'$ in the generalized Bjorken kinematics where GPD factorization is expected to hold in a collinear QCD approach has shown interesting features. 

Firstly, although any helicity state of the vector meson is populated at the same level in the twist expansion of the amplitude, the production of longitudinally polarized vector mesons  turns out to be numerically dominant. 
This mostly comes from the difference in the normalization  of chiral odd versus chiral even GPDs, as shown
in our modelization (see Figs.~\ref{Fig:GPD-H}-\ref{Fig:GPD-HT}). If our model underestimates the chiral odd GPDs (which might well be the
case, since the constraints on the transversity distributions are still quite indirect), the data rates for
$\rho_T$ production will be higher. 

Secondly, the magnitude of the cross section is large enough for the process to be analyzed in a quite detailed way by near-future experiments at JLab with photon beams originating from the 12 GeV electron beam. Detectors in Hall B, C and D seem to be perfectly suited for this study. A more detailed study is needed to decide on the feasibility of the experiment when taking into account of all detection efficiencies.

We restricted our analysis to unpolarized cross sections; this may be complemented by a computation of various polarization observables.

A NLO calculation should first confirm the validity of the  factorization hypothesis for this process, and estimate the effects on the amplitude. Let us stress that, contrary to the DVCS (and TCS) case where gluon contributions turn out to be important at this level \cite{Pire:2011st,Moutarde:2013qs}, the charge conjugation property of the process studied here protects us from these contributions. This does not exempt us to perform such a next to leading order computation, in the spirit of the study of Ref.~\cite{Nizic:1987sw,Duplancic:2006nv} in the $\gamma \gamma$ channel, but this may
help NLO corrections to be under control without the necessity of a resummation procedure.

To conclude, the cross section of our process is a factor 10 more 
than the $\gamma P \to P e^+ e^-$ process, for similar values of the hard scale, 
for which experimental proposals have been approved at JLab. Thus, the study of our process appears feasible
experimentally and promises to bring new important constraints on GPD physics.

We would like to mention that a similar study could be performed in principle
in the Compass experiment at CERN where $S_{\gamma N} \sim$ 200 GeV$^2$ and at LHC in ultraperipheral collisions~\cite{N.Cartiglia:2015gve}, as well as in future electron proton collider projects like EIC~\cite{Boer:2011fh} and LHeC~\cite{AbelleiraFernandez:2012cc}.

\section*{Acknowledgements}

We acknowledge several useful discussions with Mauro Anselmino, Marie Bo\"er, Goran Duplan\v ci\'c, Michel Guidal, Stefano Melis, Herv\'e Moutarde, St\'ephane Munier, Kornelija Passek-Kumeri\v cki, Franck Sabati\' e and Daniel Tapia-Takaki. This work is partly supported by grant No 2015/17/B/ST2/01838 by the National Science Center in Poland, by the French grant ANR PARTONS (Grant No. ANR-12-MONU-0008-01), by the COPIN-IN2P3 agreement, by the Labex P2IO and by the Polish-French collaboration agreement Polonium.

\appendix

\section{Contributions of the various diagrams} 
\label{App:diagrams}

For completeness, we present here the formulae for the contributions of the various diagrams of Fig.~\ref{Fig:diagrams}.

\subsection{Chiral-even sector}

\subsubsection{Vector case}
 
\begin{eqnarray}
\label{trCEVA_1trace}
  tr_D^V\left[ A_1\right] 
&=&   tr_D\left[   \hat p_\rho  \hat \varepsilon_k^*\frac{z \hat p_\rho +\hat k }{(z p_\rho+k)^2+i\epsilon }\hat \varepsilon_q  \frac{z \hat p_\rho +\hat k- \hat q}{(z p_\rho +k-q )^2+i\epsilon}
 \,\gamma^\mu \,\hat p \,\gamma_\mu\,\frac{1}{(-\bar{z} p_\rho-(x-\xi)p)^2+i\epsilon}\right]\!
\nonumber  \\
 &=& \frac{  8\bar z s \bar \alpha \left[   \frac{2z-1}{\alpha} (\varepsilon_{q\perp}\cdot p_{\rho \bot}) (\varepsilon_{k\perp}^*\cdot p_{\rho \bot}) - s\xi  (\varepsilon_{q\perp}\cdot    \varepsilon_{k\perp}^*)\right]            }{   (  (z p_\rho+k)^2+i\epsilon  )     ( (z p_\rho +k-q )^2+i\epsilon    )        ( (-\bar{z} p_\rho-(x-\xi)p)^2+i\epsilon    ) }
\\ \nonumber \\ \nonumber
&=& \frac{2\left[\alpha\xi sT_{A}-\left(z-\bar{z}\right)T_{B}\right]}{\alpha\bar{\alpha}\xi^{2}s^{2}z\bar{z}\left(x-\xi+i\epsilon\right)},             
\end{eqnarray}   
\begin{eqnarray}
\label{trCEVA_2trace}  
tr_D^V\left[ A_2\right]
&=&   tr_D\left[   \hat p_\rho  \hat \varepsilon_q\frac{z \hat p_\rho -\hat q }{(z p_\rho-q)^2+i\epsilon }\hat \varepsilon_k^*  \frac{z \hat p_\rho +\hat k- \hat q}{(z p_\rho +k-q )^2+i\epsilon}
 \,\gamma^\mu \,\hat p \,\gamma_\mu\,\frac{1}{(-\bar{z} p_\rho-(x-\xi)p)^2+i\epsilon}\right]
 \nonumber \\
 &=& \frac{    8\bar z s \bar \alpha \left[   \frac{2z-1}{\alpha} (\varepsilon_{q\perp}\cdot p_{\rho \bot}) (\varepsilon_{k\perp}^*\cdot p_{\rho \bot}) + s\xi \alpha  (\varepsilon_{q\perp}\cdot    \varepsilon_{k\perp}^*)\right]           }{   (  (z p_\rho-q)^2+i\epsilon   )   (   (z p_\rho +k-q )^2+i\epsilon  )    (  (-\bar{z} p_\rho-(x-\xi)p)^2+i\epsilon   ) }
\\ \nonumber \\ \nonumber
&=& \frac{2\left[\alpha^{2}\xi sT_{A}+\left(z-\bar{z}\right)T_{B}\right]}{\alpha^{2}\bar{\alpha}\xi^{2}s^{2}z\bar{z}\left(x-\xi+i\epsilon\right)} ,
\end{eqnarray}   
\begin{eqnarray}
\label{trCEVA_3trace}
  tr_D^V\left[ A_3\right] 
&=&   tr_D\left[   \hat p_\rho  \hat \varepsilon_q\frac{z \hat p_\rho -\hat q }{(z p_\rho-q)^2+i\epsilon } \gamma^\mu \frac{(x+\xi)\hat p - \hat k}{((x+\xi)p-k)^2+i\epsilon}
 \,\hat \varepsilon_k^*\,\hat p \,\gamma_\mu\,\frac{1}{(\bar{z} p_\rho+(x-\xi)p)^2+i\epsilon}\right]
 \nonumber
 \\
&=&    \frac{  8s\alpha\,\left[  -\frac{z}{\alpha} (\varepsilon_{q\perp}\cdot p_{\rho \bot})\, (\varepsilon_{k\perp}^*\cdot p_{\rho \bot}) -s\xi ( \varepsilon_{q\perp}\cdot \varepsilon_{k\perp}^*)      \right]   }{    
   (  (z p_\rho-q)^2+i\epsilon  )    ( ((x+\xi)p-k)^2+i\epsilon   )       (  (\bar{z} p_\rho+(x-\xi)p)^2+i\epsilon  )      }
\\ \nonumber \\ \nonumber
&=& \frac{-4\left[\alpha\xi sT_{A}+zT_{B}\right]}{\alpha^{2}\bar{\alpha}\xi s^{2}z\bar{z}\left(x+\xi-i\epsilon\right)\left(x-\xi+i\epsilon\right)},
  \end{eqnarray}  
\begin{eqnarray}
\label{trCEVA_4trace}
  tr_D^V\left[ A_4\right] 
&=&  \! tr_D\!\left[   \hat p_\rho  \hat \varepsilon_q\frac{z \hat p_\rho -\hat q }{(z p_\rho-q)^2+i\epsilon } \gamma^\mu \hat p  \hat \varepsilon_k^*\frac{\hat k +(x-\xi)\hat p}{(k+(x-\xi)p)^2+i\epsilon}
 \gamma_\mu \frac{1}{( z p_\rho-q -(x+\xi)p)^2\!+\!i\epsilon}\right]
 \nonumber
  \\
&=&  \frac{  8s\left[  -\bar{z}  \,( \varepsilon_{q\perp}\cdot p_{\rho \bot})\, (\varepsilon_{k\perp}^*\cdot p_{\rho \bot}) - s \xi\, \alpha \alpha_\rho\,(\varepsilon_{q\perp}\cdot \varepsilon_{k\perp}^*)    \right]         }{  (   (z p_\rho-q)^2+i\epsilon   )  (   (k+(x-\xi)p)^2+i\epsilon  )   ( ( z p_\rho-q -(x+\xi)p)^2+i\epsilon  ) }
\\ \nonumber \\ \nonumber
&=& \frac{4\left[\alpha\bar{\alpha}\xi sT_{A}+\bar{z}T_{B}\right]}{\alpha^{2}\xi s^{2}z\left(x-\xi+i\epsilon\right)\left[\left(x+\xi+i\epsilon\right)-z\left(2\alpha\xi+\bar{\alpha}\left(x+\xi+i\epsilon\right)\right)\right]} ,
\end{eqnarray}  
\begin{eqnarray}
\label{trCEVA_5trace}
  tr_D^V\left[ A_5\right] 
&=&   tr_D\left[   \hat p_\rho  \hat \varepsilon_q\frac{z \hat p_\rho -\hat q }{(z p_\rho-q)^2+i\epsilon } \gamma^\mu \hat p \gamma_\mu  \frac{- \bar z \hat p_\rho - \hat k }{(- \bar z p_\rho -k)^2+i\epsilon}\,\varepsilon_k^*
 \,\frac{1}{( z p_\rho-q -(x+\xi)p)^2+i\epsilon}\right]
 \nonumber \\
 &=& \frac{  8s\left[ \frac{ (\bar{z} - z)(z\bar \alpha -1) }{\alpha} \,(\varepsilon_{q\perp}\cdot p_{\rho \bot})\, (\varepsilon_{k\perp}^*\cdot p_{\rho \bot}) + s \xi\, \alpha \,(\varepsilon_{q\perp}\cdot \varepsilon_{k\perp}^*)    \right]             }{   (  (z p_\rho-q)^2+i\epsilon  )    ( (- \bar z p_\rho -k)^2+i\epsilon   )     (  ( z p_\rho-q -(x+\xi)p)^2+i\epsilon  )}
\\ \nonumber \\ \nonumber
&=& \frac{2\left[-\alpha^{2}\xi sT_{A}+\left(1-2z\right)\left(1-\bar{\alpha}z\right)T_{B}\right]}{\alpha^{2}\xi^{2}s^{2}z\bar{z}\left[\left(x+\xi+i\epsilon\right)-z\left(2\alpha\xi+\bar{\alpha}\left(x+\xi+i\epsilon\right)\right)\right]},
\end{eqnarray}  
\begin{eqnarray}
 &&\hspace{-.2cm} tr_D^V\left[ B_2\right] \nonumber \\
&&\hspace{-.2cm}=   tr_D\! \left[   \hat p_\rho   \gamma^\mu      \frac{\hat q + (x+\xi) \hat p - \hat k   }{(q+(x+\xi)p-k)^2+i\epsilon }
\hat \varepsilon_k^* \, \frac{\hat q +(x+\xi)\hat p}{(q+(x+\xi)p)^2\!+\!i\epsilon}\,\hat \varepsilon_q\,\hat p\,\gamma_\mu
 \frac{1}{( -\bar z p_\rho -(x-\xi)p)^2+i\epsilon}\right]\!
 \nonumber \\
 &&\hspace{-.2cm}= \frac{   4s^2(x-\xi)\bar \alpha   (\varepsilon_{q\perp}\cdot \varepsilon_{k\perp}^* )    }{     (  q+(x+\xi)p-k)^2+i\epsilon  )    ( (q+(x+\xi)p)^2+i\epsilon   )     ( ( -\bar z p_\rho -(x-\xi)p)^2+i\epsilon   )}
 \label{trCEVB_2trace}
\\ \nonumber \\ \nonumber
&&\hspace{-.2cm}= \frac{4 \,T_{A}}{\bar{\alpha}\bar{z}s\left(x+\xi+i\epsilon\right)\left(x-\xi+i\epsilon\right)},
\end{eqnarray}  
\begin{eqnarray}
\label{trCEVB_3trace}
&&\hspace{-.2cm}  tr_D^V\left[ B_3\right]  \nonumber \\
&&\hspace{-.2cm}= \! tr_D\!\left[   \hat p_\rho    \gamma^\mu      \frac{\hat q + (x+\xi) \hat p - \hat k   }{(q+(x+\xi)p-k)^2+i\epsilon }
\hat \varepsilon_q \, \frac{(x+\xi)\hat p -\hat k}{((x+\xi)p-k)^2+i\epsilon}\,\hat \varepsilon_k^*\,\hat p\,\gamma_\mu
 \frac{1}{( -\bar z p_\rho -(x-\xi)p)^2\!+\!i\epsilon}\right]
 \nonumber \\
 &&\hspace{-.2cm}=  - \,\frac{4s^2\bar \alpha \alpha (x-\xi)   ( \varepsilon_{q\perp}\cdot \varepsilon_{k\perp}^* )     }{  (  (q+(x+\xi)p-k)^2+i\epsilon )    (  ((x+\xi)p-k)^2+i\epsilon )    (  ( -\bar z p_\rho -(x-\xi)p)^2+i\epsilon  )}
 \\ \nonumber \\ \nonumber
 &&\hspace{-.2cm}= \frac{4T_{A}}{\bar{\alpha}\bar{z}s\left(x+\xi-i\epsilon\right)\left(x-\xi+i\epsilon\right)},
\end{eqnarray}  
\begin{eqnarray}
\label{trCEVB_4trace}
&&\hspace{-.2cm}  tr_D^V\left[ B_4\right] \nonumber \\
&&\hspace{-.2cm}=   tr_D\left[   \hat p_\rho \gamma^\mu \frac{ \hat q +(x+\xi)\hat p }{(q+(x+\xi)p)^2+i\epsilon } \hat \varepsilon_q \hat p  \hat \varepsilon_k^*
\frac{ (x-\xi) \hat p +\hat k }{((x-\xi)p +k)^2+i\epsilon}
 \,\gamma_\mu\,\frac{1}{(\bar  z p_\rho +k +(x-\xi)p)^2+i\epsilon}\right]
 \nonumber \\
&&\hspace{-.2cm}= \frac{   8s^2\xi \alpha  (\varepsilon_{q\perp}\cdot \varepsilon_{k\perp}^*)   }{   ( (q+(x+\xi)p)^2+i\epsilon  )     (  ((x-\xi)p +k)^2+i\epsilon  )     (   (\bar  z p_\rho +k +(x-\xi)p)^2+i\epsilon )}
 \\ \nonumber \\ \nonumber
&&\hspace{-.2cm}= \frac{8\xi T_{A}}{\left(x-\xi+i\epsilon\right)\left(x+\xi+i\epsilon\right)s\left[\left(x+\xi+i\epsilon\right)-z\left(2\alpha\xi+\bar{\alpha}\left(x+\xi+i\epsilon\right)\right)\right]},
\end{eqnarray}
\begin{eqnarray}
\label{trCEVB_5trace}
&&\hspace{-.2cm}   tr_D^V\left[ B_5\right] \nonumber \\
&&\hspace{-.2cm}=   tr_D\left[   \hat p_\rho \gamma^\mu \frac{ \hat q +(x+\xi)\hat p }{(q+(x+\xi)p)^2+i\epsilon } \hat \varepsilon_q \hat p  \gamma_\mu  
\frac{ -\bar{z}\hat p_\rho - \hat k }{(-\bar{z} p_\rho -k)^2+i\epsilon}
 \,\hat \varepsilon_k^*\,\frac{1}{(\bar  z p_\rho +k +(x-\xi)p)^2+i\epsilon}\right]
 \nonumber
  \\
&&\hspace{-.2cm}=  \frac{ 8s \left[  \frac{z}{\alpha}\, (\varepsilon_{q\perp}\cdot p_{\rho \bot})\, (\varepsilon_{k\perp}^*\cdot p_{\rho \bot}) +\, s \xi \alpha_\rho\,  (\varepsilon_{q\perp}\cdot \varepsilon_{k\perp}^*)     \right]     }{  (   (q+(x+\xi)p)^2+i\epsilon  )    (  (-\bar{z} p_\rho -k)^2+i\epsilon  )   (  (\bar  z p_\rho +k +(x-\xi)p)^2+i\epsilon  )}\,,
\\ \nonumber \\ \nonumber
&&\hspace{-.2cm}= \frac{4\left[\alpha\bar{\alpha}\xi sT_{A}+zT_{B}\right]}{\alpha\xi s^{2}\bar{z}\left(x+\xi+i\epsilon\right)\left[\left(x+\xi+i\epsilon\right)-z\left(2\alpha\xi+\bar{\alpha}\left(x+\xi+i\epsilon\right)\right)\right]} .
\end{eqnarray}
 
\subsubsection{Axial case}

\begin{eqnarray}
\label{trCEAA_1trace}  
&&\hspace{-.2cm}   tr_D^A\left[ A_1\right] \nonumber \\
&&\hspace{-.2cm} =   tr_D\left[   \hat p_\rho  \hat \varepsilon_k^*\frac{z \hat p_\rho +\hat k }{(z p_\rho+k)^2+i\epsilon }\hat \varepsilon_q  \frac{z \hat p_\rho +\hat k- \hat q}{(z p_\rho +k-q)^2+i\epsilon}
 \,\gamma^\mu \,\hat p \,\gamma^5\,\gamma_\mu\,\frac{1}{(-\bar{z} p_\rho-(x-\xi)p)^2+i\epsilon}\right]
 \nonumber \\
&&\hspace{-.2cm}=  \frac{8i\frac{\bar z}{\alpha} \left[    (1-2\alpha ) (\varepsilon_{q\perp}\cdot p_{\rho \bot}) \, \epsilon^{p\,n\,p_{\rho \bot}\,\varepsilon_{k\perp}^*}  +\, (\varepsilon_{k\perp}^*\cdot p_{\rho \bot})\,\epsilon^{p\,n\,p_{\rho \bot}\, \varepsilon_{q\perp}}    \right]  }{  (  (z p_\rho+k)^2+i\epsilon  )    ((z p_\rho +k-q)^2+i\epsilon   )   (  (-\bar{z} p_\rho-(x-\xi)p)^2+i\epsilon  )  }\nonumber
 \\ \nonumber \\ 
&&\hspace{-.2cm} = \frac{2i\left[-T_{A_5}-\left(\alpha-\bar{\alpha}\right)T_{B_5}\right]}{\alpha\bar{\alpha}^{2}\xi^{2}s^{3}z\bar{z}\left(x-\xi+i\epsilon\right)},
\end{eqnarray}   
\begin{eqnarray}
\label{trCEAA_2trace}  
&&\hspace{-.2cm}   tr_D^A\left[ A_2\right] \nonumber \\
&&\hspace{-.2cm} =   tr_D\left[   \hat p_\rho  \hat \varepsilon_q\frac{z \hat p_\rho -\hat q }{(z p_\rho-q)^2+i\epsilon }\hat \varepsilon_k^*  \frac{z \hat p_\rho +\hat k- \hat q}{(z p_\rho +k-q)^2+i\epsilon}
 \,\gamma^\mu \,\hat p \,\gamma^5\,\gamma_\mu\,\frac{1}{(-\bar{z} p_\rho-(x-\xi)p)^2+i\epsilon}\right]
 \nonumber \\
&&\hspace{-.2cm} =  -\, \frac{ 8i\bar z \,\left[    (\varepsilon_{q\perp}\cdot p_{\rho \bot}) \, \epsilon^{p\,n\,p_{\rho \bot}\,\varepsilon_{k\perp}^*}  -\,\frac{2-\alpha}{\alpha} (\varepsilon_{k\perp}^*\cdot p_{\rho \bot})\,\epsilon^{p\,n\,p_{\rho \bot}\, \varepsilon_{q\perp}}       \right]    }{   (  (z p_\rho-q)^2+i\epsilon  )    (  (z p_\rho +k-q)^2+i\epsilon  )    (  (-\bar{z} p_\rho-(x-\xi)p)^2+i\epsilon  )}\nonumber
 \\ \nonumber \\ 
&&\hspace{-.2cm} =  \frac{2i\left[-\left(\alpha-2\right)T_{A_5}+\alpha T_{B_5}\right]}{\alpha^{2}\bar{\alpha}^{2}\xi^{2}s^{3}z\bar{z}\left(x-\xi+i\epsilon\right)},
\end{eqnarray}   
\begin{eqnarray}
 \label{trCEAA_3trace}
&&\hspace{-.2cm}    tr_D^A\left[ A_3\right] \nonumber \\
&&\hspace{-.2cm} =   tr_D\left[   \hat p_\rho  \hat \varepsilon_q\frac{z \hat p_\rho -\hat q }{(z p_\rho-q)^2+i\epsilon } \gamma^\mu \frac{(x+\xi)\hat p - \hat k}{((x+\xi)p-k)^2+i\epsilon}
 \,\hat \varepsilon_k^*\,\hat p\, \gamma^5 \,\gamma_\mu\,\frac{1}{(\bar{z} p_\rho+(x-\xi)p)^2+i\epsilon}\right]
 \nonumber
  \\
&&\hspace{-.2cm} =   \frac{  8i\,\left[  \left(  -2z +\frac{1}{\alpha_\rho} \right) \, (\varepsilon_{q\perp}\cdot p_{\rho \bot}) \,\epsilon^{p\,n\,p_{\rho \bot}\,\varepsilon_{k\perp}^*} - \frac{1}{\alpha_\rho} \,
(\varepsilon_{k\perp}^*\cdot p_{\rho \bot}) \, \epsilon^{p\,n\,p_{\rho \bot} \,\varepsilon_{q\perp}}\right]   }{  (   (z p_\rho-q)^2+i\epsilon)   ( ((x+\xi)p-k)^2+i\epsilon  )   (  (\bar{z} p_\rho+(x-\xi)p)^2+i\epsilon )}\nonumber
\\ \nonumber \\ 
&&\hspace{-.2cm} = \frac{4i\left[-T_{A_5}-\left(1-2\bar{\alpha}z\right)T_{B_5}\right]}{\alpha^{2}\bar{\alpha}^{2}\xi s^{3}z\bar{z}\left(x-\xi+i\epsilon\right)\left(x+\xi-i\epsilon\right)},
\end{eqnarray}  
\begin{eqnarray}
\label{trCEAA_4trace}
&&\hspace{-.2cm}   tr_D^A\left[ A_4\right] \nonumber \\
&&\hspace{-.2cm} =   tr_D\left[   \hat p_\rho  \hat \varepsilon_q\frac{z \hat p_\rho -\hat q }{(z p_\rho-q)^2+i\epsilon } \gamma^\mu \hat p \gamma^5 \hat \varepsilon_k^*\frac{\hat k +(x-\xi)\hat p}{(k+(x-\xi)p)^2+i\epsilon}
 \,\gamma_\mu\,\frac{1}{( z p_\rho-q -(x+\xi)p)^2+i\epsilon}\right]
 \nonumber
  \\
&&\hspace{-.2cm} =    \frac{  8i\,\left[  \left(  1 -2z \right) \, (\varepsilon_{q\perp}\cdot p_{\rho \bot}) \,\epsilon^{p\,n\,p_{\rho \bot}\,\varepsilon_{k\perp}^*} + \,
(\varepsilon_{k\perp}^*\cdot p_{\rho \bot}) \, \epsilon^{p\,n\,p_{\rho \bot} \,\varepsilon_{q\perp}}\right]   }{ ( (z p_\rho-q)^2+i\epsilon  )   ( (k+(x-\xi)p)^2+i\epsilon  )   (  ( z p_\rho-q -(x+\xi)p)^2+i\epsilon ) }\nonumber
\\ \nonumber \\ 
&&\hspace{-.2cm} = \frac{4i\left[-T_{A_5}+\left(1-2z\right)T_{B_5}\right]}{\alpha^{2}\xi s^{3}z\left(x-\xi+i\epsilon\right)\left[\left(x+\xi+i\epsilon\right)-z\left(2\alpha\xi+\bar{\alpha}\left(x+\xi+i\epsilon\right)\right)\right]},
\end{eqnarray} 
\begin{eqnarray}
\label{trCEAA_5trace}
&&\hspace{-.2cm}   tr_D^A\left[ A_5\right] \nonumber \\
&&\hspace{-.2cm} =   tr_D\left[   \hat p_\rho  \hat \varepsilon_q\frac{z \hat p_\rho -\hat q }{(z p_\rho-q)^2+i\epsilon } \gamma^\mu \hat p \,\gamma^5\, \gamma_\mu  \frac{- \bar z \hat p_\rho - \hat k }{(- \bar z p_\rho -k)^2+i\epsilon}\,\varepsilon_k^*
 \,\frac{1}{( z p_\rho-q -(x+\xi)p)^2+i\epsilon}\right]
 \nonumber \\
&&\hspace{-.2cm} =  \frac{  8i\,\left[  \frac{2z(1-\alpha )-1}{1-\alpha} \, (\varepsilon_{q\perp}\cdot p_{\rho \bot}) \,\epsilon^{p\,n\,p_{\rho \bot}\,\varepsilon_{k\perp}^*} + \,
 \frac{1+(1-\alpha)(1-2z)}{\alpha(1-\alpha)}
(\varepsilon_{k\perp}^*\cdot p_{\rho \bot}) \, \epsilon^{p\,n\,p_{\rho \bot} \,\varepsilon_{q\perp}}\right]   }{   (  (z p_\rho-q)^2+i\epsilon  )    (   (- \bar z p_\rho -k)^2+i\epsilon )     (  ( z p_\rho-q -(x+\xi)p)^2+i\epsilon )} \nonumber
 \\ \nonumber \\ 
&&\hspace{-.2cm} = \frac{2i\left[-\left(2-\alpha-2\bar{\alpha}z\right)T_{A_5}-\alpha\left(1-2\bar{\alpha}z\right)T_{B_5}\right]}{\alpha^{2}\bar{\alpha}\xi^{2}s^{3}z\bar{z}\left[\left(x+\xi+i\epsilon\right)-z\left(2\alpha\xi+\bar{\alpha}\left(x+\xi+i\epsilon\right)\right)\right]},
\end{eqnarray}  
\begin{eqnarray}
\label{trCEAB_2trace}
&&\hspace{-1cm}   tr_D^A\left[ B_2\right] =   tr_D\left[   \hat p_\rho   \, \gamma^\mu      \frac{\hat q + (x+\xi) \hat p - \hat k   }{(q+(x+\xi)p-k)^2+i\epsilon }
\hat \varepsilon_k^* \, \frac{\hat q +(x+\xi)\hat p}{(q+(x+\xi)p)^2+i\epsilon}\,\hat \varepsilon_q\,\hat p\,\gamma^5\,\gamma_\mu
 \right. \nonumber \\
&&\hspace{-.2cm}\left.\times \frac{1}{( -\bar z p_\rho -(x-\xi)p)^2+i\epsilon}\right]
 \nonumber \\
&&\hspace{-1cm} =  \frac{  4i\,\frac{x-\xi}{\xi \alpha } \left[    (\varepsilon_{q\perp}\cdot p_{\rho \bot}) \, \epsilon^{p\,n\,p_{\rho \bot} \,\varepsilon_{k\perp}^*} -
  (\varepsilon_{k\perp}^*\cdot p_{\rho \bot}) \,\epsilon^{p\,n\,p_{\rho \bot}\,\varepsilon_{q\perp}}    \right]   }{  ( (q+(x+\xi)p-k)^2+i\epsilon  )    ( (q+(x+\xi)p)^2+i\epsilon  )     (  ( -\bar z p_\rho -(x-\xi)p)^2+i\epsilon  )}\nonumber
  \\ \nonumber \\ 
&&\hspace{-1cm} = \frac{4i\left[-T_{A_5}-T_{B_5}\right]}{\alpha\bar{\alpha}^{2}\xi s^{3}\bar{z}\left(x-\xi+i\epsilon\right)\left(x+\xi+i\epsilon\right)},
\end{eqnarray}  
\begin{eqnarray}
\label{trCEAB_3trace}
&&\hspace{-1cm}  tr_D^A\left[ B_3\right] 
=   tr_D\left[   \hat p_\rho   \, \gamma^\mu      \frac{\hat q + (x+\xi) \hat p - \hat k   }{(q+(x+\xi)p-k)^2+i\epsilon }
\hat \varepsilon_q \, \frac{(x+\xi)\hat p -\hat k}{((x+\xi)p-k)^2+i\epsilon}\,\hat \varepsilon_k^*\,\hat p\,\gamma^5\,\gamma_\mu \right. \nonumber \\
&&\hspace{-1cm}\left. \times \frac{1}{( -\bar z p_\rho -(x-\xi)p)^2+i\epsilon}\right]
 \nonumber \\
 &&\hspace{-1cm}= \frac{  4i\,\frac{x-\xi}{\xi } \left[    (\varepsilon_{q\perp}\cdot p_{\rho \bot}) \, \epsilon^{p\,n\,p_{\rho \bot} \,\varepsilon_{k\perp}^*} -
  (\varepsilon_{k\perp}^*\cdot p_{\rho \bot}) \,\epsilon^{p\,n\,p_{\rho \bot}\,\varepsilon_{q\perp}}    \right]   }{   (  (q+(x+\xi)p-k)^2+i\epsilon )    ( ((x+\xi)p-k)^2+i\epsilon  )    ( ( -\bar z p_\rho -(x-\xi)p)^2+i\epsilon  )}\nonumber
  \\ \nonumber \\ 
 &&\hspace{-1cm}= \frac{-4i\left[-T_{A_5}-T_{B_5}\right]}{\alpha\bar{\alpha}^{2}\xi s^{3}\bar{z}\left(x-\xi+i\epsilon\right)\left(x+\xi-i\epsilon\right)},
\end{eqnarray}  
\begin{eqnarray}
\label{trCEAB_4trace}
&&\hspace{-1.5cm}  tr_D^A\left[ B_4\right] 
=   tr_D\left[   \hat p_\rho \gamma^\mu \frac{ \hat q +(x+\xi)\hat p }{(q+(x+\xi)p)^2+i\epsilon } \hat \varepsilon_q \hat p\,\gamma^5\,  \hat \varepsilon_k^*
\frac{ (x-\xi) \hat p +\hat k }{((x-\xi)p +k)^2+i\epsilon}
 \,\gamma_\mu\right. \nonumber \\
&&\hspace{-1cm}\left. \times \frac{1}{(\bar  z p_\rho +k +(x-\xi)p)^2+i\epsilon}\right]
 \nonumber  \\
&&\hspace{-1cm}= \frac{  \frac{8i}{1-\alpha } \left[    (\varepsilon_{q\perp}\cdot p_{\rho \bot}) \, \epsilon^{p\,n\,p_{\rho \bot} \,\varepsilon_{k\perp}^*} -
  (\varepsilon_{k\perp}^*\cdot p_{\rho \bot}) \,\epsilon^{p\,n\,p_{\rho \bot}\,\varepsilon_{q\perp}}    \right]   }{ (  (q+(x+\xi)p)^2+i\epsilon )   (  ((x-\xi)p +k)^2+i\epsilon )    (   (\bar  z p_\rho +k +(x-\xi)p)^2+i\epsilon ) }
\nonumber
  \\ \nonumber \\ 
&&\hspace{-1cm}= \frac{8i\left[-T_{A_5}-T_{B_5}\right]}{\alpha\bar{\alpha}s^{3}\left(x-\xi+i\epsilon\right)\left(x+\xi+i\epsilon\right)\left[\left(x+\xi+i\epsilon\right)-z\left(2\alpha\xi+\bar{\alpha}\left(x+\xi+i\epsilon\right)\right)\right]},
\end{eqnarray} 
\begin{eqnarray}
 \label{trCEAB_5trace}
&&\hspace{-1.5cm}  tr_D^A\left[ B_5\right] 
=  tr_D\left[   \hat p_\rho \gamma^\mu \frac{ \hat q +(x+\xi)\hat p }{(q+(x+\xi)p)^2+i\epsilon } \hat \varepsilon_q \hat p \gamma^5 \gamma_\mu  
\frac{ -\bar{z}\hat p_\rho - \hat k }{(-\bar{z} p_\rho -k)^2+i\epsilon}
 \,\hat \varepsilon_k^*\right. \nonumber \\
&&\hspace{-1.5cm} \left. \times \frac{1}{(\bar  z p_\rho +k +(x-\xi)p)^2+i\epsilon}\right]
 \nonumber
  \\
&&\hspace{-1.5cm}=  -\frac{8i}{\alpha} \frac{ \left[  \left(  2\bar{z} -1 \right) \, (\varepsilon_{k\perp}^*\cdot p_{\rho \bot}) \,\epsilon^{p\,n\,p_{\rho \bot}\,\varepsilon_{q\perp}} - \,
(\varepsilon_{q\perp}\cdot p_{\rho \bot}) \, \epsilon^{p\,n\,p_{\rho \bot} \,\varepsilon_{k\perp}^*}\right]   }{  ( (q+(x+\xi)p)^2+i\epsilon  )  (  (-\bar{z} p_\rho -k)^2+i\epsilon )   (  (\bar  z p_\rho +k +(x-\xi)p)^2+i\epsilon ) }\nonumber
 \\ \nonumber \\ 
&&\hspace{-1.5cm}= \frac{4i\left[-\left(1-2z\right)T_{A_5}-T_{B_5}\right]}{\alpha\xi s^{3}\bar{z}\left(x+\xi+i\epsilon\right)\left[\left(x+\xi+i\epsilon\right)-z\left(2\alpha\xi+\bar{\alpha}\left(x+\xi+i\epsilon\right)\right)\right]}.
\end{eqnarray}

 \subsection{Chiral-odd sector} 
  
  \begin{eqnarray}
  && \hspace{-.5cm}
  tr_D^{CO}\left[ A_3\right] _j   \nonumber
  \\
&& \hspace{-.5cm} = tr_D\left[   \hat p_\rho  \hat \varepsilon^*_\rho \hat \varepsilon_q\frac{z \hat p_\rho -\hat q }{(z p_\rho-q)^2+i\epsilon } \gamma^\mu \frac{(x+\xi)\hat p - \hat k}{((x+\xi)p-k)^2+i\epsilon}
 \,\hat \varepsilon_k^*\,\hat p\, \gamma_{\bot \,j} \,\gamma_\mu\,\frac{1}{(\bar{z} p_\rho+(x-\xi)p)^2+i\epsilon}\right]
 \nonumber
 \\
  &=& \frac{16  \left[ ( p\cdot k ) \,\varepsilon^*_{k \bot j} \left(  (\varepsilon_q\cdot p_\rho) \, (q\cdot \varepsilon^*_\rho) - (q\cdot p_\rho) \,(\varepsilon_q\cdot \varepsilon^*_\rho)     \right)
   - \epsilon^{p_\rho\, \varepsilon^*_\rho\,q\, \varepsilon_q }   \,\epsilon^{p\, \nu\, k\, \varepsilon_k^*}g_{\bot \nu j}      \right]}{      (( (x+\xi)p-k )^2+i\epsilon)^2   \,((zp_\rho -q)^2+i\epsilon)^2 \,(( \bar{z}\,p_\rho + (x-\xi)p )^2+i\epsilon)^2               } \label{trCOA_3trace}
    \\ \nonumber \\ \nonumber
  &=&  \frac{T_{A\perp j}}{2\alpha^{2}\bar{\alpha}s^{3}\xi z\bar{z}\left(x-\xi+i\epsilon\right)\left(x+\xi-i\epsilon\right)}
 \end{eqnarray}

  \begin{eqnarray}
  &&\hspace{-.2cm} tr_D^{CO}\left[ A_4\right]_j   \nonumber
  \\
&& \hspace{-.2cm}=  tr_D\left[   \hat p_\rho \hat \varepsilon^*_\rho \hat \varepsilon_q\frac{z \hat p_\rho -\hat q }{(z p_\rho-q)^2+i\epsilon } \gamma^\mu \hat p \gamma_{\bot\,j} \hat \varepsilon_k^*\frac{\hat k +(x-\xi)\hat p}{(k+(x-\xi)p)^2+i\epsilon}
 \,\gamma_\mu\,\frac{1}{( z p_\rho-q -(x+\xi)p)^2+i\epsilon}\right]  \nonumber
 \\
  &&\hspace{-.2cm}= \frac{16 \left[  (p\cdot k) \,\varepsilon^*_{k \bot j}\left(  (\varepsilon_q\cdot p_\rho) \, (q\cdot \varepsilon^*_\rho) - (q\cdot p_\rho)\,(\varepsilon_q\cdot \varepsilon^*_\rho)             \right)
   - \epsilon^{p_\rho\, \varepsilon^*_\rho\,q\, \varepsilon_q }   \,\epsilon^{p\, \nu\, k\, \varepsilon_k^*}g_{\bot \nu j}      \right]}{  (( k+(x-\xi)p )^2+i\epsilon)^2   \,((zp_\rho -q)^2+i\epsilon)^2 \,(( (x+\xi)p-zp_\rho+q   )^2+i\epsilon)^2       }
 \label{trCOA_4trace}
 \\ \nonumber  \\ \nonumber
   &&\hspace{-.2cm}= -\frac{T_{A\perp j}}{2\alpha^{2}\xi s^{3}z\left(x-\xi+i\epsilon\right)\left[\left(x+\xi+i\epsilon\right)-z\left(2\alpha\xi+\bar{\alpha}\left(x+\xi+i\epsilon\right)\right)\right]}
    \end{eqnarray}

    \begin{eqnarray}
&&\hspace{-.2cm} tr_D^{CO}\left[ B_5\right]_j =  \nonumber
  \\
&&\hspace{-.2cm}   tr_D\left[   \hat p_\rho \hat \varepsilon^*_\rho\,\gamma^\mu \frac{ \hat q +(x+\xi)\hat p }{(q+(x+\xi)p)^2+i\epsilon } \hat \varepsilon_q \hat p \gamma_{\bot \, j} \gamma_\mu  
\frac{ -\bar{z}\hat p_\rho - \hat k }{(-\bar{z} p_\rho -k)^2+i\epsilon}
 \,\hat \varepsilon_k^*\,\frac{1}{(\bar  z p_\rho +k +(x-\xi)p)^2+i\epsilon}\right]
 \nonumber
 \\
 &&\hspace{-.2cm}=
2 \frac{4      
s\,\varepsilon_{q \bot\,j} \left(  (p_\rho \cdot \varepsilon_k^*)\,(\varepsilon^*_\rho\cdot k ) -   s\xi\, (\varepsilon_k^* \cdot \varepsilon^*_\rho) \right) 
 -\epsilon^{k\, \varepsilon_k^*\, p_\rho \, \varepsilon^*_\rho}\, \epsilon^{q\,\varepsilon_q\,p\,\nu} g_{\bot \nu j}
    }{((-\bar{z} p_\rho -k)^2  +i\epsilon) \,   (( z p_\rho -q -(x-\xi)p)^2+i\epsilon)      ((q+(x+\xi)p)^2+i\epsilon)  }
\label{trCOB_5trace}
\\ \nonumber \\ \nonumber
&&\hspace{-.2cm}= \frac{T_{B\perp j}}{2\xi s^{3}\bar{z}\left(x+\xi+i\epsilon\right)\left[\left(x+\xi+i\epsilon\right)-z\left(2\alpha\xi+\bar{\alpha}\left(x+\xi+i\epsilon\right)\right)\right]} .
  \end{eqnarray}

\section{Integration over $z$ and $x$} 
\label{App:z-integration}

\subsection{Building block integrals for the numerical integration over $x$}
\label{SubSec:int-I}

Here, we list the building block integrals which are involved in the numerical evaluation of the scattering amplitudes. Consider a generic GPD $f.$
We define
\beqa
\label{Def:I}
I_a[f]&=&\int_{-1}^1 \frac{1}{(-\xi +x+i \epsilon ) (2 \xi +\bar{\alpha} (-\xi +x+i \epsilon ))}f(x,\xi)\, dx\,, \\
I_b[f]&=&\int_{-1}^1 \frac{1}{(2 \xi +(1-\alpha ) (-\xi +x+i \epsilon ))^2} f(x,\xi) \, dx\,, \\
I_c[f]&=&\int_{-1}^1 
\frac{\ln \left(\frac{\xi +x+i \epsilon }{\alpha  (-\xi +x+i \epsilon )}\right)}{\left(2
   \xi +\bar{\alpha } (-\xi +x+i \epsilon )\right)^3}f(x,\xi)
 \, dx\,, \\
I_d[f]&=&\int_{-1}^1 
\frac{\ln \left(\frac{\xi +x+i \epsilon }{\alpha  (-\xi +x+i \epsilon )}\right)}{\left(2
   \xi +\bar{\alpha } (-\xi +x+i \epsilon )\right)^2} 
f(x,\xi)
\, dx\,, \\
I_e[f]&=&\int_{-1}^1 
\frac{1}{-\xi +x+i \epsilon } f(x,\xi)
\, dx\,, \\
I_f[f]&=&\int_{-1}^1 
\frac{1}{\xi +x+i \epsilon } f(x,\xi)
\, dx\,, \\
I_g[f]&=&\int_{-1}^1
\frac{1}{\xi +x-i \epsilon } f(x,\xi)
\, dx\,, \\
I_h[f]&=&\int_{-1}^1 \frac{\ln \left(\frac{\xi +x+i \epsilon }{\alpha  (-\xi +x+i \epsilon )}\right)}{2 \xi
   +\bar{\alpha } (-\xi +x+i \epsilon )}f(x,\xi)
\, dx\,, \\
I_i[f]&=&\int_{-1}^1 
\frac{1}{2 \xi +\bar{\alpha } (-\xi +x+i \epsilon )} f(x,\xi)
\, dx\,, \\
I_j[f]&=&\int_{-1}^1 
\frac{1}{(-\xi +x+i \epsilon ) (\xi +x+i \epsilon ) \left(2 \xi +\bar{\alpha } (-\xi +x+i
   \epsilon )\right)} f(x,\xi)
\, dx\,, \\
I_l[f]&=&\int_{-1}^1 
\frac{1}{(\xi +x+i \epsilon ) \left(2 \xi +\bar{\alpha } (-\xi +x+i \epsilon )\right)}
f(x,\xi)
\, dx\,, \\
I_k[f]&=&\int_{-1}^1
\frac{1}{(\xi +x+i \epsilon ) \left(2 \xi +\bar{\alpha } (-\xi +x+i \epsilon )\right)^2}
f(x,\xi)
\, dx\,.
\eqa
Each of these integrals are finite and are evaluated numerically, using our models for the various involved GPDs.
After computing this set of integrals, the evaluation of the scattering amplitude is straightforward using the decomposition given in the two next subsections.
Below, we will not indicate the function $f$, since it is obvious from the context.

\subsection{Chiral-odd case}
\label{SubSec:int-chiral-odd}

For the chiral-odd case, diagrams $A_3$ and $A_4$ contribute
to the structure $T^i_{A \perp}$ while 
diagrams $B_1$ and $B_5$ contribute
to the structure $T^i_{B \perp}.$
Thus, writing
\beqa
\label{Def:T-CO_A}
tr_D^{CO}[A_3]^i + tr_D^{CO}[A_4]^i = T^{CO}_{A} \, T^i_{A \perp}
\eqa
and
\beqa
\label{Def:T-CO_B}
tr_D^{CO}[B_1]^i + tr_D^{CO}[B_5]^i = T^{CO}_{B} \, T^i_{B \perp}\,,
\eqa
we get
\beqa
T^{CO}_{A} \phi(z) &=&\frac{1}{s^3} \left[\frac{3 (1-z)}{\alpha ^2 \xi  (\xi -x-i
   \epsilon ) (\alpha  (-\xi +x+i \epsilon
   )+(1-z) (2 \xi +(1-\alpha ) (-\xi +x+i
   \epsilon )))}
   \right.
   \nonumber \\
&&   \left.
   -\frac{3}{\alpha \bar{\alpha}
   ^2 \xi  (\xi -x-i \epsilon ) (\xi +x-i
   \epsilon )}\right] 
\eqa
and
\beqa
T^{CO}_{B} \phi(z) &=&
\frac{1}{s^3} \left[
   -\frac{3}{(1-\alpha ) \xi  (\xi -x-i
   \epsilon ) (\xi +x+i \epsilon )}
  \right.
   \nonumber \\
&&  \hspace{-1.2cm}  \left.
   +\frac{3 z}{\xi  (\xi +x+i \epsilon ) (\alpha
    (-\xi +x+i \epsilon )+(1-z) (2 \xi
   +(1-\alpha ) (-\xi +x+i \epsilon )))}
   \right].
\eqa
The integral with respect to $z$ is trivially performed in the case of a DA
expanded in the basis of Gegenbauer polynomials. We restrict ourselves to the case of an asymptotic DA $\phi(z)=6 z \bar{z}$ for which one gets
\beqa
\label{int_z-diagA-chiral-odd}
&&\hspace{-1cm} \int_{0}^1 T^{CO}_{A} \phi(z) \, dz = \frac{1}{s^3} \left[-\frac{3}{\alpha \bar{\alpha}
   ^2 \xi  (\xi -x-i \epsilon ) (\xi +x-i
   \epsilon )} \right.  \\
 &+&
 \left. \frac{3}{\alpha^2 \xi  (\xi
   -x-i \epsilon ) (2 \xi +(1-\alpha ) (-\xi
   +x+i \epsilon ))} +
\frac{3 \ln \left(\frac{\xi +x+i \epsilon
   }{\alpha  (-\xi +x+i \epsilon
   )}\right)}{\alpha \xi  (2 \xi +(1-\alpha ) (-\xi
   +x+i \epsilon ))^2}\right]\,, \nonumber
\eqa
and
\beqa
\label{int_z-diagB-chiral-odd}
&&\hspace{-1cm} \int_{0}^1 T^{CO}_{B} \phi(z) \, dz = \frac{1}{s^3} \left[-\frac{3}{(1-\alpha ) \xi  (\xi -x-i
   \epsilon ) (\xi +x+i \epsilon )} \right. \\
&-&\left.\frac{3}{\xi  (\xi
   +x+i \epsilon ) (2 \xi +(1-\alpha ) (-\xi
   +x+i \epsilon ))}  
+   
\frac{3 \ln \left(\frac{\xi +x+i \epsilon
   }{\alpha  (-\xi +x+i \epsilon
   )}\right)}{\xi  (2 \xi +(1-\alpha ) (-\xi
   +x+i \epsilon ))^2}\right]. \nonumber
\eqa
Let us note that the last term in the previous expressions (\ref{int_z-diagA-chiral-odd}) and 
(\ref{int_z-diagB-chiral-odd})
might seem to have a double pole when $x = -\frac{1+\alpha}{\bar{\alpha}}\xi -i \epsilon$. However the logarithm cancels under such conditions, so this pole is actually a simple pole.

Writing the integrals with respect to $x$ of the product of  (\ref{int_z-diagA-chiral-odd}) and 
(\ref{int_z-diagB-chiral-odd}) with the GPD $H_T^q(x\xi)$ in terms of building block integrals, we have the dimensionless coefficients
\beqa
\label{Res-int-T_CO_A}
N_{T\, A}^q \equiv  {s^3} \int_{-1}^1 \int_{0}^1 T^{CO}_{A} \phi(z) \, dz \, H_T(x,\xi) \,dx =
-\frac{3}{\alpha ^2 \xi }I_a+\frac{3}{\alpha  \xi
   }I_d+\frac{3}{2 \alpha ^2 \bar{\alpha} \xi ^2}(I_e-I_g)\,, \hspace{.5cm}
\eqa
and
\beqa
\label{Res-int-T_CO_B}
N_{T\, B}^q \equiv {s^3} \int_{-1}^1 \int_{0}^1 T^{CO}_{B} \phi(z) \, dz \, H_T(x,\xi) \, dx
=
-\frac{3 }{\xi }I_l+\frac{3 }{\xi }I_d+\frac{3 }{2 \bar{\alpha} \xi ^2}(I_e-I_f)\,
.
\eqa

\subsection{Chiral-even case}
\label{SubSec:int-chiral-even}

For the chiral-even case, we only present the result in terms of building block integrals after integration over $z$ and integration over $x$ when multiplied by GPDs.

\subsubsection{Vector part}

From the symmetry of $\phi(z)$,  the integration over $z$ of the product of diagrams $A_1$ and $A_2$ with $\phi(z)$ leads to
vanishing $T_B$ parts (their $T_B$ components are antisymmetric) and to identical  $T_A$ parts.

We decompose the trace involved in a diagram $diag$ as 
\beqa
\label{Def:T_V-diag_A-B}
tr_D^{V}[diag] = T^{V}_A[diag] \, T_A + T^{V}_B[diag] \, T_B\,,
\eqa
and we denote the dimensionless coefficients
\beqa
\label{Def:NA-vector}
N_{A}^q[diag] \equiv s \int_{-1}^1 \int_{0}^1 T^{V}_A[diag] \, \phi(z) \, dz \, H(x,\xi) \, dx\,,
\eqa
\beqa
\label{Def:NB-vector}
N_{B}^q[diag] \equiv s^2 \int_{-1}^1 \int_{0}^1 T^{V}_B[diag] \, \phi(z) \, dz \, H(x,\xi) \, dx\,.
\eqa
For further use, we define the coefficient obtained when summing over the set of diagrams $A_k$ and $B_k$
\beqa
\label{Def:N_A}
N^q_A \equiv \sum_{diag} N_{A}^q[diag]
\eqa
and
\beqa
\label{Def:N_B}
N^q_B \equiv \sum_{diag} N_{B}^q[diag]\,.
\eqa

We get for diagrams $A_k$
\beqa
\label{Res-int-T_V_Ak_A}
N_{A}^q[A_1]&=& N_{A}^q[A_2] =
 \frac{12}{\bar{\alpha} \xi}  I_e
\,, \\
N_{A}^q[A_3]&=&
-\frac{12}{\alpha \bar{\alpha} \xi}
(I_e-I_g)
\,, \\
N_{A}^q[A_4]
&=&
\frac{24 \bar{\alpha}}{\alpha} \left(I_a-\alpha 
   I_d\right)
\,, \\
N_{A}^q[A_5]
&=&
-\frac{12}{\xi }  I_h
\eqa
and
\beqa
\label{Res-int-T_V_Ak_B}
N_{B}^q[A_3]
&=&
 -\frac{6}{\alpha ^2\bar{\alpha}  \xi ^2}(I_e-I_g)
\,, \\
N_{B}^q[A_4]&=&
\frac{12}{\alpha ^2 \xi } I_a - \frac{24 }{\alpha  \xi }I_b - \frac{48}{\bar{\alpha}}I_c+\frac{24}{\bar{\alpha} \xi}I_d\,,\\
N_{B}^q[A_5]
&=&
 \frac{48}{\alpha  \xi }I_b+\frac{96}{\bar{\alpha}} I_c-\frac{24(1+\alpha)}{\alpha\bar{\alpha} \xi }
   I_d\, .
\eqa
For diagrams $B_k$ we obtain for the $T_A$ part
\beqa
\label{Res-int-T_V_Bk_A}
N_{A}^q[B_1]
&=&
-\frac{12 \alpha}{\bar{\alpha} \xi }  (I_e-I_f)
\,, \\
N_{A}^q[B_2]&=&
\frac{6}{\bar{\alpha} \xi }(I_e-I_f)
\,, \\
N_{A}^q[B_3]
&=&
\frac{6}{\bar{\alpha} \xi }(I_e-I_g)
\,, \\
N_{A}^q[B_4]
&=&
24 \xi  (I_j+2 \alpha  I_k-2 \alpha  I_c)
\,, \\ 
N_{A}^q[B_5]
&=&
24 \bar{\alpha} (I_d-I_l)\,,
\eqa
and for the non-vanishing $T_B$ part
\beqa
\label{Res-int-T_V_Bk_B}
N_{B}^q[B_1]
&=&
\frac{6}{\bar{\alpha} \alpha  \xi ^2}(I_e-I_f)\,,\\
N_{B}^q[B_5]
&=&
-\frac{36 }{\alpha  \xi }I_l +48 I_k-\frac{24 }{\xi }I_b-\frac{48}{\bar{\alpha}}I_c+\frac{24}{\alpha \bar{\alpha}  \xi }I_d\,.
\eqa

\subsubsection{Axial part}

We decompose the trace involved in a diagram $diag$ as 
\beqa
\label{Def:T_A-diag_A5-B5}
tr_D^{A}[diag] = T^{A}_{A_5}[diag] \, T_{A_5} + T^{A}_{B_5}[diag] \, T_{B_5}\,,
\eqa
and we denote the dimensionless coefficients
\beqa
\label{Def:NTildeA-axial}
\Tilde{N}_{A_5}^q[diag] \equiv s^3 \int_{-1}^1 \int_{0}^1 T^{A}_{A_5}[diag] \,\phi(z) \, dz \, \tilde{H}^q(x,\xi) \, dx\,,
\eqa
\beqa
\label{Def:NTildeB-axial}
\Tilde{N}_{B_5}^q[diag] \equiv s^3 \int_{-1}^1 \int_{0}^1 T^{A}_{B_5}[diag]\, \phi(z) \, dz \, \tilde{H}^q(x,\xi) \, dx\,.
\eqa
Similarly to the vector case, we define the coefficient obtained when summing over the set of diagrams $A_k$ and $B_k$
\beqa
\label{Def:TildeN_A}
\tilde{N}^q_{A_5} \equiv \sum_{diag} \tilde{N}_{A_5}^q[diag]
\eqa
and
\beqa
\label{Def:TildeN_B}
\tilde{N}^q_{B_5} \equiv \sum_{diag} \tilde{N}_{B_5}^q[diag]\,.
\eqa
We get for diagrams $A_k$
\beqa
\label{Res-int-T_A_Ak_A5}
\Tilde{N}_{A_5}^q[A_1]
&=&
-\frac{12 i }{\alpha \bar{\alpha}^2 \xi ^2}I_e \,,\\
\Tilde{N}_{A_5}^q[A_2]
&=&
\frac{12 i (2-\alpha ) }{\alpha ^2 \bar{\alpha}^2  \xi ^2}I_e\,,\\ 
\Tilde{N}_{A_5}^q[A_3]
&=&
-\frac{12 i }{\alpha ^2 \bar{\alpha}^2\xi ^2} (I_e-I_g)\,,\\
\Tilde{N}_{A_5}^q[A_4]
&=&
-\frac{24 i }{\alpha ^2 \xi }(I_a-\alpha  I_d)
\,,\\
\Tilde{N}_{A_5}^q[A_5]
&=&
-\frac{48 i }{\alpha \bar{\alpha}  \xi }I_d+\frac{12 i }{\alpha \bar{\alpha} \xi ^2}I_h-\frac{24 i }{\alpha ^2 \xi ^2}I_i
\eqa
and
\beqa
\label{Res-int-T_A_Ak_B5} 
\Tilde{N}_{B_5}^q[A_1]
&=&
\frac{12 i (1-2 \alpha ) }{\alpha \bar{\alpha}^2  \xi ^2}I_e \,,\\
\Tilde{N}_{B_5}^q[A_2]
&=&
\frac{12 i }{\alpha \bar{\alpha}^2  \xi ^2}I_e\,,\\ 
\Tilde{N}_{B_5}^q[A_3]
&=&
-\frac{12 i }{\alpha \bar{\alpha}^2  \xi ^2}(I_e-I_g)\,,\\
\Tilde{N}_{B_5}^q[A_4]
&=&
-\frac{48 i }{\alpha  \xi }I_b-\frac{96 i }{\bar{\alpha}}I_c+\frac{24 i(1+\alpha)}{\alpha \bar{\alpha} \xi }I_d\,
,\\
\Tilde{N}_{B_5}^q[A_5]
&=&
-\frac{48 i }{\bar{\alpha} \xi }I_d+\frac{12 i }{\alpha \bar{\alpha}  \xi
   ^2}I_h-\frac{24 i}{\alpha \xi ^2} I_i\,.
\eqa
For diagrams $B_k$ we obtain for the $T_{A_5}$ part
\beqa
\label{Res-int-T_A_Bk_A5} 
\Tilde{N}_{A_5}^q[B_1]
&=&
\frac{12 i }{\alpha \bar{\alpha}^2  \xi ^2}(I_e-I_f)
\,,\\ 
\Tilde{N}_{A_5}^q[B_2]&=&
-\frac{6i }{\alpha \bar{\alpha}^2  \xi ^2}(I_e-I_f)
\,, \\
\Tilde{N}_{A_5}^q[B_3]
&=&
\frac{6i}{\xi^2 \alpha \bar{\alpha}^2 }(I_e-I_g)
\,, \\
\Tilde{N}_{A_5}^q[B_4]
&=&
-\frac{24 i }{\alpha \bar{\alpha}}I_j-\frac{48 i }{\bar\alpha }I_k+\frac{48 i }{\bar\alpha }I_c \,,\\
\Tilde{N}_{A_5}^q[B_5]
&=& 
-\frac{48 i }{\alpha  \xi }I_b-\frac{96 i }{\bar\alpha }I_c+\frac{24 i}{\bar{\alpha} \xi } I_d+\frac{24 i }{\alpha \bar{\alpha} \xi }I_d\,,
\eqa
and for the $T_{B_5}$ part
\beqa
\label{Res-int-T_A_Bk_B5} 
\Tilde{N}_{B_5}^q[B_1]
&=&
\frac{12 i }{\bar{\alpha}^2 \xi ^2}(I_e-I_f)\,,\\
\Tilde{N}_{B_5}^q[B_2]
&=&
-\frac{6i }{\alpha \bar{\alpha}^2  \xi ^2}(I_e-I_f)\,,\\
\Tilde{N}_{B_5}^q[B_3]
&=&
\frac{6i }{\alpha \bar{\alpha}^2  \xi ^2}(I_e-I_g)\,,\\
\Tilde{N}_{B_5}^q[B_4]
&=&\Tilde{N}_{A_5}^q[B_4]
\,,\\
\Tilde{N}_{B_5}^q[B_5]
&=&
-\frac{24 i }{\alpha  \xi }(I_d-I_l)\,.
\eqa

\section{Some details on kinematics}
\label{App:kinematics}

In this section we give further useful expressions for kinematics.

\subsection{Exact kinematics}

Combining Eqs.~(\ref{transfmom}) and 
(\ref{M_gamma_rho}) one gets
\beqa
\label{M_gamma_rho-t}
M_{\gamma \rho}^2-t = 2 \xi s \left(1-\frac{2 \xi M^2}{s(1-\xi^2)} \right) + \frac{4 \xi^2 M^2}{1-\xi^2}=2 \xi s\,.
\eqa
From Eq.~(\ref{energysquared}), one gets 
\beqa
\label{s-S-M2}
s=\frac{S_{\gamma N}-M^2}{1+\xi}\,,
\eqa
so that we finally obtain
\beqa
\label{tau}
\tau \equiv \frac{M_{\gamma \rho}^2-t}{S_{\gamma N}-M^2} = \frac{2 \xi}{1+\xi} \,,
\eqa
and thus
\beqa
\label{xi-tau}
\xi = \frac{\tau}{2-\tau}\,.
\eqa

\subsection{Exact kinematics for  $\Delta_\perp=0$}

In the case 
$\Delta_\perp=0\,,$
we now provide
the exact formulas 
in order to get the set of parameters $s, \xi, \alpha, \alpha_{\rho},\vec{p}^{\,2}, (-t)_{\rm min}$ as functions of
$M_{\gamma \rho}, S_{\gamma N}, -u'\,.$

In the limit $\Delta_\perp=0,$ Eq.~(\ref{M_gamma_rho})
reads, using Eq.~(\ref{s-S-M2}),
\beqa
\bar{M}_{\gamma \rho}^2=\frac{2 \xi}{1+\xi} \left(1-\frac{2 \xi }{1-\xi}\bar{M}^2   \right)
\eqa
with 
$\bar{M}^2=M^2/(S_{\gamma N}-M^2)$ and  $\bar{M}_{\gamma \rho}^2=M_{\gamma \rho}^2/(S_{\gamma N}-M^2).$
Thus, $\xi$ is solution of the quadratic equation
\beqa
\label{eq:xi}
\xi^2 (\bar{M}_{\gamma \rho}^2-2 -4 \bar{M}^2)  + 2 \xi - \bar{M}_{\gamma \rho}^2=0
\eqa
the solution to be kept being
\beqa
\label{sol:xi}
\xi=\frac{-1+\sqrt{1+\bar{M}_{\gamma \rho}^2(\bar{M}_{\gamma \rho}^2-2 -4 \bar{M}^2)}}{\bar{M}_{\gamma \rho}^2-2 -4 \bar{M}^2}\,.
\eqa

The value of $(-t)_{\rm min}$
is obtained by setting $\dv=0$ in Eq.~(\ref{transfmom}),
{\rm i.e.}
\beq
(-t)_{\rm min}=\frac{4 \xi^2 M^2}{1-\xi^2}\,.
\eq
Combined with Eqs.~(\ref{xi-tau}) and (\ref{tau})
one easily see that $(-t)_{\rm min}$ is obtained from the solution of
\beq
\label{Eq:T-tmin}
\bar{T}^2 (1+\bar{M}^2)+ \bar{T} (2 \bar{M}^2 \, \bar{M}_{\gamma \rho}^2 +\bar{M}_{\gamma \rho}^2-1) + \bar{M}^2 \, \bar{M}_{\gamma \rho}^4=0
\eq
with $\bar{T}=(-t)_{\rm min}/(S_{\gamma N}-M^2),$ the solution to be kept being
\beqa
\label{tmin}
(-t)_{\rm min}= \frac{1-\bar{M}_{\gamma \rho}^2(1+2\bar{M}^2)-
\sqrt{1+\bar{M}_{\gamma \rho}^2(\bar{M}_{\gamma \rho}^2-2 -4 \bar{M}^2)}
}{2(1+\bar{M}^2)}(S_{\gamma N}-M^2)\,.
\eqa

From Eq.~(\ref{u'}) we have
\beqa
\label{pt2}
\pv^{\,2}=
-m_\rho^2 +\ar (m_\rho^2-u')
\eqa
so that using Eq.~(\ref{2xi}) which now reads
\beqa
\label{2xi-tmin}
2 \, \xi = \frac{\pv^{\,2}}{s \, \alpha} +
\frac{\pv^{\,2} + \mr^2}{s \, \ar}\,,
\eqa
we obtain
\beqa
\label{2xi-tmin-2}
2 \, \xi = -\frac{\ar}{\alpha} \frac{u'}s -\frac{1-\ar}{\alpha s}  m_\rho^2 -\frac{u'}s + \frac{\mr}{s}\,.
\eqa
Eq.~(\ref{exp_alpha}) reads 
\beqa
\label{alpha_rho-tmin}
\ar = 1-\alpha -\frac{2 \, \xi \, M^2}{s \, (1-\xi^2)}\,.
\eqa
so that
\beqa
\label{alpha-tmin}
\alpha = \frac{1}{2 \xi s} \left(-u' -\frac{2 \, \xi \, M^2}{s \, (1-\xi^2)}(-u'+\mr^2)\right)\,.
\eqa
Thus, computing $\xi$ through Eq.~(\ref{sol:xi}) and then $s$ through Eq.~(\ref{s-S-M2}), Eq.~(\ref{alpha-tmin}) allows to compute the value of $\alpha\,.$ The value of $\ar$ is then obtained using Eq.~(\ref{alpha_rho-tmin}). Finally, $\pv^{\,2}$
is computed using Eq.~(\ref{pt2}).

\subsection{Approximated kinematics in the Bjorken limit}
\label{SubSec:approximated_kinematics}

In this limit, $\bar{M}_{\gamma \rho}$ and $S_{\gamma N}$ are parametrically large, and $s$ is of the order of $S_{\gamma N}$. 
Neglecting $\dv^2$, $m_\rho^2$, $t$ and $M^2$  in front of $s$, (except in the definition of $\tau$ where we keep as usual $M^2$ in the denominator of Eq.~(\ref{tau})), we thus have
\beqa
\label{M-Bjorken}
M^2_{\gamma\rho} &\approx & 2 \xi s \approx \frac{\vec{p}_t^2}{\alpha\bar{\alpha}} \,,
\\
\label{alpha_rho-Bjorken}
\ar &\approx& 1-\alpha \equiv \alb \,,\\
\label{xi-Bjorken}
\xi &= & \frac{\tau}{2-\tau} ~~~,~~~~\tau \approx 
\frac{M^2_{\gamma\rho}}{S_{\gamma N}-M^2}\,,\\
\label{t'-u'-Bjorken}
-t' & \approx & \bar\alpha\, M_{\gamma\rho}^2  ~~~,~~~~ -u'  \approx  \alpha\, M_{\gamma\rho}^2 \,.
\eqa
The skewedness $\xi$ thus reads
\beqa
\label{xi-hard}
\xi= \frac{M_{\gamma\rho}^2}{2 S_{\gamma N}-2 M^2 - M_{\gamma\rho}^2}
\eqa
and
the parameter $s$ is given, using Eq.~(\ref{s-S-M2}), by
\beqa
\label{s-hard}
s = S_{\gamma N}- M^2 - \frac{M_{\gamma\rho}^2}2\,.
\eqa

\section{Phase space integration}
\label{App:phase}

\subsection{Phase space evolution}

%
\psfrag{H}{\raisebox{-.5cm}{\hspace{-.6cm}$-t$}}
\psfrag{V}{\raisebox{.1cm}{\hspace{-.5cm}$-u'$}}
\psfrag{A}{}
\psfrag{B}{}
\psfrag{J}{}
\psfrag{K}{}
\psfrag{L}{}

\def\tici{4.7cm}
\begin{figure}[H]
\begin{tabular}{ccc}
\hspace{.1cm}\includegraphics[width=\tici]{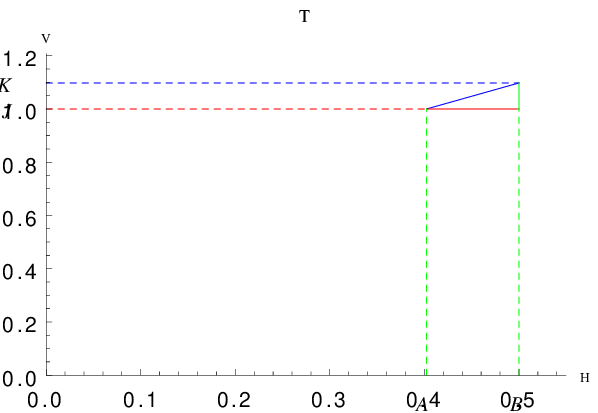} & \includegraphics[width=\tici]{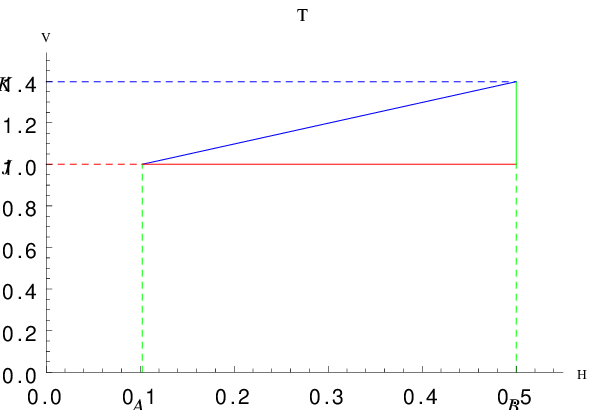} & 
\includegraphics[width=\tici]{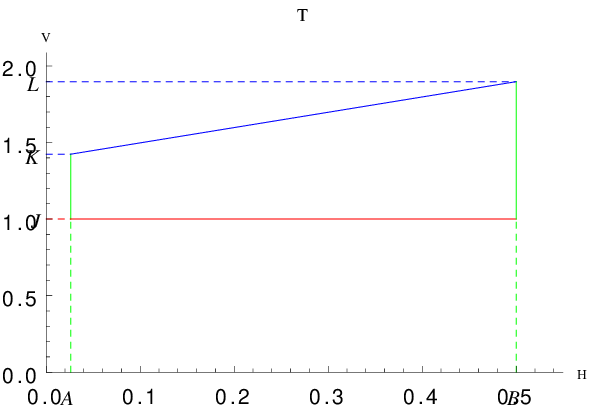}\\
\\
\hspace{.1cm}\includegraphics[width=\tici]{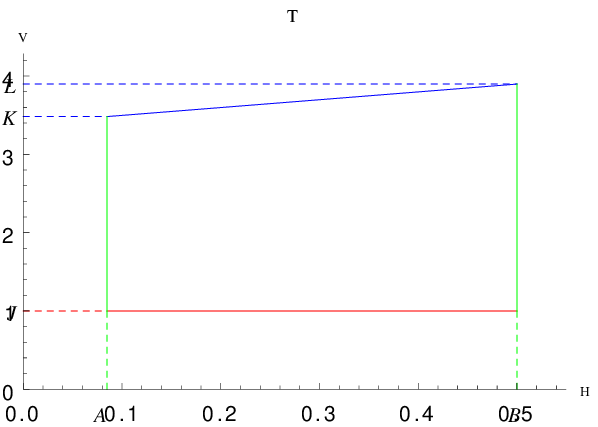} & \includegraphics[width=\tici]{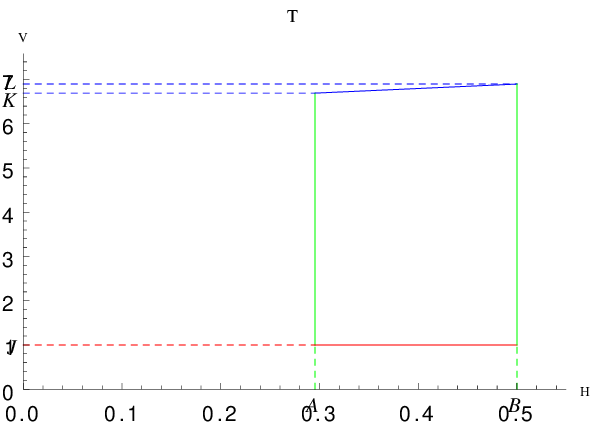} & 
\includegraphics[width=\tici]{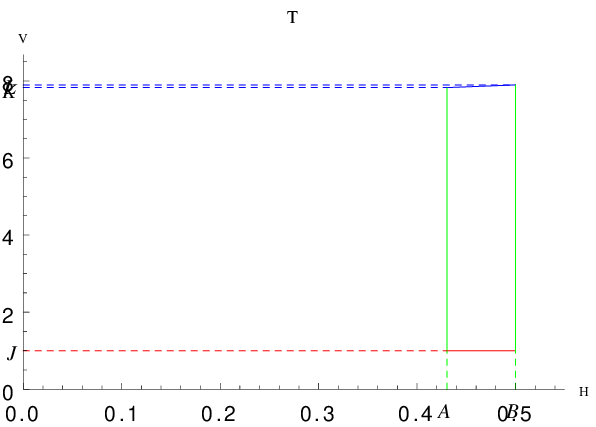}
\end{tabular}
\vsp
\caption{Evolution of the phase space for $M_{\gamma \rho}^2=2.2~{\rm GeV}^2$ (up left), $M_{\gamma \rho}^2=2.5~{\rm GeV}^2$ (up center), $M_{\gamma \rho}^2=3~{\rm GeV}^2$ (up right),
$M_{\gamma \rho}^2=5~{\rm GeV}^2$ (down left),
$M_{\gamma \rho}^2=8~{\rm GeV}^2$ (down center),
$M_{\gamma \rho}^2=9~{\rm GeV}^2$ (down right).}
\label{Fig:phase-space-evolution}
\end{figure}
\psfrag{H}{\raisebox{-.5cm}{\hspace{-.1cm}$-t$}}
\psfrag{V}{\raisebox{.1cm}{\hspace{-.6cm}$-u'$}}
\psfrag{A}{}
\psfrag{B}{\raisebox{-.4cm}{\hspace{-.2cm}$(-t)_{\rm max}$}}
\psfrag{J}{\raisebox{0cm}{\hspace{-2.5cm}$(-u')_{\rm min}$}}
\psfrag{K}{\raisebox{0cm}{\hspace{-2.5cm}$(-u')_{\rm maxMax}$}}
\psfrag{M}{\raisebox{-.4cm}{\hspace{-.3cm}$(-t)_{\rm min}(-u')$}}
\psfrag{N}{\raisebox{0cm}{\hspace{-.6cm}$-u'$}}

\begin{figure}[H]
\psfrag{A}{\raisebox{-.4cm}{\hspace{-.3cm}$(-t)_{\rm inf}$}}
\centerline{\includegraphics[width=8cm]{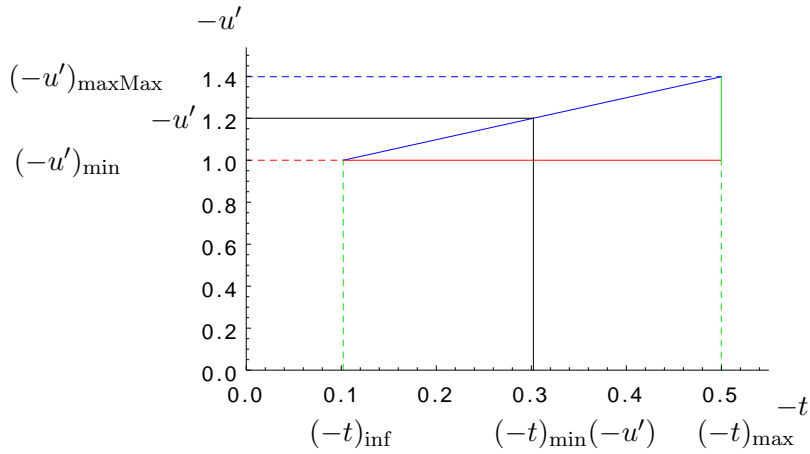}}
\vsp
\caption{Triangle-like phase space, illustrated for the case of $M_{\gamma \rho}^2=2.5~{\rm GeV}^2$.}
\label{Fig:Triangle-phase-space}
\end{figure}

\begin{figure}[h!]
\centerline{\includegraphics[width=8cm]{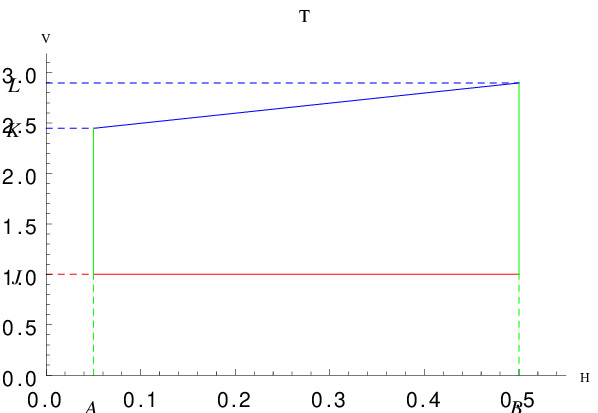}}
\vsp
\caption{Trapezoid-like phase space, illustrated for the case $M_{\gamma \rho}^2=4~{\rm GeV}^2$ and $S_{\gamma N}=20~{\rm GeV}^2.$}
\label{Fig:Trapezoid-phase-space}
\end{figure}

The phase space integration in the $( -t,-u')$ plane should take care of several cuts.
This phase space evolves with increasing $M_{\gamma \rho}^2$ from a triangle to a trapezoid, as shown in Fig.~\ref{Fig:phase-space-evolution}. 
These two cases and the corresponding parameters are displayed in Figs.~\ref{Fig:Triangle-phase-space} and \ref{Fig:Trapezoid-phase-space}.

\def\tici{4.7cm}

Let us discuss these various cuts with some details.
First, since we rely on factorization at large angle, we enforce the two constraints $-u' > (-u')_{\rm min}\,,$ and
$-t' > (-t')_{\rm min}\,,$
and take $(-u')_{\rm min}=(-t')_{\rm min}=1~{\rm GeV}^2\,.$ 
The first constraint is the red line in
Figs.~\ref{Fig:Triangle-phase-space} and \ref{Fig:Trapezoid-phase-space},
while the second, using the relation $M_{\gamma \rho}^2+t'+u'=t+m_\rho^2,$ is given by
\beq
\label{blue-line}
-u'(-t)=-t-m_\rho^2+M_{\gamma \rho}^2-(-t')_{\rm min}\,,
\eq
and shown as a blue line. 
 
The variable $(-t)$ varies from $(-t)_{\rm min}$, determined by kinematics, up to a maximal value $(-t)_{\rm max}$ which we fix to be $(-t)_{\rm max}=0.5~{\rm GeV}^2\,,$ these two boundaries being shown in green in Fig.~\ref{Fig:Trapezoid-phase-space}.

The value of $(-t)_{\rm min}$ is given by Eq.~(\ref{tmin}). 
In the domain of $M^2_{\gamma \rho}$ for which the phase-space is a triangle,
as illustrated in Fig.~\ref{Fig:Triangle-phase-space},  the minimal value of $-t$ is actually above $(-t)_{\rm min}$.
For a given value of $M^2_{\gamma \rho}\,,$ this minimal value of $-t$ is given, using Eq.~(\ref{blue-line}), by
\beqa
\label{Def:mtinf}
(-t)_{\rm inf} = m_\rho^2 - M^2_{\gamma \rho}  + (-t')_{\rm min}+ (-u')_{\rm min}\,,
\eqa
with $(-t)_{\rm min} \leqslant (-t)_{\rm inf}\,.$ 

This constraint on $-t$ leads to a minimal value of $M^2_{\gamma \rho}\,,$ denoted as $M^2_{\gamma \rho \, {\rm crit}}\,,$  when $(-t)_{\rm inf}=(-t)_{\rm max}\,,$
which thus reads
\beq
\label{Def:M2crit}
M^2_{\gamma \rho \, {\rm crit}}=(-u')_{\rm min} +(-t')_{\rm min} + m_\rho^2 - (-t)_{\rm max}\,.
\eq
With our chosen values of $(-u')_{\rm min}$, $(-t')_{\rm min}$ and $(-t')_{\rm max}$ we have $M^2_{\gamma \rho \, {\rm crit}} \simeq 2.10~{\rm GeV}^2\,,$ below which the phase-space is empty. 
We note that this value, independent of $S_{\gamma N},$ ensures that the $s-$channel Mandelstam variable $M^2_{\gamma \rho} \geqslant
M^2_{\gamma \rho \, {\rm crit}}$
is indeed large enough as it should be for large angle scattering.

For the purpose of integration, we define, for $-(u')_{\rm min} \leqslant  -u'\,,$
\beqa
\label{Def:mtmin}
(-t)_{\rm min}(-u') = m_\rho^2 - M^2_{\gamma \rho}   + (-t')_{\rm min}-u'\,.
\eqa
We denote the maximal value of $-u'$ as $(-u')_{\rm maxMax}\,,$ attained when $-t=(-t)_{\rm max}\,,$ and given by
\beqa
\label{Def:mupmaxMax}
(-u')_{\rm maxMax} = (-t)_{\rm max} -m_\rho^2 + M^2_{\gamma \rho}   - (-t')_{\rm min}\,,
\eqa
see Fig.~\ref{Fig:Triangle-phase-space}.

The phase-space becomes a trapezoid when $(-t)_{\rm inf}
= (-t)_{\rm min}\,,$ {\it i.e.} according to Eq.~(\ref{Def:mtinf}) when
\beqa
M^2_{\gamma \rho}= -(-t)_{\rm min} + (-t')_{\rm min} +  (-u')_{\rm min} +  m_\rho^2\,.
\eqa
Combined with Eq.~(\ref{Eq:T-tmin}), this leads to
\beq
\label{Def:MinM2Exacte}
M^2_{\gamma \rho \, {\rm trans}} = (S_{\gamma N} - M^2) \, \bar{m}^2 \frac{1 - \bar{m}^2 (1 + \bar{M}^2)}{1-\bar{m}^2}\,,
\eq
where
\beqa
\bar{m}^2 = \frac{(-u')_{\rm min} + (-t')_{\rm min} + m_\rho^2}{S_{\gamma N} - M^2}\,.
\eqa
With our choice of parameters, we get 
$M^2_{\gamma \rho \, {\rm trans}} \simeq 2.58~{\rm GeV}^2$ in the case of $S_{\gamma N}=20~{\rm GeV}^2\,.$

\psfrag{T}{}
\psfrag{H}{\raisebox{-.4cm}{\hspace{-.2cm}$-t$}}
\psfrag{V}{\hspace{-.6cm}$-u'$}
\psfrag{A}{\raisebox{-.4cm}{\hspace{-.1cm}$(-t)_{\rm min}$}}
\psfrag{B}{\raisebox{-.4cm}{\hspace{-.1cm}$(-t)_{\rm max}$}}
\psfrag{J}{\raisebox{0cm}{\hspace{-2.5cm}$(-u')_{\rm min}$}}
\psfrag{K}{\raisebox{0cm}{\hspace{-2.5cm}$(-u')_{\rm maxMin}$}}
\psfrag{L}{\raisebox{0cm}{\hspace{-2.5cm}$(-u')_{\rm maxMax}$}}

Above this value, the phase-space is a trapezoid, illustrated 
in Fig.~\ref{Fig:Trapezoid-phase-space}. This trapezoid
reduces to an empty domain when $(-t)_{\rm min} = (-t)_{\rm max}\,.$ From the solution of Eq.~(\ref{Eq:T-tmin}), this occurs
for
\beqa
\label{Def:MaxtminmaxM2}
M^2_{\gamma \rho \, {\rm Max}} = (S_{\gamma N} - M^2)
\frac{-(1+2 \bar{M}^2)(-\bar{t})_{\rm max} + \sqrt{(-\bar{t})_{\rm max}((-\bar{t})_{\rm max}+ 4 \bar{M}^2)}}{2 \bar{M}^2}\,,
\eqa
with 
$\bar{M}^2=M^2/(S_{\gamma N}-M^2)$
and
$(-\bar{t})_{\rm max}=(-t)_{\rm max}/(S_{\gamma N}-M^2)\,.$
With our choice of parameters, we get 
$M^2_{\gamma \rho \, {\rm Max}} \simeq 9.47~{\rm GeV}^2$ in the case of $S_{\gamma N}=20~{\rm GeV}^2\,.$ This value decreases with decreasing values of $S_{\gamma N}\,.$

The minimal value of $S_{\gamma N}$ is obtained from the constraint
$M^2_{\gamma \rho \, {\rm crit}}=M^2_{\gamma \rho \, {\rm Max}}$ and equals $S_{\gamma N {\rm crit}} \simeq 5.87~{\rm GeV}^2\,.$

Finally, let us briefly discuss the invariant mass
$M_{\rho N'}^2$, which should be restricted to be far above any possible resonance.
Using Eq.~(\ref{M_rho_N}), for a given value of $S_{\gamma N}$, a careful study of the allowed phase space shows that $M_{\rho N'}^2$ is minimal when $-u'=(-u')_{\rm maxMax}$ and $M^2_{\gamma \rho}=M^2_{\gamma \rho \, {\rm Max}},$
and for $\dv$ and $\pv$ anti collinear, with $|\dv|$ being the value corresponding to $-t=(-t)_{\rm max}\,.$ This minimal value increases with $S_{\gamma N}.$
Its minimal value is thus obtained when $S_{\gamma N}=S_{\gamma N {\rm crit}},$ this value being $M_{\rho N' {\rm Min}}^2 \simeq 3.4~{\rm GeV}^2$ which is far above the resonance region.

\subsection{Method for the phase space integration}

Using the above described phase-space, the integrated cross section reads
\beqa
\label{integration-phase-space}
&&\frac{d \sigma}{d M_{\gamma \rho}^2} =
\theta(M_{\gamma \rho\, {\rm crit}}^2 < M_{\gamma \rho}^2 < 
M_{\gamma \rho\, {\rm trans}}^2) \\
&&\times
\int_{(-u')_{\rm min}}^{(-u')_{\rm maxMax}} d(-u') 
\int_{(-t)_{\rm min}(-u')}^{(-t)_{\rm max}} d(-t) \,
F(t)^2 
\left.\frac{d \sigma}{d M^2_{\gamma \rho} d(-u') d(-t)}\right|_{(-t)_{\rm min}} \nonumber \\
&& + \, \theta(M_{\gamma \rho\, {\rm trans}}^2 < M_{\gamma \rho}^2 < 
M_{\gamma \rho\, {\rm Max}}^2) \nonumber \\
&& \times \left\{ \int_{(-u')_{\rm min}}^{(-u')_{\rm maxMin}} d(-u') 
\int_{(-t)_{\rm min}}^{(-t)_{\rm max}} d(-t) \,
F(t)^2 
\left.\frac{d \sigma}{d M^2_{\gamma \rho} d(-u') d(-t)}\right|_{(-t)_{\rm min}} \right.\nonumber \\
&& + \left. 
\int_{(-u')_{\rm maxMin}}^{(-u')_{\rm maxMax}} d(-u') 
\int_{(-t)_{\rm min}(-u')}^{(-t)_{\rm max}} d(-t) \,
F(t)^2 
\left.\frac{d \sigma}{d M^2_{\gamma \rho} d(-u') d(-t)}\right|_{(-t)_{\rm min}} \right\} \nonumber \,.
\eqa
Using our explicit dipole ansatz for $F(t),$ see
Eq.~(\ref{dipole}), we obtain
\beqa
\label{integration-phase-space-final}
&&\frac{d \sigma}{d M_{\gamma \rho}^2} = \frac{C^4}3 \bigg[
\theta(M_{\gamma \rho\, {\rm crit}}^2 < M_{\gamma \rho}^2 < 
M_{\gamma \rho\, {\rm trans}}^2)  \\
&&\hspace{-.3cm}\times \!\int_{(-u')_{\rm min}}^{(-u')_{\rm maxMax}} \! d(-u') \!
\left[\frac{1}{(-(-t)_{\rm max}- C)^3}
- \frac{1}{(-(-t)_{\rm min}(-u')- C)^3}
\right] \left.\frac{d \sigma}{d M^2_{\gamma \rho} d(-u') d(-t)}\right|_{(-t)_{\rm min}} \!\!\nonumber \\
&&
\hspace{-.3cm}\left. + \,
\theta(M_{\gamma \rho\, {\rm trans}}^2 < M_{\gamma \rho}^2 < 
M_{\gamma \rho\, {\rm Max}}^2)
 \right. \nonumber \\
&&\hspace{-.3cm}\left. \times
\left\{
\left[\frac{1}{(-(-t)_{\rm max} - C)^3} - \frac{1}{(-(-t)_{\rm min}- C)^3}\right]
\int_{(-u')_{\rm min}}^{(-u')_{\rm maxMin}} d(-u') 
\left.\frac{d \sigma}{d M^2_{\gamma \rho} d(-u') d(-t)}\right|_{(-t)_{\rm min}} 
\right.\right.\nonumber \\
&&\hspace{-.3cm} + 
 \left. \left. \!\!\!
\int_{(-u')_{\rm maxMin}}^{(-u')_{\rm maxMax}}\! \!\! \! \! \! \! \! d(-u') \!
\left[\frac{1}{(-(-t)_{\rm max}- C)^3}
- \frac{1}{(-(-t)_{\rm min}(-u')- C)^3}
\right] \!
\left.\frac{d \sigma}{d M^2_{\gamma \rho} d(-u') d(-t)}\right|_{(-t)_{\rm min}}\!
\right\} \!\right]\! ,\! \nonumber
\eqa
which is our building formula for the numerical evaluation of integrated cross sections.

\section{Angular cut over the outgoing photon}
\label{App:theta-cut}

%
\psfrag{H}{\hspace{-1.5cm}\raisebox{-.6cm}{\scalebox{.7}{$\theta$}}}
\psfrag{V}{\raisebox{.3cm}{\scalebox{.7}{$\hspace{-.4cm}\displaystyle\frac{1}{\sigma_{even}}\frac{d \sigma_{even}}{d \theta}$}}}
\begin{figure}[h!]
\begin{center}
\includegraphics[width=7cm]{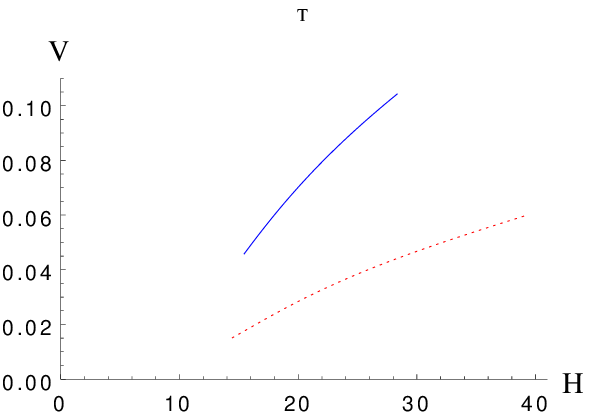}
\includegraphics[width=7cm]{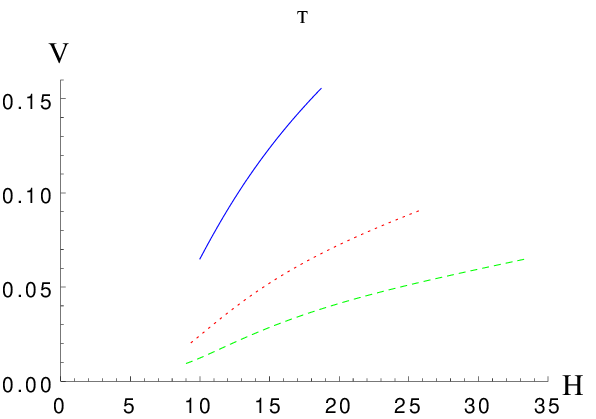}

\vspace{.8cm}
\includegraphics[width=7cm]{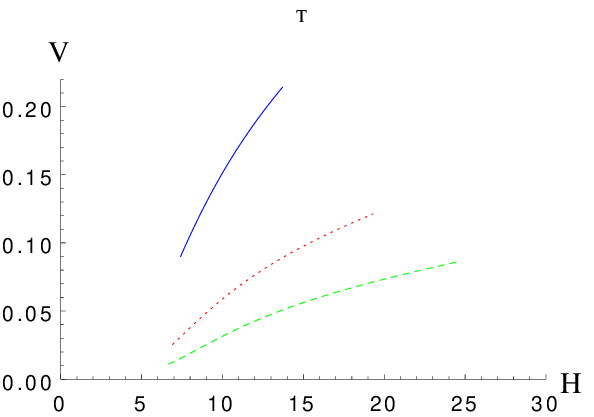}
\vspace{.4cm}
\caption{Angular distribution in the chiral-even case.
Up, left: $S_{\gamma N}=10~{\rm GeV}^2$, for $M^2_{\gamma \rho}=
3~{\rm GeV}^2$ (solid blue) and  $M^2_{\gamma \rho}=
4~{\rm GeV}^2$ (dotted red).  Up, right:
$S_{\gamma N}=15~{\rm GeV}^2$, for $M^2_{\gamma \rho}=
3~{\rm GeV}^2$ (solid blue), $M^2_{\gamma \rho}=
4~{\rm GeV}^2$ (dotted red) and 
$M^2_{\gamma \rho}=
5~{\rm GeV}^2$ (dashed green).
Down: 
$S_{\gamma N}=20~{\rm GeV}^2$, for $M^2_{\gamma \rho}=
3~{\rm GeV}^2$ (solid blue), $M^2_{\gamma \rho}=
4~{\rm GeV}^2$ (dotted red) and 
$M^2_{\gamma \rho}=
5~{\rm GeV}^2$ (dashed green).}
\label{Fig:thetacut-even}
\end{center}
\end{figure}
In order to take into account limitations of detection of the produced photon, it is necessary to know the photon scattering angle in the rest frame of the nucleon target. 
The incoming nucleon momentum $p_1^\mu$ in Eq.~(\ref{impini}) and the one in its rest frame $p_{1 rf}^\mu=(M,0,0,0)$ are related by the longitudinal boost along $z$ axis characterized by
the rapidity $\zeta$ such that, in the Bjorken limit,
\begin{eqnarray}
\cosh \zeta = \frac{1}{2}\left[  \frac{M}{\sqrt{s}(1+\xi)}   +  \frac{\sqrt{s}(1+\xi)}{M}\right]\;.
\label{boost}
\end{eqnarray}
The incoming photon flies almost towards the $-z$ axis, in the light-cone and in the rest frame, so that the scattering angle 
$\theta$
of the produced photon in the nucleon rest frame with respect to this direction satisfies 
\begin{eqnarray}
\label{tan_theta_rf-Delta}
\tan \theta = - \frac{2M s(1+\xi)\alpha\parallel \pv - \frac{\dv}{2}\parallel}{  -\alpha^2 (1+\xi)^2 s^2 +(\pv - \frac{\dv}{2})^2 M^2 }\;.
\end{eqnarray}
Using the relation $\alpha = M^2_{\gamma \rho}/(-u')$, see Eq.~(\ref{t'-u'-Bjorken}), one gets from this expression $\tan \theta$ as a function of $-u'$, which we formally write
\beqa
\label{def:f}
\tan \theta = f(-u')\,.
\eqa
%
%
\psfrag{H}{\hspace{-1.5cm}\raisebox{-.6cm}{\scalebox{.7}{$\theta$}}}
\psfrag{V}{\raisebox{.3cm}{\scalebox{.7}{$\hspace{-.4cm}\displaystyle\frac{1}{\sigma_{odd}}\frac{d \sigma_{odd}}{d \theta}$}}}
\begin{figure}[h!]
\begin{center}
\includegraphics[width=7cm]{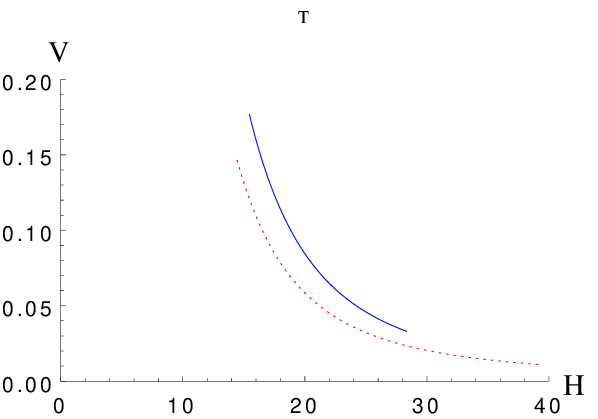}
\includegraphics[width=7cm]{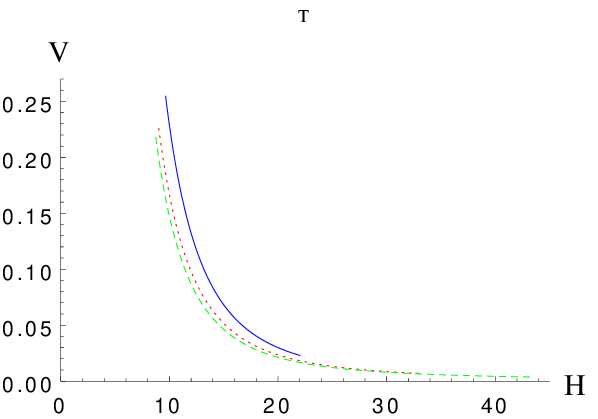}

\vspace{.8cm}
\includegraphics[width=7cm]{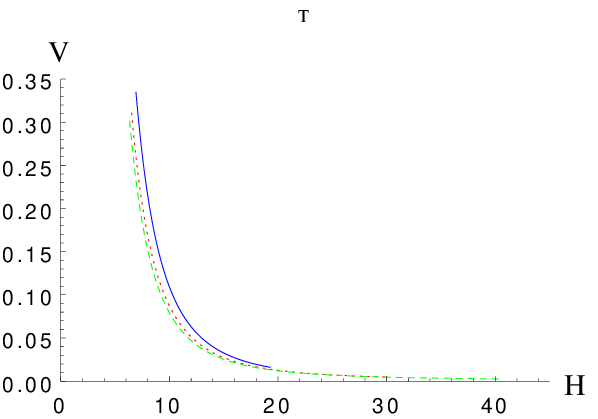}
\vspace{.4cm}
\caption{Angular distribution in the chiral-odd case.
Up, left: $S_{\gamma N}=10~{\rm GeV}^2$, for $M^2_{\gamma \rho}=
3~{\rm GeV}^2$ (solid blue) and  $M^2_{\gamma \rho}=
4~{\rm GeV}^2$ (dotted red).  Up, right:
$S_{\gamma N}=15~{\rm GeV}^2$, for $M^2_{\gamma \rho}=
3.5~{\rm GeV}^2$ (solid blue), $M^2_{\gamma \rho}=
5~{\rm GeV}^2$ (dotted red) and 
$M^2_{\gamma \rho}=
6.5~{\rm GeV}^2$ (dashed green).
Down: 
$S_{\gamma N}=20~{\rm GeV}^2$, for $M^2_{\gamma \rho}=
4~{\rm GeV}^2$ (solid blue), $M^2_{\gamma \rho}=
6~{\rm GeV}^2$ (dotted red) and 
$M^2_{\gamma \rho}=
8~{\rm GeV}^2$ (dashed green).}
\label{Fig:thetacut-odd}
\end{center}
\end{figure}
From this relation, $\theta$ being positive, one should take
\beqa
\label{theta_rf1}
{\rm for} \ \tan \theta>0, \quad \theta &=& \arctan(\tan \theta),\\
\label{theta_rf2}
{\rm for} \ \tan \theta<0, \quad  \theta &=& \pi + \arctan(\tan\theta)\,,
\eqa
where $\tan \theta$ is given by Eq.~(\ref{tan_theta_rf-Delta}).

For simplicity, we now perform our analysis in the case $\dv=0,$ and thus
write
\begin{eqnarray}
\label{tan_theta}
\tan \theta = - \frac{2M s(1+\xi) \alpha \, p_t }{  -\alpha^2 (1+\xi)^2 s^2 +\pv^{\,2} M^2 },
\end{eqnarray}
where $p_t=\parallel \!\pv \!\parallel .$

Using the formulas given in Sec.~\ref{SubSec:approximated_kinematics},
one can compute $\alpha$ as a function of $\theta\,.$
One gets
\beqa
\label{alpha-theta1}
{\rm for} \ \tan \theta>0, \quad \alpha &=& \frac{(1+\xi+ \tilde{\tau}) \, \tilde{\tau} \, \tan^2 \theta + a\left(1+\sqrt{1+\tan^2 \theta}\right)}{(1+\xi+ \tilde{\tau})^2 \tan^2 \theta+2a} ,\\
\label{alpha-theta2}
{\rm for} \ \tan \theta<0, \quad \alpha &=& \frac{(1+\xi+ \tilde{\tau}) \, \tilde{\tau} \, \tan^2 \theta + a\left(1-\sqrt{1+\tan^2 \theta}\right)}{(1+\xi+ \tilde{\tau})^2 \tan^2 \theta+2a} ,
\eqa
where
\beqa
\label{def:a_G_H}
a &=& \frac{4 M_{\gamma \rho}^2}s \,,\\
\tilde{\tau} &=& \frac{2 \xi}{1+\xi} \frac{M_{\gamma \rho}^2}s = \tau \frac{M_{\gamma \rho}^2}s\,,
\eqa
thus providing $-u'$ as a function of $\theta$ using $-u'= \alpha M^2_{\gamma \rho}$, see Eq.~(\ref{t'-u'-Bjorken}).

The angular distribution of the produced photon can easily be obtained from the differential cross-section by using the relation
\beqa
\label{dsigma-du'}
\frac{d \theta}{d (-u')}= \frac{f'(-u')}{1+f^2(-u')}
\eqa
so that we get 
\beqa
\label{dsigma-dtheta}
\frac{1}{\sigma}\frac{d \sigma}{d \theta}= \frac{1}{\sigma}\frac{d \sigma}{d (-u')} \frac{d (-u')}{d \theta}=\frac{1}{\sigma}
\frac{d \sigma}{d (-u')} \frac{1+f^2(-u'[\theta])}{f'(-u'[\theta])}\,.
\eqa

The obtained angular distribution is shown in Fig.~\ref{Fig:thetacut-even} for the chiral-even case, and in Fig.~\ref{Fig:thetacut-odd}
for the chiral-odd case. In the chiral-even case, the obtained angular distribution is an increasing 
function of $\theta,$ while in the chiral-odd case, it decreases with increasing $\theta.$ In both cases, the distributions are dominated by moderate values of $\theta.$

%
\psfrag{H}{\hspace{-1.5cm}\raisebox{-.6cm}{\scalebox{.7}{$M^2_{\gamma \rho}~({\rm GeV}^{2})$}}}
\psfrag{V}{\raisebox{.3cm}{\scalebox{.7}{$\hspace{-.4cm}\displaystyle\frac{d \sigma_{\rm even}}{d M^2_{\gamma \rho}}~({\rm pb} \cdot {\rm GeV}^{-2})$}}}
\begin{figure}[h!]
\begin{center}
\includegraphics[width=7cm]{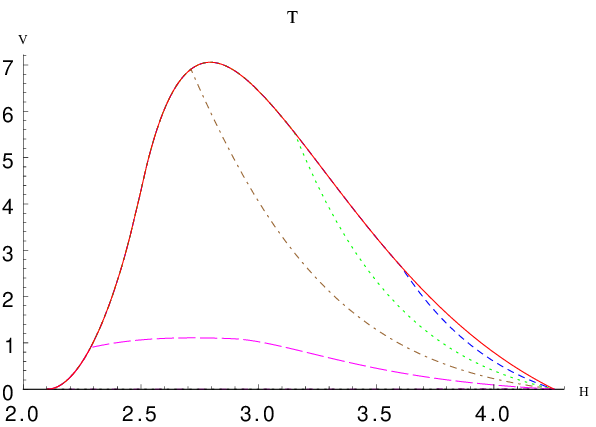}
\includegraphics[width=7cm]{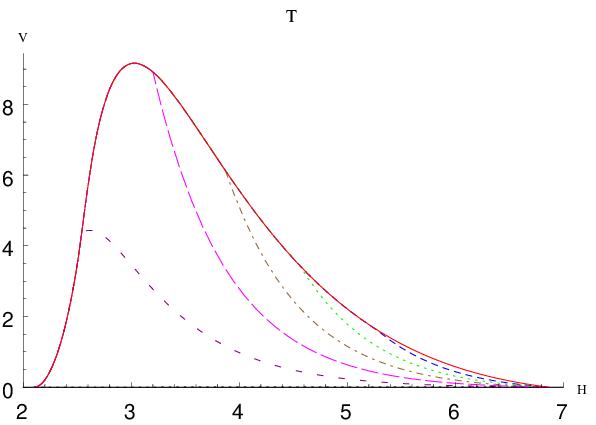}

\vspace{.8cm}
\includegraphics[width=7cm]{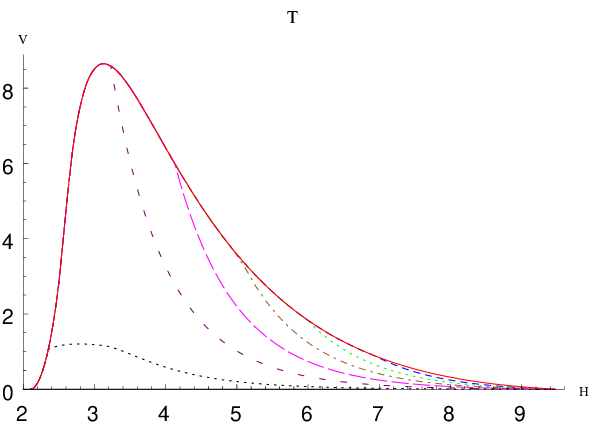}
\vspace{.4cm}
\caption{The differential cross section
$\frac{d \sigma_{\rm even}}{d M^2_{\gamma \rho}}$.
Solid red: no angular cut. Other curves show the effect of an upper angular cut $\theta$ for the out-going $\gamma$: $35\degree$ (dashed blue), $30\degree$ (dotted green),  $25\degree$ (dashed-dotted brown), $20\degree$ (long-dashed magenta),  $15\degree$ (short-dashed purple) and $10\degree$ (dotted black).
Up, left: $S_{\gamma N}=10~{\rm GeV}^2$.  Up, right:
$S_{\gamma N}=15~{\rm GeV}^2$.
Down: 
$S_{\gamma N}=20~{\rm GeV}^2$.
}  
\label{Fig:dsigmathetacut-even}
\end{center}
\end{figure}
%
%
%
\psfrag{H}{\hspace{-1.5cm}\raisebox{-.6cm}{\scalebox{.7}{$M^2_{\gamma \rho}~({\rm GeV}^{2})$}}}
\psfrag{V}{\raisebox{.3cm}{\scalebox{.7}{$\hspace{-.4cm}\displaystyle\frac{d \sigma_{\rm odd}}{d M^2_{\gamma \rho}}~({\rm pb} \cdot {\rm GeV}^{-2})$}}}
\begin{figure}[h!]
\begin{center}
\includegraphics[width=7cm]{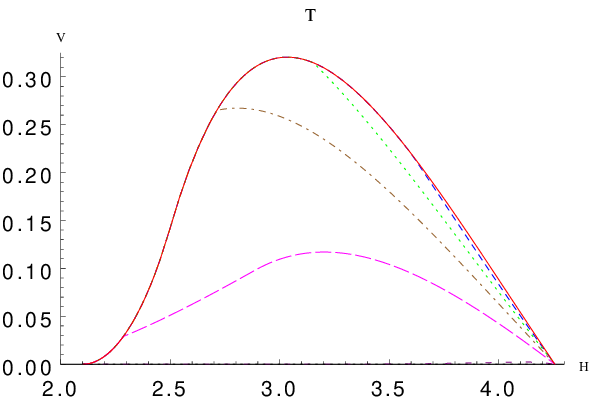}
\includegraphics[width=7cm]{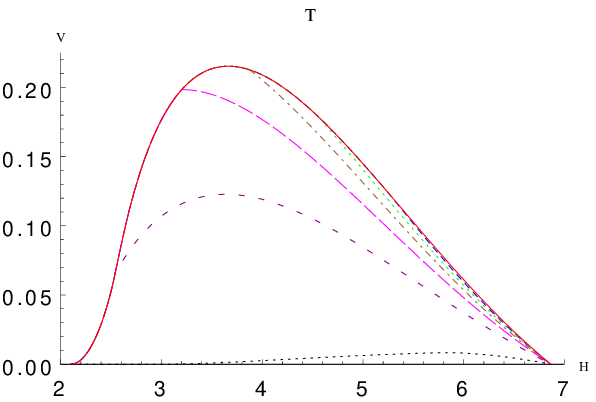}

\vspace{.8cm}
\includegraphics[width=7cm]{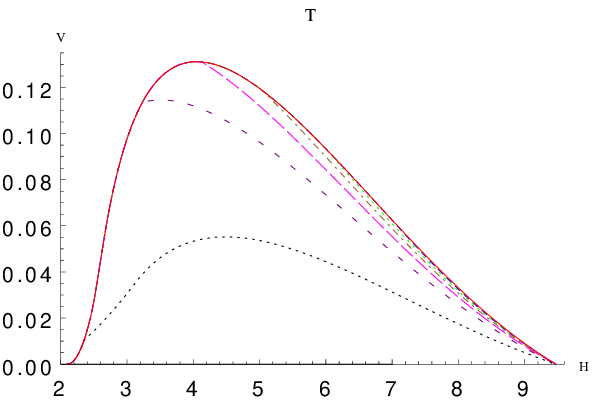}
\vspace{.4cm}
\caption{The differential cross section
$\frac{d \sigma_{\rm odd}}{d M^2_{\gamma \rho}}$. Solid red: no angular cut. Other curves show the effect of an upper angular cut $\theta$ for the out-going $\gamma$: $35\degree$ (dashed blue), $30\degree$ (dotted green),  $25\degree$ (dashed-dotted brown), $20\degree$ (long-dashed magenta),  $15\degree$ (short-dashed purple) and $10\degree$ (dotted black).
Up, left: $S_{\gamma N}=10~{\rm GeV}^2$.  
Up, right:
$S_{\gamma N}=15~{\rm GeV}^2$.
Down: 
$S_{\gamma N}=20~{\rm GeV}^2$.
}  
\label{Fig:dsigmathetacut-odd}
\end{center}
\end{figure}

In practice, at JLab, in Hall B, 
the outgoing photon could be detected with an angle between $5\degree$ and $35\degree$ from the incoming beam. 

The effect of an upper angular cut can be seen in Fig.~\ref{Fig:dsigmathetacut-even} for the chiral-even case, and in Fig.~\ref{Fig:dsigmathetacut-odd} for the chiral-odd case. As seen from Figs.~\ref{Fig:thetacut-even}  and \ref{Fig:thetacut-odd}, it mainly affects the low $S_{\gamma N}$
domain. In particular, the effect of the JLab $35\degree$ upper cut remains negligible as shown in Figs.~\ref{Fig:dsigmathetacut-even} and \ref{Fig:dsigmathetacut-odd}, both for the chiral-even and chiral-odd cases. 

One should note that using cuts on $\theta$, it is possible to reduce dramatically the contribution of the chiral-even contribution, in particular in the region of $S_{\gamma N}$ around 20 GeV$^2$, while moderately reducing the chiral-odd contribution. Putting additional cuts on $M^2_{\gamma \rho}$, like  $M^2_{\gamma \rho} > 6~{\rm GeV}^2,$
allows to increase the ratio odd versus even from $\sim$ 1/25 to $\sim$ 2/3, keeping about 3\% of the chiral-odd contribution, for typically $S_{\gamma N}$ between 18 GeV$^2$ and the maximal value 21.5 GeV$^2.$
This in principle would lead, dealing with observables sensitive to the interference between the chiral-odd and the chiral-even contributions, to a relative signal
of the order of 80\%.


\providecommand{\href}[2]{#2}\begingroup\raggedright\endgroup

\end{document}